\documentclass[aps,prd,twocolumn,superscriptaddress,nofootinbib,10pt]{revtex4-1}

\pdfoutput=1
\usepackage[utf8]{inputenc}
\usepackage{graphicx,amsmath,amsfonts,amssymb,aas_macros,slashed}
\usepackage{bbold}
\usepackage{datetime}
\usepackage{numprint}
\usepackage{enumitem}
 
\usepackage{graphicx,color,amsmath}
\usepackage{hyperref}

\allowdisplaybreaks

\newcommand{\cm}{\ensuremath{\mathrm{cm}}}

\newcommand{\keV}{\ensuremath{\mathrm{keV}}}
\newcommand{\MeV}{\ensuremath{\mathrm{MeV}}}
\newcommand{\GeV}{\ensuremath{\mathrm{GeV}}}
\newcommand{\TeV}{\ensuremath{\mathrm{TeV}}}

\newcommand{\ket}[1]{\ensuremath   | #1 \rangle   }

\newcommand{\bra}[1]{\ensuremath   \langle #1 |   }

\newcommand{\me}[3]{\ensuremath \langle #1 | #2 | #3 \rangle   }

\DeclareMathOperator{\imag}{Im}
\DeclareMathOperator{\Tr}{Tr}

\usepackage{pgfplots}
\usepackage{bm}
\usepackage{mathtools}

\let\oldhat\hat
\let\vvec\vec
\renewcommand{\vec}[1]{\bm{#1}}
\renewcommand{\hat}[1]{\oldhat{\bm{#1}}}

\newcommand{\hham}{{\mathcal{H}}}

\newcommand{\sdark}{s_{\chi\bar\chi}} %
\newcommand{\sXdark}{s_{3\chi\bar\chi}} %
\newcommand{\Rdark}{R\chi\bar\chi} %
\newcommand{\RXdark}{R3\chi\bar\chi} %
\newcommand{\shadr}{s_X} %

\newcommand{\tAD}{t_{14}} %
\newcommand{\tBF}{t_{2\chi}} %
\newcommand{\MCDE}{s_{34\bar \chi}} %
\newcommand{\MDE}{s_{4\bar \chi}} %
\newcommand{\qBCF}{q_{23\chi}} %

\usetikzlibrary{positioning,arrows,patterns}
\usetikzlibrary{decorations.markings}
\usetikzlibrary{calc}

\tikzset{
    photon/.style={decorate, decoration={snake}, draw=black},
    vector/.style={decorate, decoration={snake}, draw},
	provector/.style={decorate, decoration={snake,amplitude=2.5pt}, draw},
	antivector/.style={decorate, decoration={snake,amplitude=-2.5pt}, draw},
    fermion/.style={draw=black, postaction={decorate},
        decoration={markings,mark=at position .55 with {\arrow[draw=black]{>}}}},
    fermionbar/.style={draw=black, postaction={decorate},
        decoration={markings,mark=at position .55 with {\arrow[draw=black]{<}}}},
    fermionnoarrow/.style={draw=black},
    gluon/.style={decorate, draw=black,
        decoration={coil,amplitude=4pt, segment length=5pt}},
    scalar/.style={dashed,draw=black, postaction={decorate},
        decoration={markings,mark=at position .55 with {\arrow[draw=black]{>}}}},
    scalarbar/.style={dashed,draw=black, postaction={decorate},
        decoration={markings,mark=at position .55 with {\arrow[draw=black]{<}}}},
    scalartwo/.style={dotted,draw=black, postaction={decorate},
        decoration={markings,mark=at position .55 with {\arrow[draw=black]{>}}}},
    scalartwobar/.style={dotted,draw=black, postaction={decorate},
        decoration={markings,mark=at position .55 with {\arrow[draw=black]{<}}}},
    scalarnoarrow/.style={dashed,draw=black},
    electron/.style={draw=black, postaction={decorate},
        decoration={markings,mark=at position .55 with {\arrow[draw=black]{>}}}},
	bigvector/.style={decorate, decoration={snake,amplitude=4pt}, draw},
    vertex/.style={draw,shape=circle,fill=black,minimum size=1pt,inner sep=0pt},
    fermion2/.style={double, draw=black, postaction={decorate},
		decoration={markings,mark=at position .55 with {\arrow[draw=black]{>}}}},
    momentum/.style={draw=black,line width=0.15mm, postaction={decorate},
        decoration={markings,mark=at position 1 with {\arrow[draw=black]{>}}}}
}

\begin{document}

\title{Light dark states with electromagnetic form factors}

\author{Xiaoyong Chu}
\email{xiaoyong.chu@oeaw.ac.at}
\affiliation{Institute of High Energy Physics, Austrian Academy of Sciences, Nikolsdorfergasse 18, 1050 Vienna, Austria}
\author{Josef Pradler}
\email{josef.pradler@oeaw.ac.at}
\affiliation{Institute of High Energy Physics, Austrian Academy of Sciences, Nikolsdorfergasse 18, 1050 Vienna, Austria}
\author{Lukas Semmelrock}
\email{lukas.semmelrock@oeaw.ac.at}
\affiliation{Institute of High Energy Physics, Austrian Academy of Sciences, Nikolsdorfergasse 18, 1050 Vienna, Austria}
\affiliation{Department of Physics and Astronomy, Johns Hopkins University,
3400 N. Charles St., Baltimore, MD 21218, USA}

\begin{abstract}
  New particles $\chi$ that are electrically neutral but couple to the
  electromagnetic current via higher-dimensional operators and that
  are sufficiently light, at or below the GeV-mass scale, can be
  produced in pairs in a number of dedicated high-intensity
  experiments. In this work we consider the production of $\chi$
  through  magnetic- and electric-dipole moments as well as through
  anapole moment and charge radius interactions in electron beams. We
  derive new constraints from BaBar, NA64 and mQ and forecast the
  future sensitivity on the existence of such states, from Belle-II,
  LDMX and BDX.  We present for the first time a detailed treatment of
  the off-shell production of photons in electron beams with
  subsequent decay into a $\chi\bar\chi$ pair in a 2-to-4
  process. These direct limits are then compared to the effects on SM
  precision observables, as well as to bounds from flavor physics and
  high energy colliders.  Finally, we consider the possibility that
  $\chi$ is dark matter and study ensuing astrophysical and
  cosmological constraints. We find that a combination of all
  considered probes rule out $\chi$ particles with mass-dimension five
  and six photon interactions as dark matter when assuming a standard
  freeze-out abundance.
\end{abstract}

\maketitle

\section{Introduction}
\label{sec:introduction}

Through a long history of astronomical and cosmological observations
that date back into the first half of the previous century, it has now
firmly been established that Newton's laws --- when applied to the
observed distribution of luminous matter --- fail on galactic (kpc)
length scales and beyond.  Whereas there is an ongoing debate whether
the gravitational force on small acceleration scale requires
modification or if the Standard Model (SM) of particle physics is
incomplete, predictions that are based on the existence of a new,
neutral, and cold matter component stand unchallenged in
explaining the precision observations of the cosmic microwave
background (CMB). Indeed, the success of a model in which the Universe
is filled with baryons, photons, neutrinos, dark matter (DM), and dark
energy is so distinct that it has become a cornerstone of modern
physics, termed the ``standard cosmological model'' or
$\Lambda$CDM~\cite{Hinshaw:2012aka,Ade:2015xua}. This new paradigm
that purports the existence of a new form of matter, the dark matter, has become the center of much experimental and theoretical
activity~\cite{Bertone:2004pz}.

Even if gravity as it stands today will pass the test of time, new
forces are likely at play in a dark sector of particles.  While the
electromagnetic interaction and the photon as its carrier stand out as
the most important messengers in astronomy, dark
matter --- as the name suggests --- must be to large degree electrically neutral and hence be
non-luminous.  However, even if DM carries no charge, it may still be
coupled to the photon \textit{e.g.}~through various moments such as
magnetic- or electric- dipole moments (MDM or EDM), through an anapole
moment (AM) or through a charge radius interaction (CR). Indeed, ``how
dark'' neutral DM needs to be in terms of its coupling to
photons is a quantifiable question that has been addressed previously
in \cite{Pospelov:2000bq,Sigurdson:2004zp,Ho:2012bg}. In this work we
will revisit those ideas in light of the much recent interest in
sub-GeV dark sector searches, notably at the intensity
frontier~\cite{Essig:2013lka,Battaglieri:2017aum}. This interest not
only spans DM detection, but more generally aims to probe light new
particles and in this spirit we do not require that $\chi$ is DM
\textit{per
  se} but only long-lived on time-scales pertinent to terrestrial experiments.

In this paper we shall consider Dirac fermions $\chi$ that interact
with an external electromagnetic field or current through MDM, EDM, AM 
and CR. In the non-relativistic limit, the corresponding 
 respective Hamiltonian operators for the particle $\chi$ are
$\hham_\text{MDM}=- \mu_{\chi} (\vec B \cdot \vec \sigma_{\chi})$, 
$\hham_\text{EDM}=- d_{\chi} (\vec E \cdot \vec \sigma_{\chi} )$,
$\hham_\text{AM}= - a_{\chi} (\vec J \cdot \vec \sigma_{\chi}) $ and 
$\hham_\text{CR}=- b_{\chi}( \vec \nabla \cdot \vec E)$. 
We will refer to  $\mu_{\chi}$, $d_{\chi}$, and $a_{\chi}$ as the magnetic
dipole\mbox{-},  electric dipole\mbox{-}, and anapole moment, respectively, and to
$b_{\chi}$ as the charge radius. These operators differ most notably
in the symmetry properties with respect to space reflection (P),
charge conjugation (C), and time-reversal (T).
Since the magnetic field $\vec B$ and the spin
$\vec S_{\chi} = \vec\sigma_\chi/2$ are axial vectors, an MDM
interaction is P- and T-even. Hence CP is a good symmetry, and, as is
inherent in the field-theoretical description of particles, any state
$\chi$ that has electromagnetic charge $Q_{\chi}$ will automatically
carry a magnetic moment
$\vec\mu = g Q_{\chi} /( 2 m_{\chi} )\,\vec S_\chi$ with $g=2$
up to anomalous contributions; $m_{\chi}$ is the mass of $\chi$.
In turn, any EDM interaction with the electric field $\vec E$
maximally violates P and T and is a new source of flavor-diagonal CP
violation --- one that has not been observed in any SM particle
yet. Hence, searches for EDMs are considered essentially ``background
free'' probes of new physics~\cite{Pospelov:2005pr} and any detection
of a dark particle through its EDM would be particularly
striking. If $\chi$ has a non-vanishing anapole moment, it will interact
with external electromagnetic currents $\vec J$; in that case $\chi$
may even be a Majorana particle. Unlike MDM and EDM, AM does not
correspond to a multipolar distribution of charge. It was first
proposed by Zel'dovich~\cite{1958JETP....6.1184Z} as a P-violating (but
CP conserving) electromagnetic interaction, and has since been
discovered in atomic nuclei~\cite{Wood:1997zq}. Finally, the CR
interaction appears at the second order in the expansion of a charge form
factor, and as such it is a quantity \textit{e.g.}~well present in composite
particles of the SM, and it is a searched for property of
neutrinos~\cite{Giunti:2014ixa}. Its transformation property with
respect to discrete symmetries is one of a scalar.

The effective interactions of neutral $\chi$ with the photon field may
arise in a variety of UV-complete descriptions. An imminent
possibility is that $\chi$ is a composite particle so that a dipole
moment arises through the particle's internal structure, such as,
\textit{e.g.}, in technicolor
theories~\cite{Bagnasco:1993st,Foadi:2008qv,Antipin:2015xia}. Another possibility
are frameworks with an extended set of particles such that
electromagnetic moments are produced perturbatively, by the loops of
charged states~\cite{Raby:1987ga}. Here, a sizable MDM can
be generated \textit{e.g.}~through axial or vector type Yukawa
interactions $y_{A,V}$ with a scalar and a fermion at a common
mass-scale $M$ in the loop, so that parametrically
$\mu_{\chi} \sim Q |y_{A,V}|^2/M$ where $Q$ is the electric charge of 
the mediators. The appearance of an EDM would then of course be tightly connected to
CP violation in the new physics sector and
$d_{\chi}\sim Q \imag [y_V y_A^{*}]/M$.
Finally, AM and CR interactions may also arise radiatively; in the
above example, $a_{\chi},\, b_{\chi} \sim Q |y_{A,V}|^2/ M^2$,
possibly be enhanced through a DM-mediator
mass-degeneracy~\cite{Pospelov:2008qx}.
Indeed, a significant amount of attention has been directed to the
phenomenology of those interactions, see also
\cite{Schmidt:2012yg,Kopp:2014tsa,Ibarra:2015fqa,Sandick:2016zut,Kavanagh:2018xeh}
besides the above mentioned works and references therein. In this
work, we shall remain agnostic about the origin of the~MDM, EDM, AM 
and CR of $\chi$.  While any embedding into a concrete setup likely
implies further constraints on the existence of the effective values
of $ \mu_{\chi}\,,  d_{\chi}$, $a_{\chi}$ and $b_{\chi}$, this way we
remain model-independent.

Much of the effort on the intensity frontier goes into the search of
light, sub-GeV mass dark sector states. New particles can be produced
in electron or positron beams on fixed targets, at $e^+e^-$ colliders,
in hadron beams from protons on fixed targets or in proton-proton
collisions at LHC.
Here, much attention has been devoted to the on-shell or the
$s$-channel resonant production of particles that either decay to
electrons or muons inside the sensitive detector volume or escape or
decay invisibly. A prominent example is the search for dark
photons~\cite{Essig:2009nc,Reece:2009un,Bjorken:2009mm,Batell:2009di}
in fixed target experiments, \textit{e.g.}~at
APEX~\cite{Abrahamyan:2011gv} or NA48/2~\cite{Batley:2015lha} and at
$e^+e^-$ colliders, \textit{e.g.}~at BaBar~\cite{Lees:2014xha} just to
 name a few; for a recent comprehensive compilation
see~\cite{Alexander:2016aln}.

In this work we study the pair-production of $\chi\bar\chi$ through
the MDM, EDM, AM  and CR interaction with photons in experiments involving
\textit{electrons} as projectiles. For dark sector experiments, there are
three principal detection methods:
\begin{enumerate}[leftmargin=*] %
\item search for missing momentum in $e^+e^-$ colliders such as
  BaBar~\cite{Aubert:2001tu}, Belle~\cite{Abashian:2000cg},
  Belle-II~\cite{Abe:2010gxa} and BESIII~\cite{Ablikim:2009aa},
\item search for missing energy in $e^-$ fixed-target experiments such as
  NA64~\cite{Banerjee:2017hhz} and the proposed LDMX
  detector~\cite{Akesson:2018vlm},%
  \footnote{In an analogous category, positron fixed-target
    experiments include PADME~\cite{Raggi:2014zpa} and
    VEPP-3~\cite{Wojtsekhowski:2009vz,Wojtsekhowski:2012zq}.}
\item direct search for $\chi N$ or $\chi e^-$ scattering of $\chi$
  particles produced in $e^-$ fixed-target experiments such as the
  previous mQ~\cite{Prinz:1998ua} detector and E137 at
  SLAC~\cite{Bjorken:1988as}, as well as the proposed beam dump
  experiments at JLab (BDX)~\cite{Battaglieri:2016ggd} or at the MESA
  accelerator~\cite{Doria:2018sfx}.
\end{enumerate}
We consider all three detection methods in turn, work out current
constraints and estimate future sensitivity. As the production in
fixed targets proceeds via the emission of off-shell photons, we also
present a detailed exposition of the underlying 2-to-4 scattering
processes, thereby filling a gap in the existing literature on BSM
intensity frontier searches.

We then contrast these direct limits with predictions and constraints
from SM precision observables and from limits originating from flavor
physics. The experimental searches for dark states on the intensity
frontier have in no little part been motivated by the observation that
light new physics can induce a shift in the anomalous magnetic moment
of the muon, see
\textit{e.g.}~\cite{Gninenko:2001hx,Pospelov:2008zw,Chen:2015vqy,Batell:2016ove}
among other works, and hence explain a long-standing discrepancy
between the SM prediction and experiment~\cite{Jegerlehner:2009ry}. We
investigate this possibility in the light of dark-sector
electromagnetic form factors and finally, we return to the possibility
that $\chi$ is cosmologically long-lived and may constitute dark
matter.  In this case, further restrictions on the
  parameter space of $\chi$ from cosmological and astrophysical
  observations take hold --- such as energy injection to the SM
  sector, altering the CMB or influencing structure formation, as well
  as a local dark matter abundance that could lead to direct and
  indirect detection signals.

The paper is organized as follows. In Sec.~\ref{sec:electr-form-fact}
we present the relativistic versions of the interaction operators for
MDM, EDM, AM and CR.
In Sec.~\ref{sec:pair-prod-electr} we present our cross-section
calculations on the production of $\chi\bar\chi$ in $e^+ e^-$ collisions as well as $e^-$-scattering on a nuclear target with details
relegated to Appendix~\ref{sec:matrix-el}. In
Sec.~\ref{sec:intens-front-search} we then specialize to the various
experiments, work out constraints and sensitivity projections and
discuss other complementary probes. In Sec.~\ref{sec:precision-tests}
we discuss indirect probes of $\chi$-particles and in
Sec.~\ref{sec:dark-matt-electr} we comment on the possibility that
$\chi$ is DM and highlight the additional restrictions on a successful
model of DM. In Sec.~\ref{sec:conclusions} we present our conclusions
and provide an outlook to further going works.

\section{Electromagnetic form factor interactions}
\label{sec:electr-form-fact}

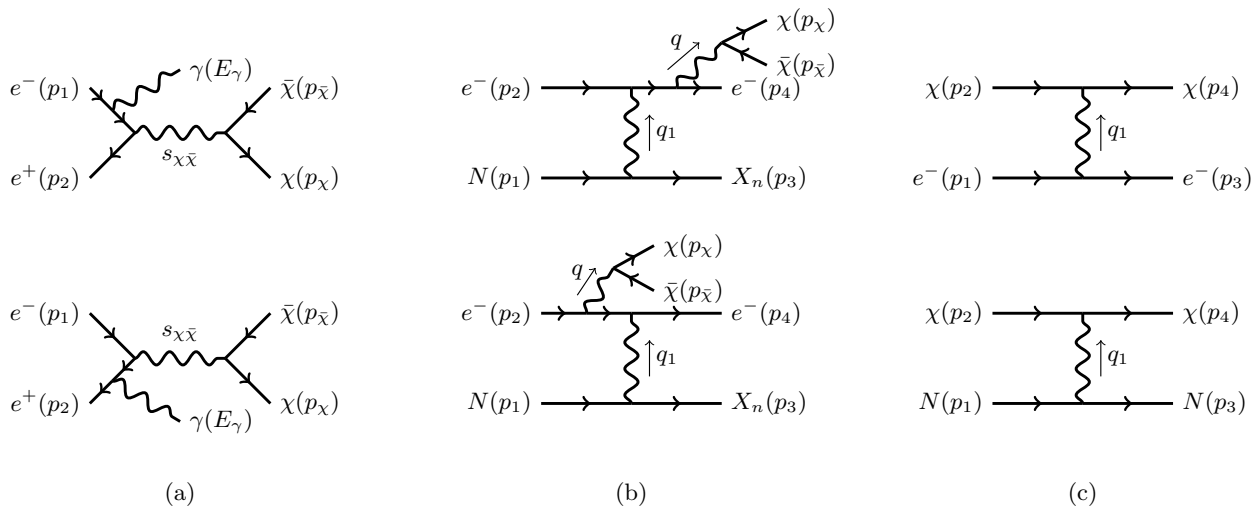
\begin{figure*}[ht]
  \centering
\begin{tikzpicture}[line width=1.1 pt, scale=1.2]

\begin{scope}[shift={(0,-0.5)}]
\coordinate[label=above : ] (v1) at (1.5,0);
		\coordinate (v3) at (0.25,0.25);
		\coordinate[label=right : $\gamma(E_\gamma)$] (v4) at (1,0.7);
		\coordinate[label=above : ] (v2) at (0.5,0);
		\coordinate[label=right : $\bar\chi(p_{\bar\chi})$] (e1) at (2,0.5);
		\coordinate[label=left : $e^-(p_1)$] (e2) at (0,0.5);
		\coordinate[label=right : $\chi(p_\chi)$] (e3) at (2,-0.5);
		\coordinate[label=left : $e^+(p_2)$] (e4) at (0,-0.5);
		\draw[fermion] (e1) -- (v1);
		\draw[fermion] (e2) -- (v3);
		\draw[fermion] (v3) -- (v2);
		\draw[fermion] (v1) -- (e3);
		\draw[fermion] (v2) -- (e4);
		\draw[vector] (v2) -- (v1) node[midway,below=0.1cm] {$\sdark$};
		\draw[vector] (v3) -- (v4);
\end{scope}

\begin{scope}[shift={(0,-3)}]
\coordinate[label=above : ] (v1) at (1.5,0);
		\coordinate (v3) at (0.25,-0.25);
		\coordinate[label=right : $\gamma(E_\gamma)$] (v4) at (1,-0.7);
		\coordinate[label=above : ] (v2) at (0.5,0);
		\coordinate[label=right : $\bar\chi(p_{\bar\chi})$] (e1) at (2,0.5);
		\coordinate[label=left : $e^-(p_1)$] (e2) at (0,0.5);
		\coordinate[label=right : $\chi(p_\chi)$] (e3) at (2,-0.5);
		\coordinate[label=left : $e^+(p_2)$] (e4) at (0,-0.5);
		\draw[fermion] (e1) -- (v1);
		\draw[fermion] (e2) -- (v2);
		\draw[fermion] (v2) -- (v3);
		\draw[fermion] (v1) -- (e3);
		\draw[fermion] (v3) -- (e4);
		\draw[vector] (v2) -- (v1) node[midway,above=0.1cm] {$\sdark$};
		\draw[vector] (v3) -- (v4);

\draw (1,-1.5) node{(a)};
\end{scope}

\begin{scope}[shift={(5,0)}]
\coordinate (v1) at (1,0);
\coordinate (v2) at (1,-1);
\coordinate (v3) at (1.5,0);
\coordinate  (v4) at (2,0.5);
\coordinate[label=left :$e^-(p_2)$]  (e1) at (0,0);
\coordinate[label=left :$N(p_1)$]  (e2) at (0,-1);
\coordinate[label=right :$e^-(p_4)$] (e3) at (2,0);
\coordinate[label=right :$X_n(p_3)$] (e4) at (2,-1);
\coordinate[label=right :$\chi(p_\chi)$]  (e5) at (2.5,0.75);
\coordinate[label=right :$\bar\chi(p_{\bar\chi})$]  (e6) at (2.5,0.25);
\draw[fermion] (e1) -- (v1);
\draw[fermion] (e2) -- (v2);
\draw[fermion] (v1) -- (v3);
\draw[fermion] (v3) -- (e3);
\draw[fermion] (v2) -- (e4);
\draw[fermion] (v4) -- (e5);
\draw[fermionbar] (v4) -- (e6);
\draw[photon] (v2) -- (v1) node[midway,right=0.2cm] {$q_1$};
\coordinate (m1) at (1.2,-0.7);
\coordinate (m2) at (1.2,-0.3);
\draw[momentum] (m1)-- (m2);

\draw[photon] (v3) -- (v4) ;
 \coordinate (m3) at (1.4,0.2);
 \coordinate (m4) at (1.75,0.5);
 \draw[momentum] (m3)-- (m4) ;
 \draw (1.5,0.5) node{$q$};

\end{scope}

\begin{scope}[shift={(5,-2.5)}]
\coordinate (v1) at (1,0);
\coordinate (v2) at (1,-1);
\coordinate (v3) at (0.5,0);
\coordinate  (v4) at (0.8,0.5);
\coordinate[label=left :$e^-(p_2)$]  (e1) at (0,0);
\coordinate[label=left :$N(p_1)$]  (e2) at (0,-1);
\coordinate[label=right :$e^-(p_4)$] (e3) at (2,0);
\coordinate[label=right :$X_n(p_3)$] (e4) at (2,-1);
\coordinate[label=right :$\chi(p_\chi)$]  (e5) at (1.25,0.75);
\coordinate[label=right :$\bar\chi(p_{\bar\chi})$]  (e6) at (1.25,0.25);
\draw[fermion] (e1) -- (v3);
\draw[fermion] (e2) -- (v2);
\draw[fermion] (v1) -- (e3);
\draw[fermion] (v2) -- (e4);
\draw[fermion] (v3) -- (v1);
\draw[fermion] (v4) -- (e5);
\draw[fermionbar] (v4) -- (e6);
\draw[photon] (v2) -- (v1) node[midway,right=0.2cm] {$q_1$};
\draw[photon] (v3) -- (v4) ;
\coordinate (m1) at (1.2,-0.7);
\coordinate (m2) at (1.2,-0.3);
\draw[momentum] (m1)-- (m2);
 \coordinate (m3) at (0.4,0.2);
 \coordinate (m4) at (0.6,0.5);
 \draw[momentum] (m3)-- (m4) ;
 \draw (0.4,0.45) node{$q$};

\draw (1,-2) node{(b)};

\end{scope}

\begin{scope}[shift={(10,0)}]
\coordinate (v1) at (1,0);
\coordinate (v2) at (1,-1);
\coordinate (v3) at (1.5,0);
\coordinate  (v4) at (2,0.5);
\coordinate[label=left :$\chi(p_2)$]  (e1) at (0,0);
\coordinate[label=left :$e^-(p_1)$]  (e2) at (0,-1);
\coordinate[label=right :$\chi(p_4)$] (e3) at (2,0);
\coordinate[label=right :$e^-(p_3)$] (e4) at (2,-1);
\draw[fermion] (e1) -- (v1);
\draw[fermion] (e2) -- (v2);
\draw[fermion] (v1) -- (e3);
\draw[fermion] (v2) -- (e4);
\draw[photon] (v2) -- (v1) node[midway,right=0.2cm] {$q_1$};
\coordinate (m1) at (1.2,-0.7);
\coordinate (m2) at (1.2,-0.3);
\draw[momentum] (m1)-- (m2);

\end{scope}

\begin{scope}[shift={(10,-2.5)}]
\coordinate (v1) at (1,0);
\coordinate (v2) at (1,-1);
\coordinate (v3) at (0.5,0);
\coordinate  (v4) at (0.8,0.5);
\coordinate[label=left :$\chi(p_2)$]  (e1) at (0,0);
\coordinate[label=left :$N(p_1)$]  (e2) at (0,-1);
\coordinate[label=right :$\chi(p_4)$] (e3) at (2,0);
\coordinate[label=right :$N(p_3)$] (e4) at (2,-1);
\draw[fermion] (e1) -- (v1);
\draw[fermion] (e2) -- (v2);
\draw[fermion] (v1) -- (e3);
\draw[fermion] (v2) -- (e4);
\draw[photon] (v2) -- (v1) node[midway,right=0.2cm] {$q_1$};
\coordinate (m1) at (1.2,-0.7);
\coordinate (m2) at (1.2,-0.3);
\draw[momentum] (m1)-- (m2);
\draw (1,-2) node{(c)};
\end{scope}

\end{tikzpicture}
\caption{Pair production of $\bar\chi\chi$ in (a) $e^+e^-$ collisions
  and (b) in electron scattering on a target nucleus $N$. The mQ and
  BDX experiments search for (c) the elastic scattering of $\chi$ on
  electrons and nuclei. Momenta flow from left (bottom) to right (top).}
  \label{fig:feyn_diag}
\end{figure*}

We consider the following interaction terms in a Lagrangian of a Dirac
fermion~$\chi$ that interacts with the photon gauge field $A_{\mu}$ or
its field strength tensor $F_{\mu\nu}$,
\begin{subequations}
  \label{eq:Lagrangians}
\begin{align} 
\label{eq:LmQ}  &\text{millicharge ($\epsilon$Q):      }        & & +   \epsilon e \, \bar\chi \gamma^{\mu} \chi A_{\mu}  , \\
\label{eq:Lmdm} &\text{magnetic dipole (MDM): }   & & + \frac{1}{2} \mu_\chi \, \bar\chi \sigma^{\mu\nu} \chi F_{\mu\nu}  , \\
\label{eq:Ledm} &\text{electric dipole (EDM): }   & &  + \frac{i}{2} d_\chi \, \bar\chi  \sigma^{\mu\nu}\gamma^5 \chi F_{\mu\nu} , \\
\label{eq:Lam}  &\text{anapole moment (AM):   }     & &  - a_\chi \,  \bar\chi \gamma^{\mu} \gamma^5\chi \partial^{\nu} F_{\mu\nu} ,\\
\label{eq:Lcr}  &\text{charge radius (CR):    }      & &  + b_\chi \,  \bar\chi \gamma^{\mu} \chi \partial^{\nu} F_{\mu\nu} .
\end{align}
\end{subequations}
Here $e>0$ is the electric charge,\footnote{The sign convention in (\ref{eq:LmQ}) is such that $\chi$
  ($\bar \chi$) carries a fraction~$\epsilon$ of the electron (positron) charge.} 
$\mu_\chi$ and $d_\chi$ are the dimensionful coefficients
of the mass-dimension~5 dipole interactions, and $a_\chi$ and
$b_{\chi}$ are the dimensionful coefficients of the mass-dimension~6
anapole moment and charge radius interaction; as usual,
$\sigma^{\mu\nu} = \frac{i}{2} [\gamma^{\mu}, \gamma^{\nu}]$. In what
follows we shall measure the EDM and MDM moments in units of the Bohr
magneton, $\mu_B \equiv e/(2m_e) = 1.93\times 10^{-11}\, e\,\cm $;
$m_e$ is the mass of the electron. All coupling strengths are real.

The interactions in~(\ref{eq:Lagrangians}) are assembled in the matrix
element of the effective electromagnetic current of~$\chi$,
\begin{align}
  \me{\chi(p_f)}{J_{\chi}^{\mu}(0)} {\chi(p_i)} =  \bar u(p_f) \Gamma^{\mu}(q) u(p_i)  ,
\end{align}
where $p_{i,f}$ and $q = p_i-p_f$ are  four-momenta.  For a neutral
particle $\chi$ the vertex function reads,
\begin{align} \label{eqn:effective_vertex}
\begin{split}
	 \Gamma^{\mu}(q) & =  \left( q^2 \gamma^{\mu} - q^{\mu} \slashed q  \right) \left[ V(q^2) - A(q^2)\gamma^5  \right] \\
   & + i \sigma^{\mu\nu} q_{\nu} \left[  M(q^2)  + i D(q^2) \gamma^5  \right] . 
\end{split}
\end{align}
We regard the various moments as being generated at an energy scale
that is well above the typical center-of-mass (CM) energies of the
high-intensity beams. The coefficients in (\ref{eq:Lagrangians}) are
hence the static limits of the form factors above,
\begin{align*}
  \mu_{\chi} = M(0), \quad d_{\chi} = D(0), \quad a_{\chi} = A(0), \quad b_{\chi} = V(0) .
\end{align*}

A millicharged particle $\chi$ will have an additional form factor
$Q(q^2)\gamma^\mu$ with $Q(0)=\epsilon e$. The next term in the
$q^2$-expansion of $Q(q^2)$ is then the charge radius interaction and
the equivalence with $V(0)$ above can be seen when the Dirac equation
is applied for on-shell $\chi$,
$\bar u(p_f) \left( q^2 \gamma^{\mu} - q^{\mu} \slashed q \right)
u(p_i) = \bar u(p_f) q^2 \gamma^\mu u(p_i) $. As should be clear from
above, an electric charge of $\chi$ is not mandatory for $\chi$ to
possess CR interactions, and we shall consider both interactions
separately below.  We note in passing that the Dirac and anapole form
factors $V(q^2)$ and $A(q^2)$ are related to the so-called vector and
axial vector charge radii as $\langle r_V^2 \rangle = - 6 V(0)$ and
$\langle r_A^2 \rangle = - 6 A(0)$.

Finally, in addition to the operators listed above, at
mass-dimension-7 there are the Rayleigh (or susceptibility) operators
where scalar $\bar\chi \chi$ or pseudoscalar $\bar\chi\gamma^5\chi$
bilinears multiply either  the CP-even or CP-odd squared field
strength tensors $F_{\mu\nu}F^{\mu\nu}$ and
$F_{\mu\nu}\tilde F^{\mu\nu}$. Such terms imply two-photon
interactions with the $\chi$-field and processes with a single photon
in the final state proceed through a loop. In the processes arising
from (\ref{eq:Lagrangians}) that we shall consider in the paper, the
latter interactions are then not only suppressed by the higher
dimensionality of the coupling, but also by a loop-factor. As the
constraints on dim-6 operators already turn out to be weak, we shall
not consider Rayleigh operators any further in this work.

\section{Pair-production in electron beams}
\label{sec:pair-prod-electr}
In this section we present the main expressions that are used to set
bounds on new particles $\chi$ interacting through electromagnetic
form factors. These involve the calculation of the $\chi\bar\chi$
pair-production cross section, first, from $e^+e^-$ annihilation
accompanied by initial state radiation as shown in
Fig.~\ref{fig:feyn_diag}a and, second, from Bremsstrahlung in the
scattering of electrons on nuclear targets as shown in
Fig.~\ref{fig:feyn_diag}b. They are the dominant production modes in
BaBar or Belle II and in NA64, LDMX, mQ or BDX, respectively.
Finally, since mQ and BDX search for new particles via their elastic
scattering within the detector, Fig.~\ref{fig:feyn_diag}c, we
furthermore establish the nuclear and electron recoil cross sections
in $\chi$-scattering.

\subsection{Pair-production at \texorpdfstring{$\mathbf{ e^+e^-}$}{ee} colliders}
\label{sec:dm-prod-belle}
In B-factories such as BaBar, Belle, and Belle II, fermion pairs can be
produced in $e^+e^-$ collisions accompanied by initial state radiation
(ISR). Production in association with final state radiation (FSR) is
suppressed with respect to ISR by $\epsilon^2$ for $\epsilon$Q and
scales as
$\mu_\chi^2 s/(4 \pi \alpha) \sim (10^4\mu_\chi/\mu_B)^2$ for MDM and
EDM. For the values of $\mu_\chi$ and $d_\chi$ that are constrained
through ISR by the various considered experiments,
this gives a suppression of $10^{-1}$ and smaller for FSR with MDM and
EDM interactions. For AM and CR, the FSR diagrams vanish identically
at tree level. Therefore, we are allowed to neglect FSR and only
consider $\chi\bar\chi$ production with associated ISR.  Finally,
mono-photon resonant production through a putative process
$\Upsilon (nS)\to \gamma \gamma^* \to \gamma \chi\bar\chi$ vanishes by
Furry's theorem since $\Upsilon (nS)$ are C-odd states.

The ISR cross section approximately factorizes into the process
without ISR, ${e^+e^-\rightarrow \bar \chi \chi}$, times
the improved Altarelli-Parisi radiator function that takes into
account all soft and collinear radiative corrections up to order
$m_e^2/s$~\cite{NICROSINI1989487,MONTAGNA1995161},
	\begin{align} \label{e_annih_cs}
		\frac{d\sigma_{e^+e^- \to \chi\bar\chi \gamma}}{d x_\gamma d\cos{\theta_\gamma}} =  \sigma_{e^+e^- \to \chi\bar\chi}\left(s,\sdark\right) \mathcal R^{(\alpha)} (x_\gamma, \theta_\gamma, s)
	\end{align}
with the radiator function
	\begin{align}\label{eqn:radiator_function}
		\mathcal R^{(\alpha)} (x_\gamma, \theta_\gamma, s) =\frac{\alpha}{\pi} \frac{1}{x_\gamma} \left[\frac{1+(1-x_\gamma)^2}{1+\frac{4 m_e^2}{s}-\cos ^2\theta_ \gamma }-\frac{x_\gamma^2}{2} \right] .
	\end{align}
        Here, $x_\gamma=E_\gamma/\sqrt{s}$ is the energy fraction
        carried away by the ISR, $\sqrt{s}\simeq 10\,\GeV$  and $\sqrt{\sdark}$ are the CM energies
        of the $e^+ e^-$ and $\chi\bar\chi$ system, respectively, with
        $\sdark = (1-x_\gamma) s$; $\theta_ \gamma$ is the CM angle of
        the ISR photon.  The $\chi$-pair production cross section without ISR
        reads
\begin{align}  \label{eqn:collider_prod_cs}
	\sigma_{e^+e^-\to \chi\bar\chi}=\frac{\alpha}{4}\frac{  f\left(\sdark\right) }{\sdark^2} \left(1+\frac{2 m_e^2}{s}\right) \sqrt{\frac{\sdark-4 m_\chi^2}{s-4 m_e^2}} \, .
\end{align}
Here, $f\left(\sdark\right)$ is the squared amplitude of
$\gamma^*\to \chi\bar\chi$, integrated over the 2-body phase space of the fermion
pair, as defined in (\ref{eqn:int_DM_piece}). This factor
will appear repeatedly and reads,
\begin{subequations}\label{eqn:LDMX_dark_matter_part}
\begin{align} 
\label{eqn:LDMX_dark_matter_part_mQ}
	\text{$\epsilon$Q:\quad } 
	f(\sdark) &= \frac{4}{3} \epsilon^2 e^2 \sdark \left( 1+\frac{2 m_\chi^2}{\sdark}\right) , \\
\label{eqn:LDMX_dark_matter_part_MDM}
	\text{MDM:\quad } 
	f(\sdark) &= \frac{2}{3} \mu_\chi^2 \sdark^2 \left( 1+\frac{8 m_\chi^2}{\sdark}\right) , \\
\label{eqn:LDMX_dark_matter_part_EDM}
	\text{EDM:\quad } 
	f(\sdark) &= \frac{2}{3}d_\chi^2 \sdark^2 \left( 1-\frac{4 m_\chi^2}{\sdark}\right) , \\
\label{eqn:LDMX_dark_matter_part_AM}
	\text{AM:\quad } 
	f(\sdark) &=  \frac{4}{3} a_\chi^2  \sdark^3 \left(1-\frac{4 m_\chi^2}{\sdark}\right) ,\\
\label{eqn:LDMX_dark_matter_part_CR}
	\text{CR:\quad } 
	f(\sdark) &=  \frac{4}{3} b_\chi^2  \sdark^3 \left(1+\frac{2 m_\chi^2}{\sdark}\right) .
\end{align}
\end{subequations}
Plugging (\ref{eqn:LDMX_dark_matter_part}) into (\ref{e_annih_cs}), for $m_e\rightarrow 0$, we recover the expression in
\cite{Choi:2015zka} for particles carrying $\epsilon$Q and the expressions in
\cite{Fortin:2011hv} for MDM and EDM.

\begin{figure}[tb]
\centering
\includegraphics[width=\columnwidth]{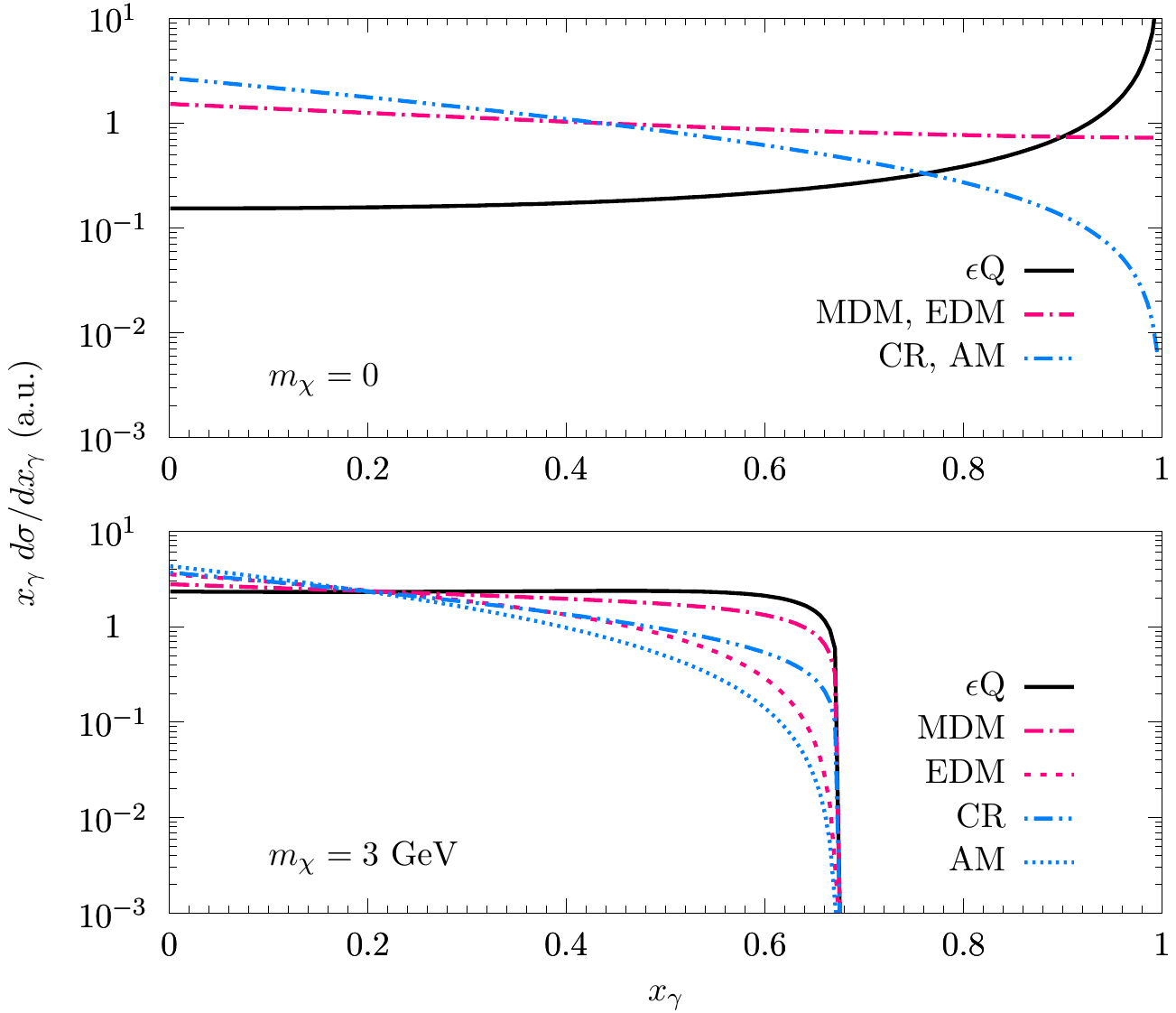}
\caption{Differential cross section for a CM energy of 10 GeV obtained
  from (\ref{e_annih_cs}) by integrating over $\cos\theta_{\gamma}$ as
  a function of the ISR energy fraction,
  $x_\gamma = E_\gamma/\sqrt{s}$, multiplied by $x_\gamma$ for
  $\epsilon$Q (black), MDM/EDM (pink) and CR/AM (blue). All curves are
  normalized to the same total cross
  section.}\label{fig:diff_cs_xgamma}
\end{figure}

In order to illustrate the difference between the various
interactions, in Fig.~\ref{fig:diff_cs_xgamma} we
show the differential cross sections with respect to the ISR
energy for $m_{\chi} = 0 $ (top) and
$m_{\chi}= 3\,\GeV$ (bottom).
The different behaviors of the cross sections with respect to the
fractional photon energy $x_{\gamma}$ suggest, that stronger limits on
$\epsilon$Q can be expected from high-energy mono-photon searches, while for AM
and CR the low energy region appears more prospective in constraining
these interactions (compare Fig.~\ref{fig:babar_events}; to be discussed in more detail in
Sec.~\ref{sec:intens-front-search}).
In the limit of $m_\chi \to 0$  the models with the same dimensionality
of the interaction operators show the same behavior, {\it i.e.}~for
$\mu_{\chi}= d_{\chi}$ ($a_{\chi}=b_{\chi}$) the cross sections for
MDM and EDM  (for AM and CR) as a function of $x_\gamma$ are identical. 
Finally, we note in passing that in the presented numerical results we
neglect the running of $\alpha$ since it is a few percent effect for
energies up to 10~GeV.

\begin{figure}[tb]
\centering
\includegraphics[width=\columnwidth]{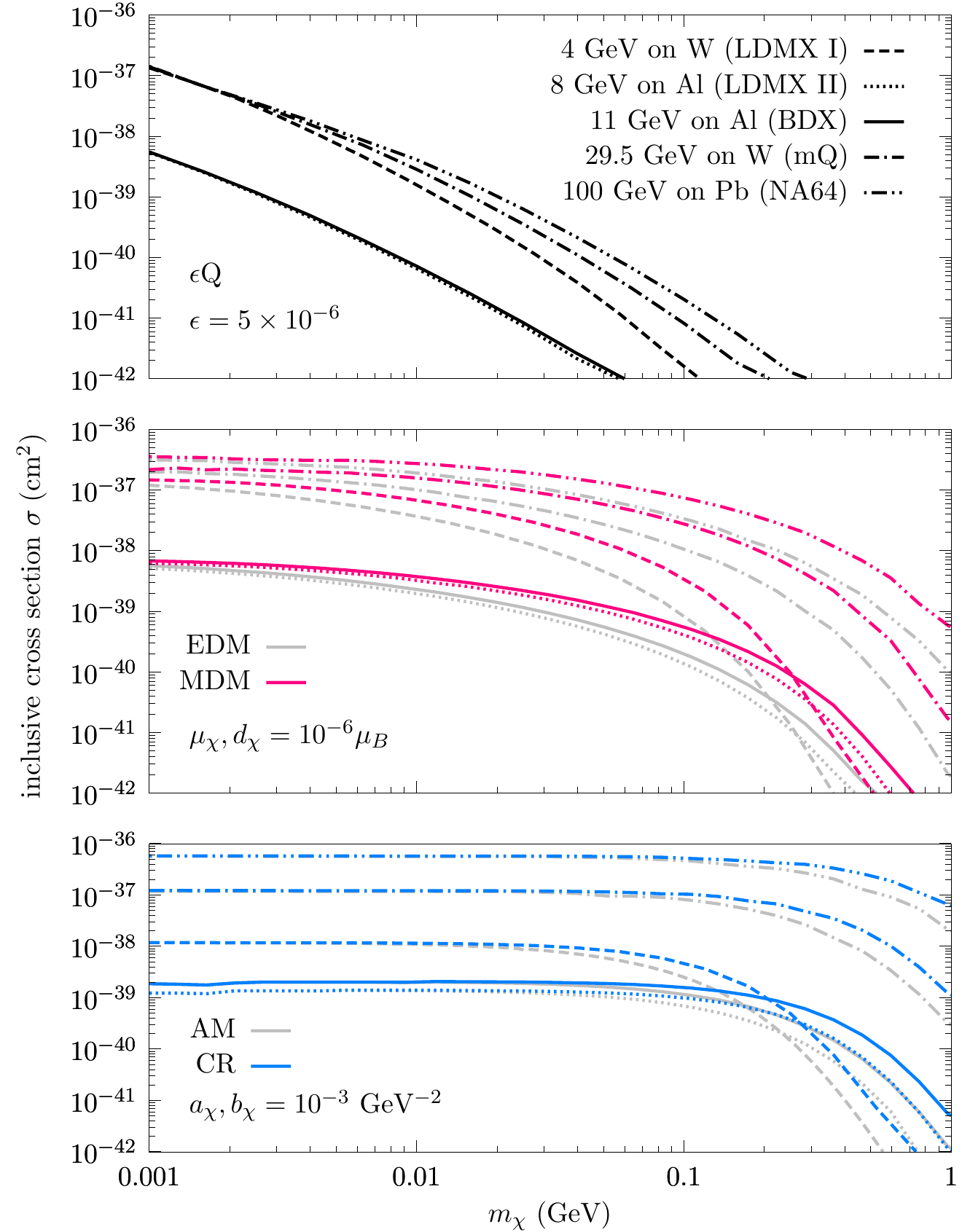} 
\caption{Inclusive cross section of $\chi$-pair production in
  $e^-$-scattering on a fixed target, the process depicted in
  Fig.~\ref{fig:feyn_diag}b. The beam energies and target nuclei correspond to the experiments discussed in this paper.
  In the limit of small DM mass, MDM and EDM in the central panel as
  well as CR and AM in the bottom panel yield identical cross sections. The mQ results agree with \cite{Prinz:2001qz}.
 }\label{fig:cs_incl}
\end{figure}

\subsection{Pair-production at fixed target experiments }
In the fixed target experiments we consider in this work, a pair of
fermions is produced via an off-shell photon emitted by an electron as
it hits a nuclear target
($e^- N \rightarrow e^- N \gamma^{*} \rightarrow e^- N \chi \bar
\chi$). The Feynman diagrams of the process are depicted in
Fig.~\ref{fig:feyn_diag}b. In principle, also inelastic scatterings to
an excited nuclear state $N^{*}$ or to an all-inclusive final state
$X$ are possible. In the following we neglect such contributions, but
present the general formalism to account for the latter in
App.~\ref{sec:matrix-el} and \ref{sec:had-ten}.

Note that we only consider the emission of the photon from the
electron leg. The emission from the nuclear leg is parametrically
smaller by a factor of $(Z m_e/m_N)^2 \sim 10^{-6}$ for coherent
photon emission and receives a further suppression from the nuclear
form factor for photon momenta above a few 100~MeV.\footnote{Photon
  emission from final-state light hadrons such as pions that accompany
  the primary scattering is suppressed by
  $(m_e/m_\pi)^2\sim 10^{-5}$.}  Finally, in order to properly cover
the full kinematic range that is accessible to $\chi$ and $\bar \chi$
at a given CM energy, we compute the underlying 2-to-4 process
exactly, without relying on approximations such as the
Weizsäcker-Williams method.

It is still possible, however, to relate the 2-to-4 cross section to a
2-to-3 process with the emission of a vector boson of virtual squared
mass $\sdark$,
\begin{align} \label{eqn:2to4_cs}
	\frac{d\sigma_{2\rightarrow 4}}{d\sdark} = \sigma_{2\rightarrow 3} \left(\sdark\right) \frac{f\left(\sdark\right)}{16 \pi^2 \sdark^2}\sqrt{1-\frac{4m_\chi^2}{\sdark}}\,,
\end{align}
in which the directions of $\chi$-particles have been integrated out in $f\left(\sdark\right)$.%
\footnote{In the calculation of $\sigma_{2\to3}$ one may still make
  the replacement
  $\sum_{\rm pol} \epsilon_{\mu} \epsilon_{\nu} \to -g_{\mu\nu}$ in
  the polarization sum of the massive vector as the
  $q^{\mu}$-dependent piece drops out by virtue of the Ward identity.}
Equation~(\ref{eqn:2to4_cs}) is exact, and can in principle be
generalized for higher differential cross sections that are not
sensitive to the direction of $\chi$ or $\bar \chi$.
%
%
Our actual calculations require higher differential cross sections
than (\ref{eqn:2to4_cs}).  For NA64 and LDMX the full 4-body phase
space is presented in App.~\ref{sec:4-body-phase-ldmx}
[Eq.~\ref{eqn:LDMX_phas_space}] and for mQ and BDX where the momentum
of $\chi$ enters in the elastic detection channel it is presented in
App.~\ref{sec:4-body-phase-bdx} [Eq.~\ref{eqn:bdx_phase_space}].


In principle, also pions and $\eta$-mesons (among other mesons) can be
produced in this scattering process, which can subsequently decay into
$\chi \bar \chi$ if kinematically allowed. However, while meson
production dominates in fixed target experiments with a proton beam,
it presents a subleading contribution for the electron beam
experiments we consider. In addition, note that NA64 and LDMX should veto
events that involve produced hadrons in the final state. For these
reasons we neglect the contribution of hadronic production channels of
$\chi\bar\chi$ to the event rate.

In Fig.~\ref{fig:cs_incl} we show the inclusive
$\chi\bar\chi$-production cross sections as a function of $m_\chi$ for
several target materials and beam energies corresponding to the
experiments discussed in Sec.~\ref{sec:intens-front-search}. As
expected from~(\ref{eqn:LDMX_dark_matter_part}), the dimensionality of the operator  determines the
behavior of the cross section for $m_\chi^2 \ll s$. Since models with lower dimensional operators show a flatter dependence on $\sdark$ in~(\ref{eqn:LDMX_dark_matter_part}), there is no visible difference in the inclusive cross section for $\epsilon$Q at low masses between 4, 30 and 100~GeV beam energy in the top panel of Fig.~\ref{fig:cs_incl}, whereas for all other models, shown in the center and bottom panel, the cross section visibly rises with increasing beam energy.
It is noteworthy that the cross section for an aluminium (Al) target is generally smaller than for tungsten (W) and lead (Pb) targets since the cross section scales with $Z^2$ and the charge of the Al nucleus is smaller by roughly a factor of  5--6. However, due to its smaller mass, the suppression at the kinematic edge is less pronounced in comparison to the heavier targets. The charge and mass differences between W and Pb are negligible in this case.
Finally, while the cross
section for millicharged particles rises with decreasing mass, it
becomes independent of $m_{\chi}$ for MDM and EDM, and even quicker so
for AM and CR. When $m_{\chi}$ increases towards the kinematic
endpoint, the operators containing a factor of $\gamma^5$ yield
velocity factors that result in a stronger suppression of the cross
sections.

\subsection{Detection via elastic scattering on nuclei}
\label{sec:elastic-dm-nuc}

In mQ and BDX the produced fermions can directly be detected via
elastic $\chi$-nucleus and/or $\chi$-electron scattering. In this case, the signal of $\chi$ particles would be a number of  recoil events in the detector beyond the background, instead of missing momentum/energy.  A
presentation of the expressions of the recoil cross section
$d\sigma/dE_R$ in the lab frame as a function of the recoil energy
$E_{R}$ and valid for relativistic $\chi$-energies $E_{\chi}$ is
relegated to App.~\ref{sec:elast-scatt-cross}.
The cross sections are functions of the nuclear mass (magnetic
moment) $m_N$ ($\mu_N$) and the nuclear spin $I_N$. In the
small-velocity limit, \textit{i.e.}, for~$|\bm p_\chi|=m_\chi v$ and
$E_\chi \simeq m_\chi+m_\chi v^2/2$ with $v\ll 1$, the expressions in
(\ref{eq:el-cs}) are in agreement with the formul\ae\ found in
Refs.~\cite{Chang:2010en,Geytenbeek:2016nfg}. Since the expressions
are exact with respect to phase space, they apply to both 
$\chi N$ and $\chi e^-$ scatterings. %

\section{Intensity frontier searches with electron beams}
\label{sec:intens-front-search}
In this section we place constraints on the discussed fermion models
with electromagnetic form factors from BaBar, NA64 and mQ and present
projections for Belle II, LDMX and BDX. All limits in the presented
plots are obtained at 90$\%$ confidence level (C.L.).  The limits and
projections for the models with dim-5 (dim-6) operators are shown in
the summary plots in Fig.~\ref{fig:limits_dim5}
(Fig.~\ref{fig:limits_dim6}).\footnote{Even stronger (projected)
  bounds may be achievable from beam dump experiments with high-energy
  proton beams, as suggested by \cite{Mohanty:2015koa}. We leave such
  studies for a future work; see Sec.~\ref{sec:conclusions}.}

\subsection{BaBar and Belle 2}
 
BaBar at SLAC~\cite{Aubert:2001tu} and Belle-II at SuperKEKB~\cite{Abe:2010gxa} are experiments at asymmetric $e^+ e^-$ colliders. %
Long-lived states $\chi$ that are produced
through their (feeble) electromagnetic form factor interaction can escape from 
the detectors. The primary signature is hence the missing energy in
$\chi\bar\chi $ production in association with a single photon
(mono-photon missing energy search).  The main background processes
are SM final states in which electrons or photons leave the detector
undetected, \textit{i.e.},~$e^+e^-\rightarrow \gamma \slashed \gamma$, 
which produces a peak at $\sdark=0$ and
$e^+e^-\rightarrow \gamma \slashed e^+ \slashed e^-$ as well as
$e^+e^-\rightarrow \gamma \slashed \gamma \slashed \gamma$ which constitute
 a continuum background that rises with $\sdark$. The irreducible
background $e^+e^-\rightarrow \gamma \nu\bar\nu$ is  suppressed by
$m_Z^{-4}$ and can be neglected.

\begin{figure}[tb]
\centering
\includegraphics[width=\columnwidth]{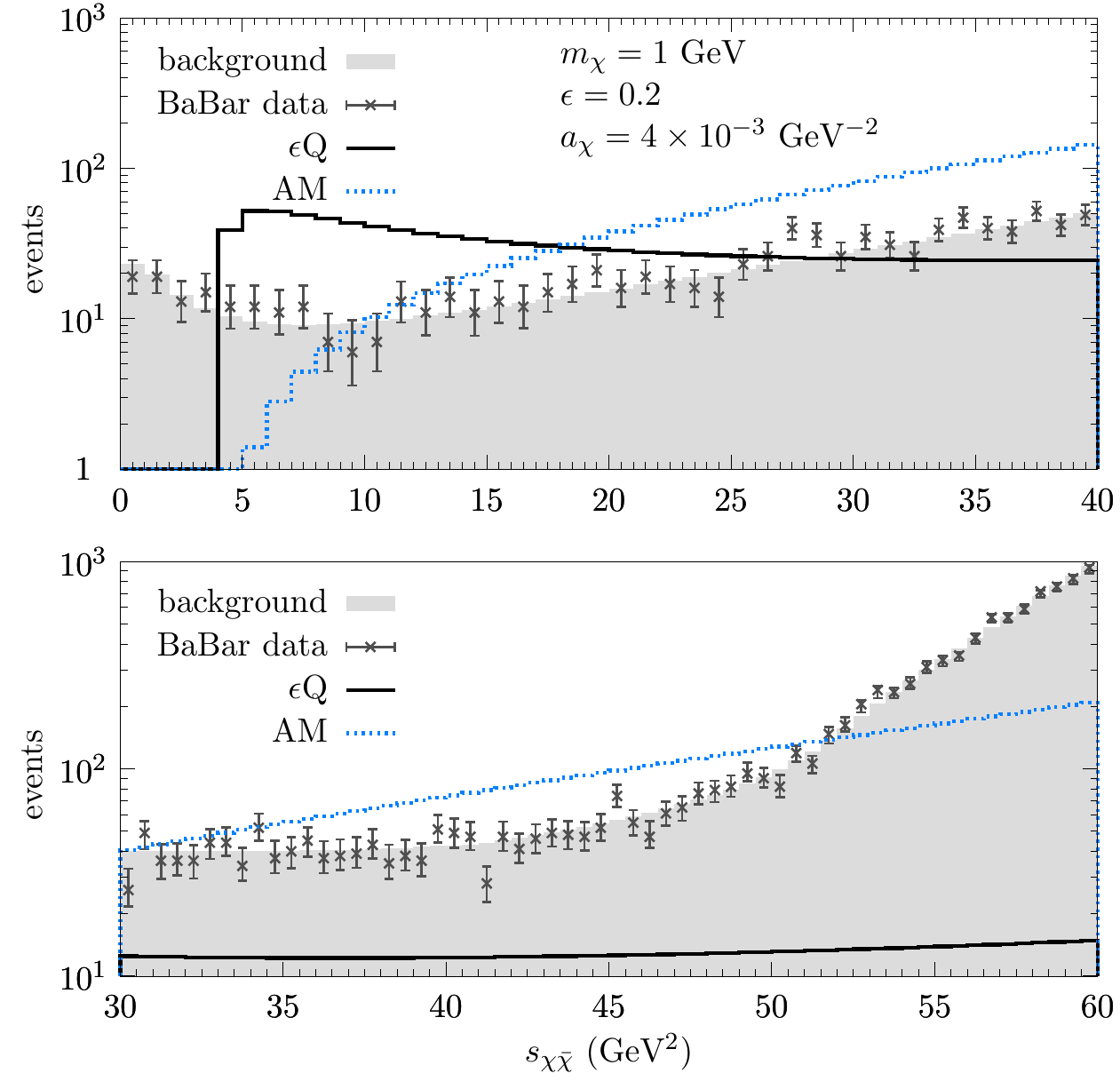}
\caption{High energy ($3.2 \; \GeV <E_\gamma<5.5\;\GeV$) and low
  energy ($2.2 \; \GeV <E_\gamma<3.7\;\GeV$) mono-photon search region
  of BaBar in the top and bottom panels, respectively. The data points
  and background events taken from~\cite{Aubert:2008as} are shown
  as a function of squared invariant missing mass and are binned per
  GeV$^2$ (top) and per half GeV$^2$ (bottom). Also shown are
  exemplary event shapes for $\epsilon$Q and AM interactions with
  $\epsilon = 0.2$ and $a_{\chi} = 4\times 10^{-3}\GeV^{-2}$ for
  $m_{\chi} = 1\,\GeV$ in solid (black) and dotted (blue),
  respectively.  \label{fig:babar_events} }
\end{figure}

For deriving constraints from BaBar, we use the data of the analysis
of mono-photon events performed by the collaboration in a search for
invisible decays of a light scalar at the $\Upsilon$(3S)
resonance~\cite{Aubert:2008as,Lees:2017lec}.\footnote{In more recent
  \cite{Lees:2017lec}, Boosted Decision Trees (BDT), trained on
  simulated massive dark photon signal events, were used for event
  selection. Our cases of main interest have different event shapes as
  shown in Fig.~\ref{fig:babar_events}, and hence differ significantly
  from the training set used in~\cite{Lees:2017lec}. For this reason,
  we use the smaller data set~\cite{Aubert:2008as} with cut-based
  event selection.}  The CM energy is
thus 10.35\,GeV. Two search regions with different trigger and cut
efficiencies, a high photon energy region with
$3.2 \; \GeV <E_\gamma<5.5\;\GeV$ and low energy region with
$2.2 \; \GeV <E_\gamma<3.7\;\GeV$ in the CM frame, were presented. The high energy
search uses the full data set of 28 fb$^{-1}$ integrated luminosity
and the low energy search a subsample of 19 fb$^{-1}$.  The geometric
cuts of the search regions are $-0.31<\cos \theta_\gamma<0.6$ (high
energy) and $-0.46<\cos \theta_\gamma<0.46$ (low energy),  where
$\theta_{\gamma}$ is the photon angle in the CM frame with respect to
the beam line as in (\ref{e_annih_cs}).

With the respective angular cuts applied, the number of signal events
is given by %
\begin{align}
  N^{(i)}_\text{sig} = \epsilon_\text{eff} \; \mathcal L 
  \int_{\text{bin}\ i} \frac{d\sdark}{s} \int_{\cos\theta_{\gamma}^{\rm min}}^{\cos\theta_{\gamma}^{\rm max}}  d\cos\theta_{\gamma}  \;
  \frac{d\sigma_{e^+e^-\to \chi\bar\chi\gamma}}{d x_\gamma d\cos{\theta_\gamma}} ,
\end{align}
where $\epsilon_\text{eff}$ is the total trigger and cut efficiency,
$\mathcal L$ is the integrated luminosity. The boundaries of $\sdark$
are given by the bins in Fig.~\ref{fig:babar_events}, and the
production cross section $d\sigma_{e^+e^-\to \chi\bar\chi\gamma}$ is
given in~(\ref{e_annih_cs}).
Following Ref.~\cite{Essig:2013vha}, in our analysis we apply a
non-geometric cut $\epsilon_{\text{eff}}$ of $30\%$ ($55\%$) in the
high (low) energy region for BaBar and place conservative constraints
on the models in (\ref{eq:Lagrangians}) by requiring that the expected
number of signal events does not exceed the observed number
$N^{(i)}_\text{obs} $ at 90$\%$~C.L.~in any bin $i$.  Concretely, we
require
$ N^{(i)}_\text{sig} < N^{(i)}_\text{obs} + 1.28 \,
\sigma^{(i)}_\text{obs}$,  where $\sigma^{(i)}_\text{obs}$ is given by
the statistical and systematical uncertainties of $N^{(i)}_\text{obs}$
added in quadrature.%
\footnote{As is evident from Fig.~\ref{fig:babar_events}, a signal
  will simultaneously be present at similar strength in many bins. It
  is hence safe to neglect the look-elsewhere effect. }

For the Belle II projections, we follow Ref.~\cite{Essig:2013vha} and
scale up the Babar background from Ref.~\cite{Aubert:2008as} (shown by
the gray histogram in Fig.~\ref{fig:babar_events}) to an integrated
luminosity of 50 ab$^{-1}$ with the CM energy of 10.57\,GeV, employ a
constant efficiency cut of $50\%$ in both search regions and take
identical geometric cuts for Belle II and
BaBar. %
While the exact background subtraction in the mono-photon search of
Belle II needs to be provided by event generators~\cite{Kou:2018nap},
here we adopt the two projection-methods of Ref.~\cite{Essig:2013vha}: first
is a ``statistics limited'' projection, where we assume a full
subtraction of the background yielding a limit that is only governed
by statistical fluctuations. It comprises the best case scenario with
the strongest ensuing limits on the new couplings.  Second is a
``systematics limited'' projection, where we assume a $10\%$
systematic uncertainty for the $\gamma\slashed \gamma$ peak and a
$20\%$ uncertainty for the continuum backgrounds and exclude parameter regions 
where the signal exceeds the the systematic plus statistical
uncertainties at $90\%$~C.L.. With good control over the backgrounds,
the actual reach of Belle II should lie between those two projections.

In Fig.~\ref{fig:babar_events} we show exemplary event shapes as a
function of $\sdark$ at BaBar for $\epsilon$Q with $\epsilon = 0.2$
and for AM with $a_{\chi} = 4\times 10^{-3} \,\GeV^{-2}$. As can be
seen, the limit on millicharged particles is driven by the high
mono-photon energy search region (top panel of
Fig.~\ref{fig:babar_events}) while for particles carrying an anapole
moment (and in analogy the other operators) the limits are set in the
low mono-photon energy search region as well as from the bins with
lower photon energy in the high energy region. The ensuing
constraints, as well as projected ones from Belle II, are shown as the
parameter space of $\chi$-mass and couplings in
Fig.~\ref{fig:limits_dim5} for MDM and EDM and in
Fig.~\ref{fig:limits_dim6} for CR and AM models.

For the $\epsilon$Q model, we find that in the kinematically
unsuppressed region, $m_\chi\lesssim 1
\GeV$, %
the BaBar data excludes $\epsilon \ge 6.4 \times 10^{-2}$ at 90\%
C.L. Moreover, the reach of the Belle II experiment is
$\sim 2 \times 10^{-2}$ and $\sim 4 \times 10^{-3}$ for the systematics and
statistics limited scenarios, respectively.

\subsection{NA64 and LDMX}

The NA64 experiment at the CERN SPS~\cite{Banerjee:2017hhz} and LDMX at SLAC~\cite{Akesson:2018vlm}  are current and proposed fixed target experiments using electron beams,  where a $\chi\bar\chi$ pair
can be produced via off-shell photon emission. An active veto system
together with cuts on the search region makes both experiments 
essentially background-free.  Therefore, in the following we place limits
on the operators in (\ref{eq:Lagrangians}) by assuming zero background
in the regions of interest.

In the NA64 experiment, an  electron
beam with $E_0= 100$\,GeV hits an electromagnetic calorimeter (ECAL) which is a matrix of
$6 \times 6$ modules consisting of lead (Pb) and scandium (Sc) plates,
each module having a size of approximately 40 radiation lengths
($X_0$), with $X_0 = 0.56$~cm for Pb. The radiation length of  Sc is about one 
order of magnitude larger, so the scattering of electrons with Sc is sub-leading, and will not be considered here. In
the initial part of the ECAL there is a pre-shower (PS) detector of $4 \, X_0$ in length
with the same setup, functioning as target and detector
simultaneously. Missing energy signals are required to have a starting point of
the EM shower localized within the first few radiation lengths of the detector~\cite{Gninenko:2016kpg}. In our analysis, we 
conservatively take a thin target limit, with
$L_{\text{target}}=X_0$, meaning that the fermion pair is essentially produced within the first radiation length of the target material.
After scattering, the final state electrons are detected in the main ECAL with a search region  of $0.3 \; \GeV < E_4 < 50 \; \GeV$, where the lower
boundary is due to the electron identification threshold in the PS detector and the upper boundary requires a missing energy larger than $0.5E_0$~\cite{Banerjee:2017hhz}. We adopt a polar
angular coverage of $\theta_4 < 0.23$\,rad so that the final state electron should pass through the whole ECAL. The total cross section, however, is strongly forward-peaked and therefore the contribution of large angle scattering is negligible, meaning that a variation of $\theta_4^\text{max}$ does not affect our limits.  %

The number of signal events in the thin target limit and with the geometric
and angular cuts as specified above is given by
\begin{align}
  \begin{split}
    \label{eq:NsigFixedTarget}
	N_{\text{sig}} &= N_{\text{EOT}} \frac{\rho_{\text{target}}}{m_N} L_{\text{target}} \int_{E_\text{min}}^{E_\text{max}} d E_4 \; \epsilon_\text{eff}(E_4)  \\
	&\times \int_{\cos\theta_4^\text{min}}^{\cos\theta_4^\text{max}} d\cos\theta_4 \frac{d\sigma_{\text{prod}}}{dE_4 d\cos\theta_4} ,
\end{split}
\end{align}
where $d\sigma_{\text{prod}}$ is the $\chi$-pair production cross
section, $e^- N \rightarrow e^- N \chi \bar\chi$, given by
(\ref{eqn:diff_ldmx_cs}), $N_{\rm EOT} = 4.3 \times 10^{10}$ is the
number of electrons on target (EOT), $\rho_{\text{target}}$ is the
target density and $m_N$ the target mass. The detector efficiency for
NA64 only marginally depends on the electron energy, and we take it to
be constant with $\epsilon_\text{eff}=0.5$.  The constraints from NA64
on the interactions in (\ref{eq:Lagrangians}) are then derived by
requiring that parameters that generate more than 2.3 events in the
detector are excluded.

\begin{figure}[tb]
\centering
\includegraphics[width=\columnwidth]{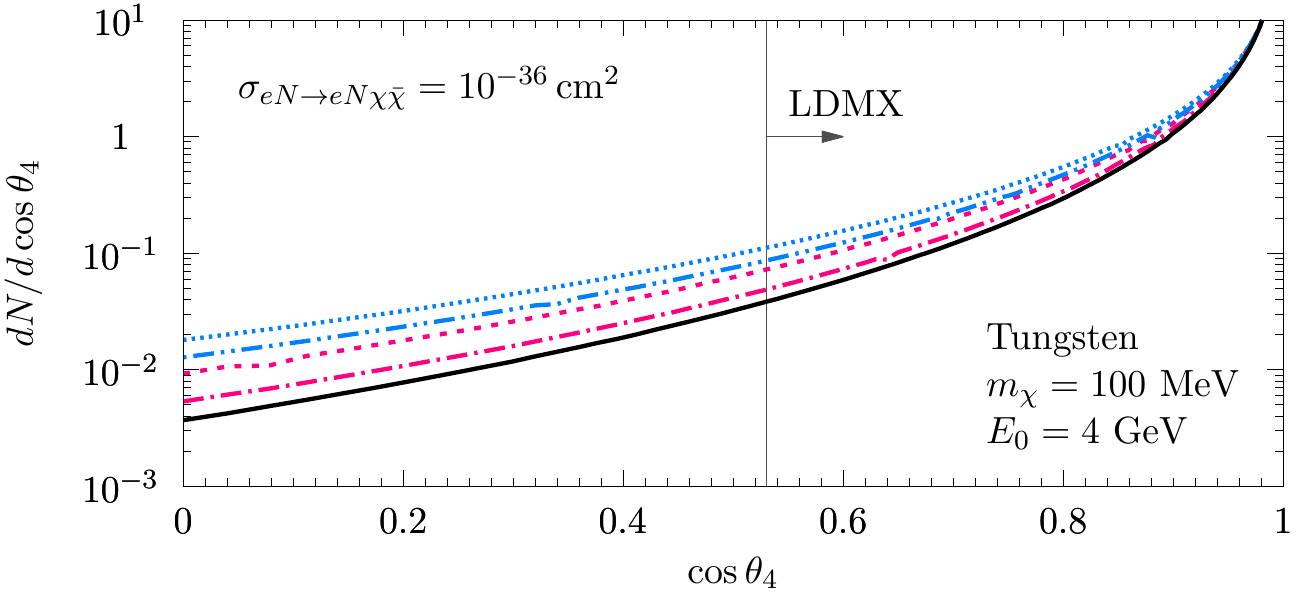} 
\includegraphics[width=\columnwidth]{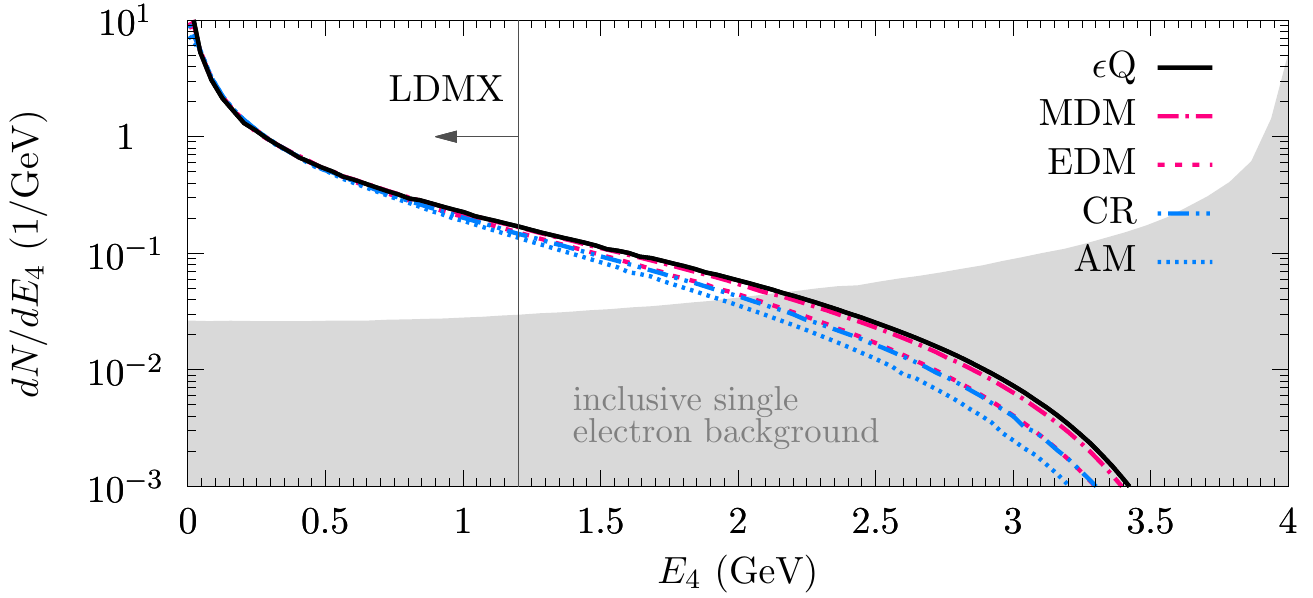} 
\caption{Angular (top) and energy (bottom) distribution of the scattered electron in $\chi$-pair
  production in $e^-$-scattering on a tungsten target
  for $m_\chi=100$~MeV and electron beam energy of 4~GeV, before imposing detector efficiency.  In the bottom panel the inclusive single electron background taken from \cite{Akesson:2018vlm} is shown in gray.  \label{fig:LDMX_distr}}
\end{figure}

The LDMX proposal consists of a silicon tracker system, an ECAL and a
hadron calorimeter (HCAL). 
In our analysis we use the benchmark points of the
proposed LDMX phase~I with $4\times 10^{14}$ EOT at $E_0=4$\,GeV, and
phase~II with $3.2 \times 10^{15}$ EOT at $E_0=8$\,GeV~\cite{Akesson:2018vlm}. LDMX will use a
tungsten (W) target ($X_0=0.35$~cm) with a thickness of $0.1\,X_0$ in
phase~I and an aluminium (Al) target ($X_0=8.9$~cm) with $0.4\,X_0$ in
phase~II. The first four track sensors with an area of 4\,cm$\times $10\,cm are located $7.5-52.5$~mm behind the target along the beam line.  We take a polar angular coverage in the forward direction
of $\theta_4<\pi/4$  and a search region for the final state electron
energy of $50 \,\MeV < E_4<0.3 \, E_0$.  While the actual signal efficiency varies mildly with the dark state mass and track requirements~\cite{Akesson:2018vlm},   a constant $\epsilon_\text{eff} =0.5$ is assumed here.  In the end, the number of
signal events is  calculated according
to~(\ref{eq:NsigFixedTarget}), and the projected constraints from LDMX are 
derived analogously, requesting at least 2.3 events in the detector. 

In Fig.~\ref{fig:LDMX_distr} we show the angular (top panel) and
energy distributions (bottom panel) of the final state electrons for
LDMX phase~I obtained from (\ref{eq:NsigFixedTarget}).  The event
rates are rather similar for all models when normalized to the same
total cross section; we normalize it to $\sigma = 10^{-36} \, \text{cm}^2$ yielding approximately 1 predicted signal event in LDMX phase I. 
 The signal is strongly forward peaked for
$\sdark\gg m_\chi^2$, but in the transverse direction and at the
kinematic edge, the curves start to deviate slightly.  We also show the geometric acceptance region of
LDMX in the top panel of Fig.~\ref{fig:LDMX_distr} and the energy cut
in the bottom panel, indicating that LDMX is searching in a region in
the ($\theta_4,\, E_4$)-plane where the event shapes of all models are
similar.

Both the current limits from NA64 and projections for the LDMX phases
are shown in Fig.~\ref{fig:limits_dim5} for MDM and
EDM and in Fig.~\ref{fig:limits_dim6} for CR and AM models. Whereas the NA64
limit is weaker than the limit derived from BaBar data, LDMX will
significantly improve the sensitivity to $\mu_{\chi}$ and $d_{\chi}$
but will be on par with BaBar for $a_{\chi}$ and $b_{\chi}$.  We note
in passing that for the $\epsilon$Q model, our projections for LDMX agree with
Ref.~\cite{Akesson:2018vlm} and for NA64 are slightly weaker, by about a factor 2, than the limits presented in Ref.~\cite{Gninenko:2018ter}. This is most likely due to a combination of a different assumption on detection efficiencies and the neglect of an energy threshold in  the final state electron  in \cite{Gninenko:2018ter}.

\begin{figure*}[tb]
\centering
\includegraphics[width=\columnwidth]{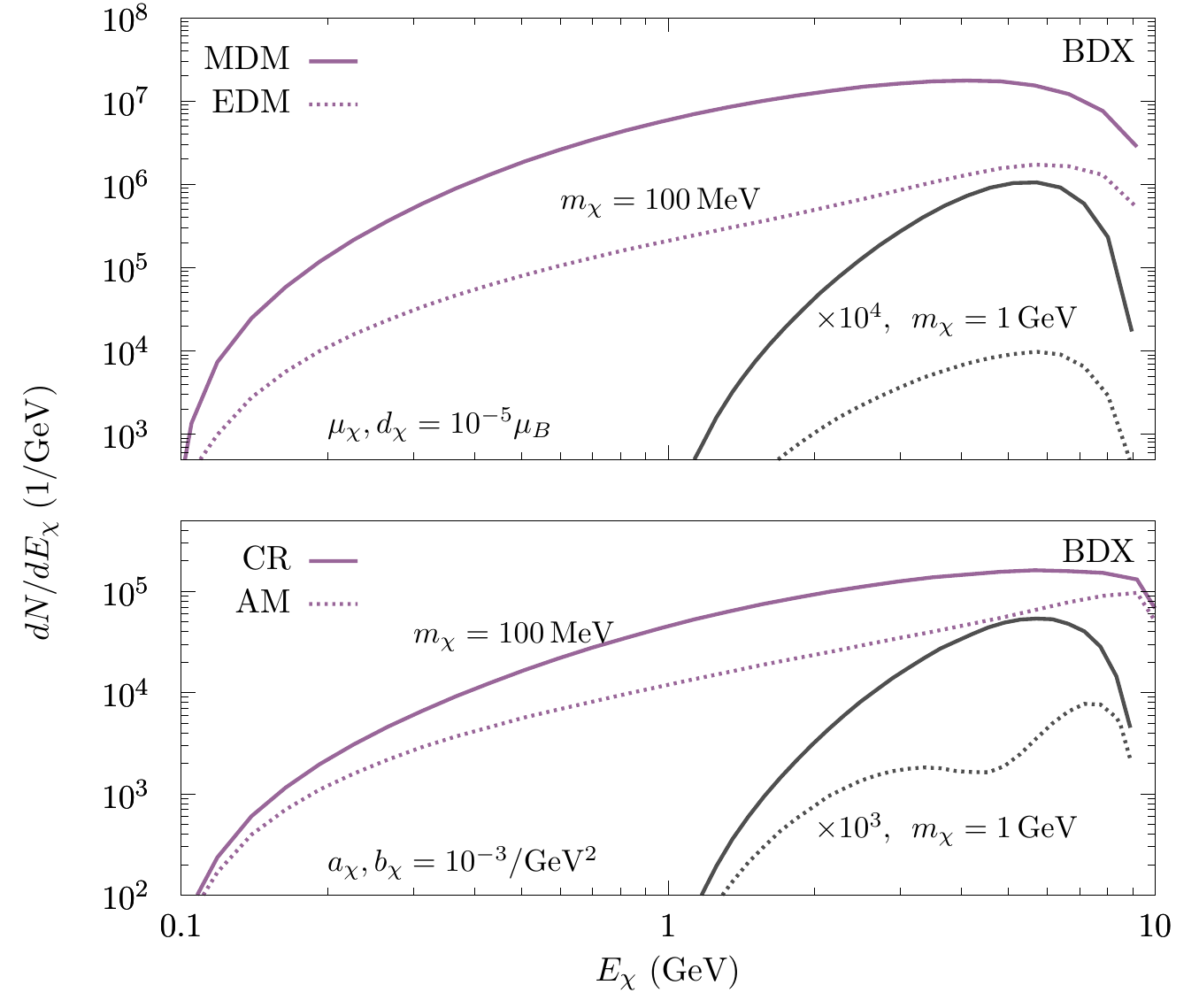}%
\includegraphics[width=\columnwidth]{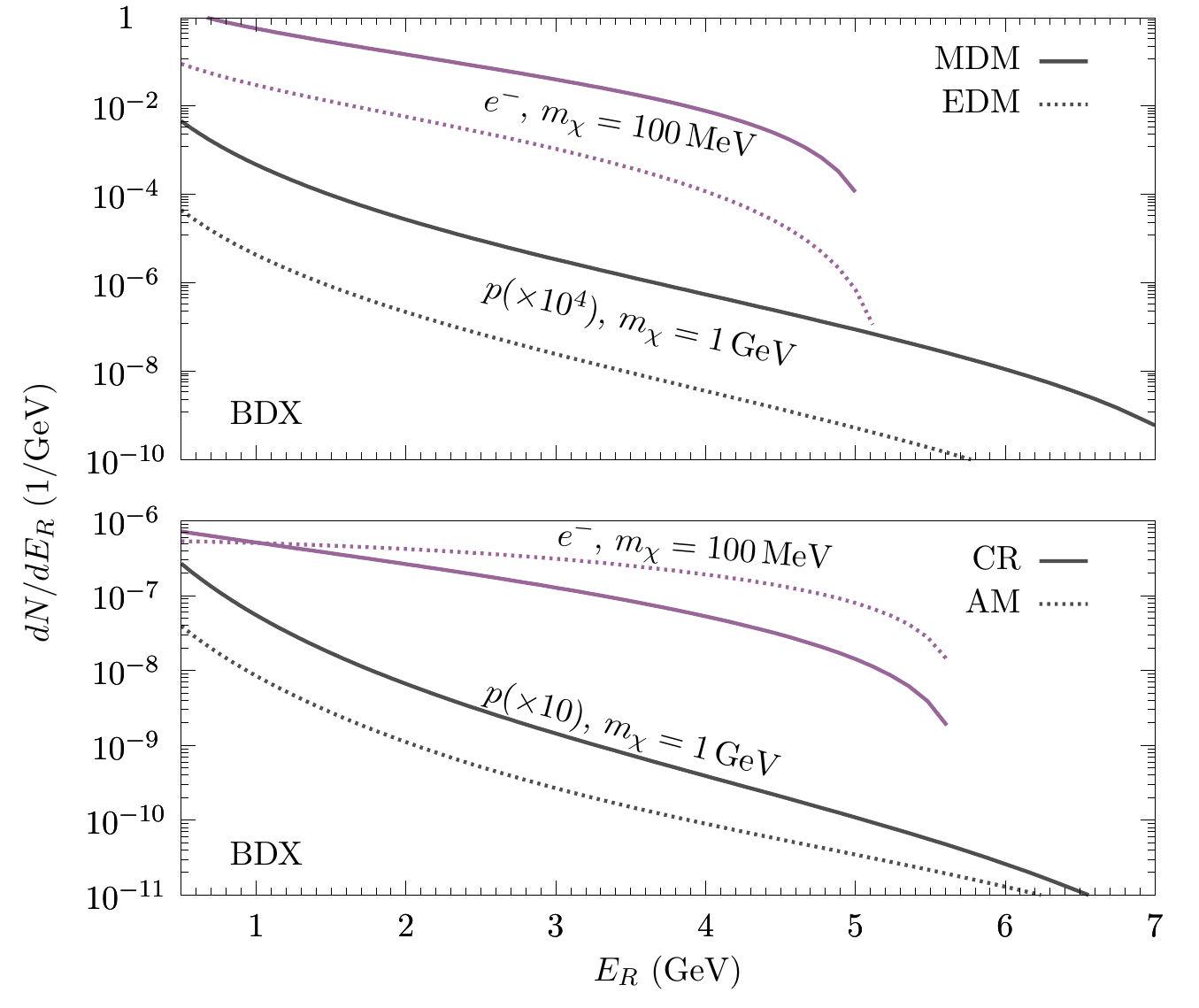}
\caption{\textit{Left:} Energy-differential total number
  $dN_{\chi}/dE_{\chi}$ of $\chi$-particles in the BDX detector as
  obtained from~(\ref{eq:dNchidEchi}). The top (bottom) panel shows
  the mass dimension-5(6) interactions for $m_{\chi} = 100\, \MeV$ and
  $1\,\GeV$ and coupling strengths as labeled.  \textit{Right: }
  Corresponding electron and proton recoil spectra $dN_{\chi}/dE_R$ as
  obtained from~(\ref{fixedT:events}) for the same parameters.}
\label{BDX:Echispectrum}
\end{figure*}

\subsection{mQ and BDX}

In the mQ experiment at SLAC~\cite{Prinz:1998ua, Prinz:2001qz}, 
$N_{\rm EOT} =8.4\times 10^{18}$ electrons with $E_0=29.5$ GeV energy
were incident on a W target of thickness $6 \, X_0$.
The produced dark particles are then searched for
in their scattering off electrons and nuclei in the main detector,
approximately 110~m down the beam line. The detector, made of plastic
scintillators, has a polar angular coverage of $\theta_\chi<2$~mrad, and
a total path length of $L_{\rm det} = 131~\cm$. The experiment found
$207$ recoil events above the estimated background, but below the
estimated background uncertainty
$\sigma_\text{bkg} = \sqrt{N_{\rm bkg}} \simeq 382$, within the signal time
window.
While millicharged particles would mostly produce single
photo-electron\,(PE) events, such assumption is not justified for the
recoil spectra of the models considered in this work.
Therefore, similarly to~\cite{Diamond:2013oda}, we focus on
nuclear/electron recoil events with a deposited energy larger than
0.1~MeV. %
Following \cite{Prinz:2001qz}, 90\% C.L. exclusion limits can then be
obtained by requiring the number of signal events induced by $\chi$,
$N_\text{sig} < 207 + 1.28\sigma_\text{bkg} \simeq 
697$. Here we will not consider a further possible reduction of
single-PE backgrounds~\cite{Diamond:2013oda}; they may strengthen our
limits by a factor of 2--3.

The proposed BDX experiment  at Jefferson Lab~\cite{Battaglieri:2016ggd}
 will use an electron beam of $E_0=11~\GeV$ incident on an
Al target that comprises 80 layers with thickness of 1-2~cm
each.  The $\chi$ particles produced in the bump pass through thick
shields and reach an ECAL 20~m
downstream. The ECAL, composed of CSI(Tl) crystals, will have an area
of 50$\times$55\,cm$^2$ and a depth of $L_{\rm det} = 260\,\cm$, implying a polar angle
coverage of $\theta_\chi \lesssim 12.5$~mrad. The BDX collaboration has
estimated that for $N_{\rm EOT} = 10^{22}$ the number of background
events with deposited energy above 500~MeV is approximately 10, coming
mostly from high energy neutrinos that scatter off
electrons~\cite{Battaglieri:2017qen}.
In the following, we conservatively take the uncertainty in the
background estimation $\sigma_\text{bkg} = N_\text{bkg} = 10$, and
assume a benchmark value for the projected number of observed events
$N_\text{obs}= 15$. It allows to set 90\% C.L. limits on models
studied here by requiring $N_\text{sig} \leq 18$.

In both experiments, the energy-differential expected number of
particles $\chi$ that reach the detector is given by
\begin{align}
  \label{eq:dNchidEchi}
	\frac{dN_{\chi}}{dE_\chi} &= N_{\text{EOT}} \frac{\rho_{\text{target}}}{m_{N}} X_0 \int^{E_0}_{E_\chi} dE \;  I(E)  \frac{d\sigma_{\text{prod}}(E)}{dE_\chi}\,,
\end{align}
where  the production cross section $d\sigma_{\text{prod}}$ given by (\ref{eqn:BDX_cross_section}) has been integrated over the angular coverage of the detector, and 
\begin{equation*}
	 I(E) = \frac{1+ (\frac{E_0- E}{E_0})^{\frac{4t_0}{3}} \left(\frac{4t_0}{3} \ln (\frac{E_0-E}{E_0})-1\right)}{\frac{4}{3} (E_0-E) \ln ^2(\frac{E_0-E}{E_0})}
\end{equation*}
is the integrated energy distribution of electrons during its 
propagation in the target~\cite{Bjorken:2009mm}. For the mQ
experiment, $t_0=L_\text{target}/X_0= 6$  ($X_0=0.35\,$cm for W), and for BDX we take
$t_0 = 15$ as fiducial value ($X_0=8.9\,$cm for Al). Finally,
$m_{N}$ and $\rho_{\rm target}$ are the target mass and mass
density, respectively.  The energy spectra~(\ref{eq:dNchidEchi}) are
shown in Fig.~\ref{BDX:Echispectrum} for several benchmark parameter
values. One can observe that most of the $\chi$ particles entering the
detector carry kinetic energies in the GeV-ballpark.

The total number of signal events in the downstream detectors that are
produced in the scattering of $\chi$ on target $i$ is then given by
\begin{align}	\label{fixedT:events}
\begin{split}
  N^{(i)}_{\text{sig}}&= N_T^{(i)}  L_\text{det}\int_{m_\chi}^{E_\chi^\text{max}}
                        \!\!\!\! d E_\chi\int_{E_R^{\text{th}}}^{E_R^\text{max}}\!\!\!\! d E_R \,
                        \epsilon^{(i)}_\text{eff}(E_R) \frac{dN_{\chi}}{dE_\chi}\frac{d\sigma^{(i)}_{\text{det}}}{dE_R}\,, 
\end{split}
\end{align}
where $\epsilon^{(i)}_\text{eff}$ is the detection efficiency of an event
with recoil energy $E_R$. $N_T^{(i)}$ is the target number
density. For mQ, the signal is dominated by the scattering on $e^-$
and by the (coherent) scattering on $C$ atoms. Hence $i = e^-,\, C$
and we compute $N_T^{(i)}$ according to the chemical composition of
the detector. For BDX, the detector is made of CsI, but
given a threshold in deposited energy of $E_R^{\rm th} = 500~\MeV$,
we consider $i = p,  e^-$ as targets.
The corresponding detection cross sections
$d\sigma^{(i)}_{\text{det}}/dE_R$ are given by the elastic recoil
cross sections found in App.~\ref{sec:elast-scatt-cross}; the
form-factors $F_{E,M}$ that are to be used for the respective targets
$i$ are provided in App.~\ref{sec:had-ten}.
The energy of $\chi$ particles, $E_\chi$, is limited by the electron
beam energy, $E_{\chi}^{\rm max}$, as given in \eqref{eq:inputmax}.  In turn, $E_\chi $ determines the maximal
recoil energy of target $i$, $E_R^\text{max}$, see \eqref{eq:recoilmax}.

\begin{figure*}[tb]
\centering
\includegraphics[width=0.48\textwidth]{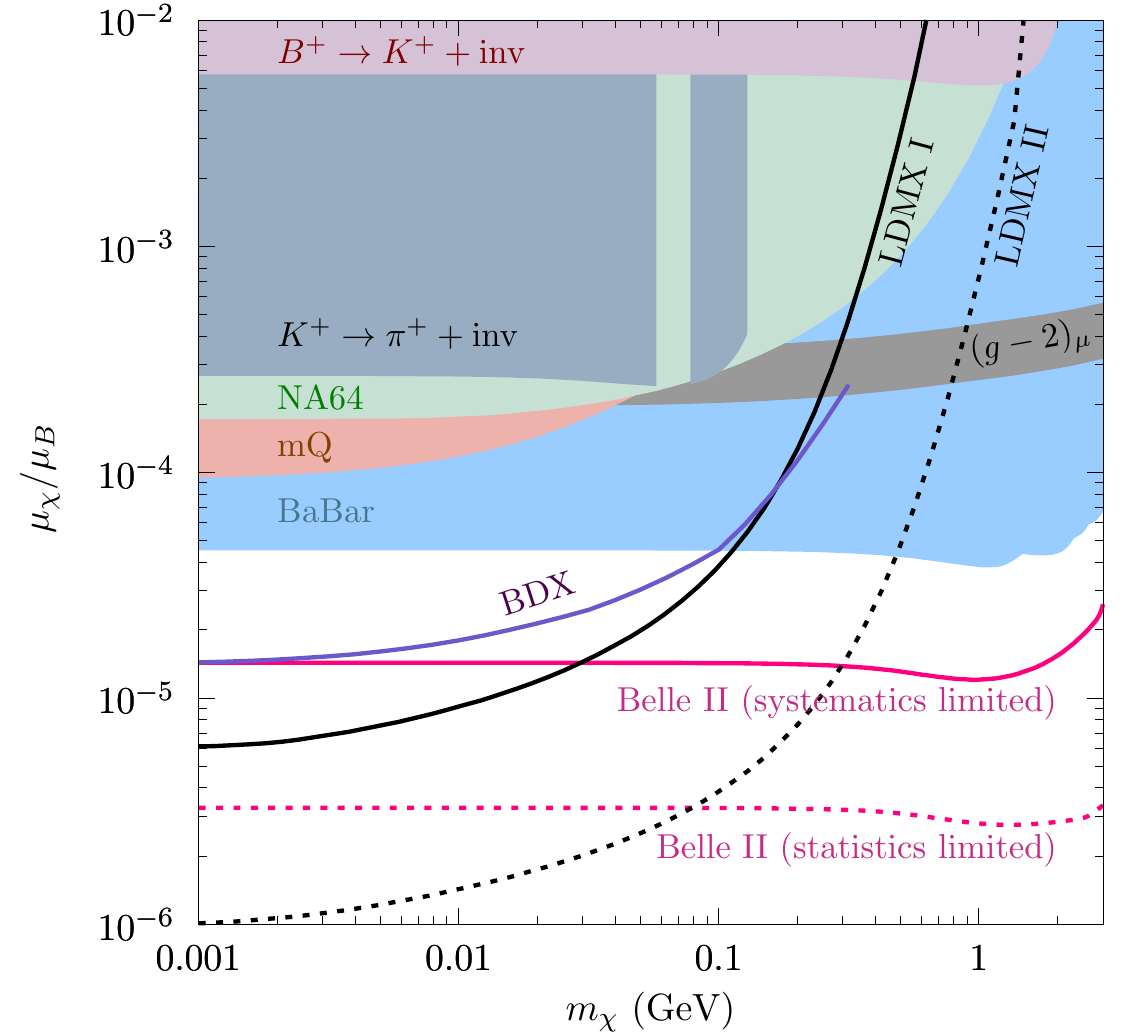} \quad
\includegraphics[width=0.48\textwidth]{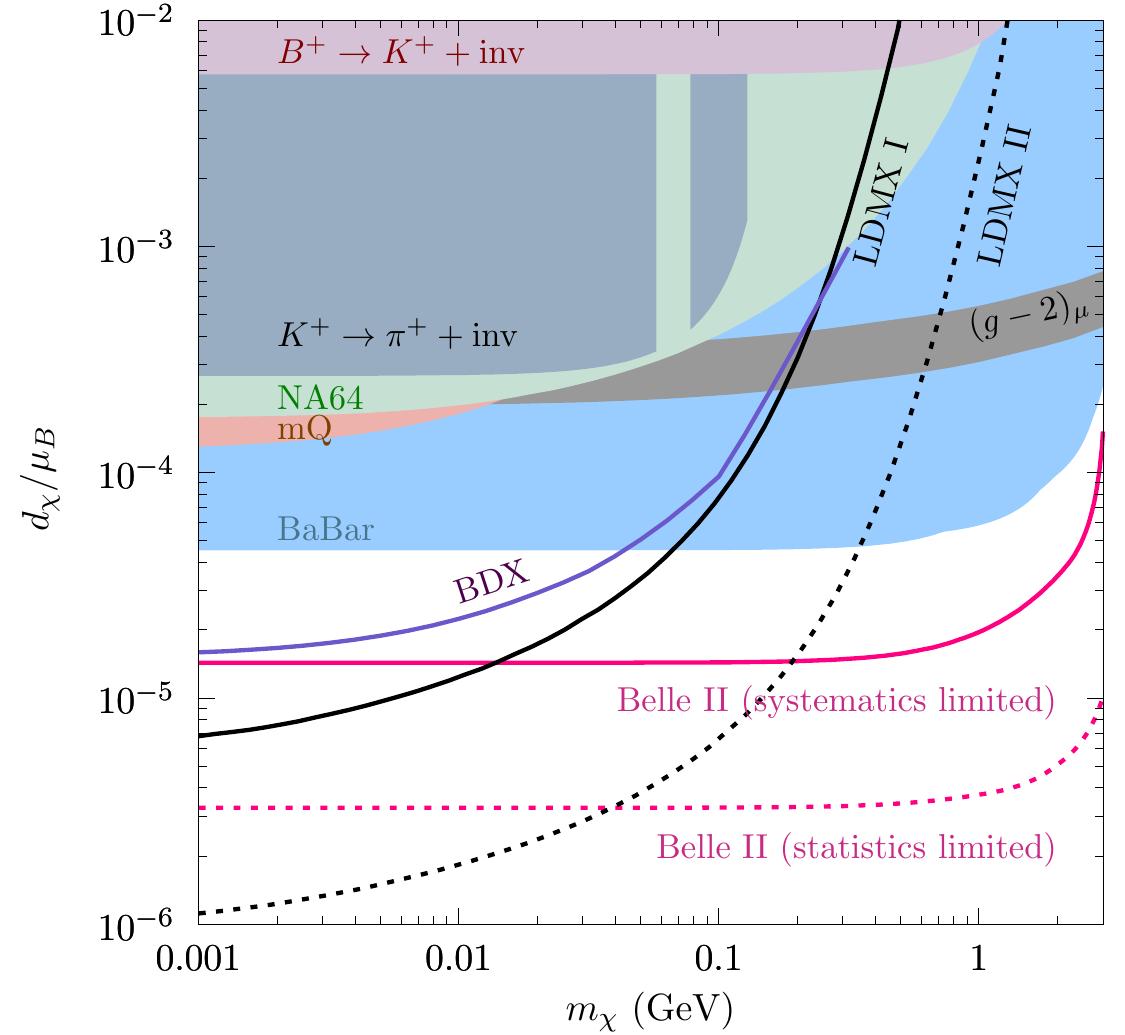} 
\caption{Constraints (shaded regions) and projections (lines) on the electromagnetic form factors for fermions with an MDM (left) and an EDM (right).}\label{fig:limits_dim5}
\end{figure*}

\begin{figure}[tb]
\centering
\includegraphics[width=0.48\textwidth]{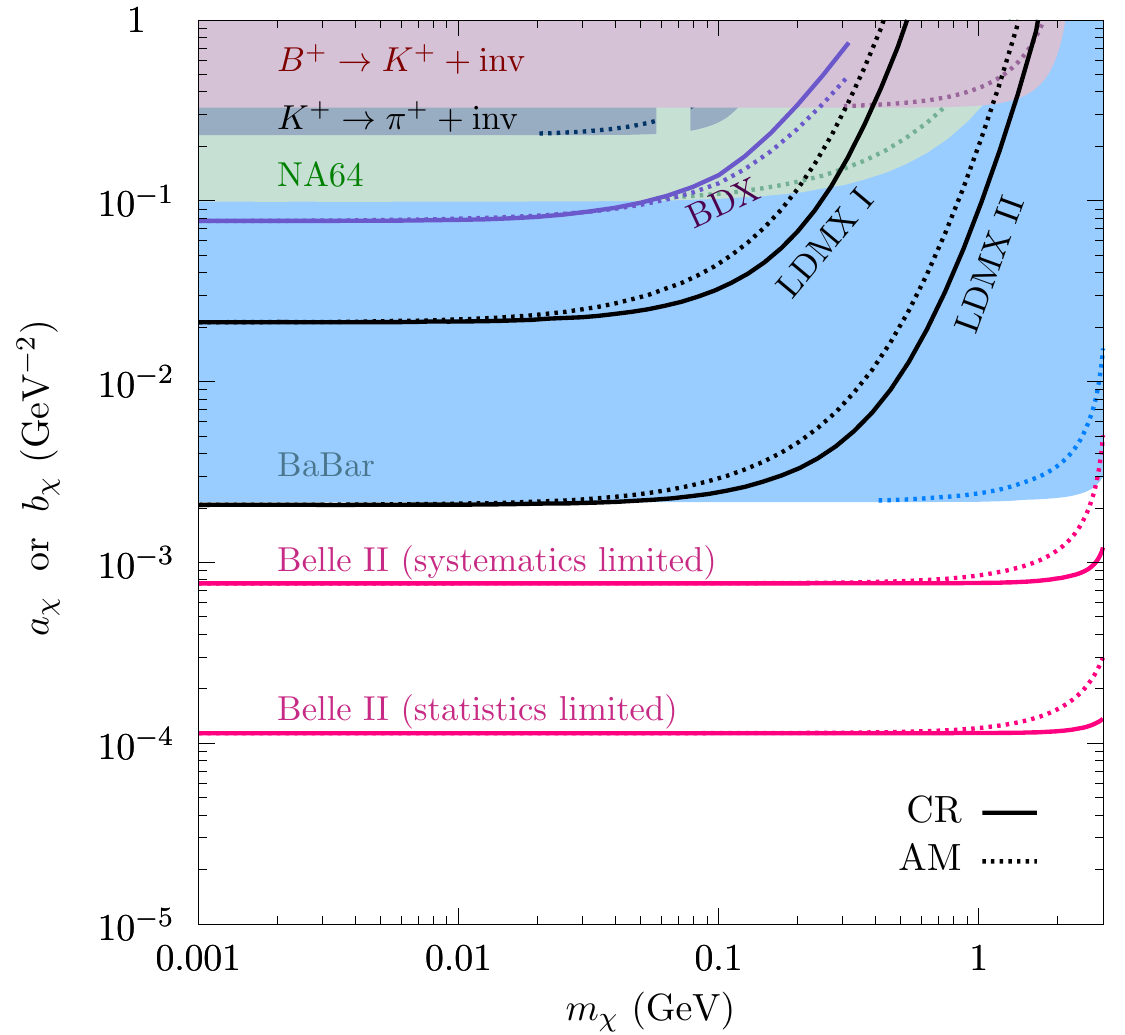} 
\caption{Same as Fig.~\ref{fig:limits_dim5} for fermions with dim-6
  operators. The solid (dotted) lines show the
  constraints and projections from the experiments for CR (AM) as
  labeled.}\label{fig:limits_dim6}
\end{figure}

In the right panel of Fig.~\ref{BDX:Echispectrum} we show the
theoretical recoil spectra $dN^{(i)}/dE_R$ that are obtained from
(\ref{fixedT:events}) by dropping the integration over $E_R$ and
before accounting for detection efficiencies, \textit{i.e.}~by setting
$\epsilon_{\text{eff}}=1$ for the same values of parameters as in the
left panel of Fig.~\ref{BDX:Echispectrum}. For
$m_{\chi}\lesssim 1\,\GeV$ the scattering on electron is most
important for setting any limit on the interactions. Once
$m_{\chi} \gtrsim 1\,\GeV$ the maximum electron recoil energy
$E_R^{\rm max}$ falls below the detector threshold $E_R^{\rm th} $ and
any further sensitivity relies on recoil of protons (and, in
principle, on neutrons). For completeness, we also note that for
Fig.~\ref{BDX:Echispectrum} we have neglected the momentum 
distribution of protons inside the target nucleus. The Fermi momentum 
of nucleons is $p_F\sim 250\,\MeV$ and it will hence affect the shape
of the proton recoil spectrum at energies $E_R < 1\,\GeV$, but be
negligible for the bulk of the shown spectra.

For the value of the detection efficiency $\epsilon_\text{eff}$, we
use $0.5$ for mQ~\cite{Prinz:2001qz}. Based on the simulations in
\cite{Battaglieri:2016ggd}, a numerical function of $E_R$ has been
adopted for electron recoils in BDX, giving
$\epsilon_\text{eff}\sim 5\%$ for $E_R=0.5$\,GeV and
$\epsilon_\text{eff}\sim 1\%$ for $E_R=1.5$\,GeV. However, there is
currently no data available for proton recoils in BDX with
$E_R\ge 0.5$\,GeV, so we derive the BDX projections based on electron
recoils only.

By requiring $N_{\rm sig}$, calculated from \eqref{fixedT:events},
to be below the values given above, the 90\% C.L. (projected) limits
on the electromagnetic factors of $\chi$ have been derived and are
shown in Figs.~\ref{fig:limits_dim5} and \ref{fig:limits_dim6}. For $m_\chi \ge 0.5$\,GeV and $E_0= 11\,$GeV, no 
incoming $\chi$ can deposit energy more than 0.5\,GeV by its scattering off an
electron. So that the high-$m_\chi$ end in  the parameter space of Figs.~\ref{fig:limits_dim5}--\ref{fig:limits_dim6} will not be constrained by electron recoils in BDX. In general, BDX will have stronger
sensitivity than mQ to the models studied here. In contrast,  the mQ
  experiment remains more sensitive to millicharged particles due to
  its low threshold in recoil energy.

The summary plots, Figs.~\ref{fig:limits_dim5} and \ref{fig:limits_dim6}, show that while the projections of BDX and LDMX for
dim-5 interactions 
reach beyond Belle II for small $m_\chi$, for dim-6 interactions they
flatten out below $\sim 0.1 \,$\,GeV without surpassing Belle II in
sensitivity. The difference of slopes of the BDX and LDMX curves in
Figs.~7 and 8 is understood as follows: first, note that the
$\chi$-pair production rate receives significant contributions from
the e-N forward scattering part which in turn is regulated by
$m_{\chi}$; the lower boundary of $s_{\chi\bar\chi}$ is $4
m_\chi^2$. Second, as seen in Fig.~3, the $m_\chi$-dependence of the
production rate becomes weaker for higher mass-dimension
interactions, since the $s_{\chi\bar\chi}$ factors in the emission
pieces, (\ref{eqn:LDMX_dark_matter_part_MDM}--\ref{eqn:LDMX_dark_matter_part_CR}), tend to reduce the contribution of the forward
scattering part. Taken together, this explains the difference in
slope. It is also for these reasons that the millicharged case (dim-4)~\cite{Prinz:2001qz} has an even bigger slope than the dim-5 ones.
 
\section{SM Precision Observables and High Energy Searches}
\label{sec:precision-tests}

Focusing on a mass of $\chi$ below several GeV, SM precision
observables and LEP measurements in general play an important role in
constraining the existence of the various electromagnetic
interactions. In the following we update bounds on EDM and MDM SM
precision tests first obtained in~\cite{Sigurdson:2004zp}, going
beyond parametric estimates where possible, and provide new ones on CR
and AM. In addition we compute the constraints from flavor physics, as
well as from missing energy searches at LEP.

\subsection{Running of \texorpdfstring{\boldmath$\alpha$}{alpha}}
\label{sec:running-of-alpha}

In the electroweak theory, Fermi's constant $G_F$ and the masses of
$W$ and $Z$ bosons are related and receive calculable quantum
corrections, summarized in the $\Delta r$
parameter~\cite{Sirlin:1980nh}.  The SM prediction is
$\Delta r_{\rm SM} = 0.03672 \pm 0.00019$~\cite{Tanabashi:2018oca}
and the observed value can be obtained from
  \begin{align}
    \label{eq:Deltarobs}
    \Delta r_{\rm obs} = 1 - \frac{A_0^2}{M_W^2 s_w^2} = 0.03492\pm 0.00097 , 
  \end{align}
  where $A_0 = (\pi \alpha/\sqrt{2}G_F)^{1/2}$ and
  $s_w^2 = 1-M_W^2/M_Z^2$. For the actual number we have used the
  global averages of $M_W$, $M_Z$, and $A_0$ and their errors quoted
  in~\cite{Tanabashi:2018oca}.
  The range in which new physics can
  contribute is hence limited at 95\% C.L. as%
  \footnote{The tension between measurement and observation is
    1.8$\sigma$ and any positive contribution to $\Delta r_{\rm new}$
    is hence already tightly constrained.}
  \begin{align}
    \label{eq:DeltarRange}
    -0.0038 < \Delta r_{\rm new} <   0.00018 .
  \end{align}

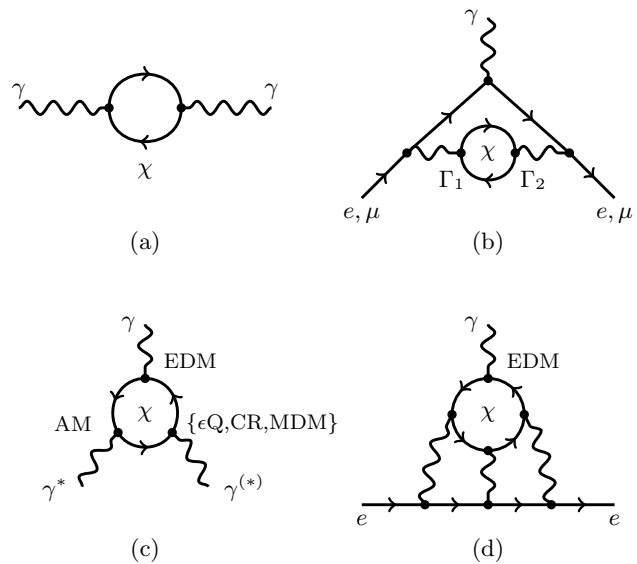
\begin{figure}[tb]
  \centering
\begin{tikzpicture}[line width=1.1 pt, scale=1.2]

\begin{scope}[shift={(0,0)}]
\coordinate[label=above :$\gamma$]  (e1) at (-0.4,0);
\coordinate[label=above :$\gamma$]  (e2) at (2.4,0);
\coordinate[label=below :] (v1) at (0.6,0);
\coordinate[label=below right:] (v2) at (1.4,0);
		
\draw[photon] (e1) to (v1);
\draw[photon] (e2) to (v2);
\draw[fermion, out=90,in=90, distance=0.5cm] (v1) to (v2);
\draw[fermion, out=-90,in=-90, distance=0.5cm] (v2) to (v1);
		
\filldraw (v1) circle (1pt);
\filldraw (v2) circle (1pt);

\draw (1,-0.7) node{$\chi$};
\draw (1,-1.5) node{(a)};
\end{scope}

\begin{scope}[shift={(3.8,0)}]
\coordinate[label=above :]  (e1) at (0.1,-0.5);
\coordinate[label=above :]  (e2) at (1.9,-0.5);
\coordinate[label=below :] (v1) at (0.7,-0.5);
\coordinate[label=below right:] (v2) at (1.3,-0.5);
\coordinate[label=below :{$e,\mu$}] (e3) at (2.4,-1);
\coordinate[label=below :{$e,\mu$}] (e4) at (-0.4,-1);
\coordinate[label=below :] (e5) at (1,0.3);
\coordinate[label=left :$\gamma$] (e6) at (1,1);
		
\draw[photon] (e1) to (v1);
\draw[photon] (v2) to (e2);
\draw[photon] (e6) to (e5);
\draw[fermion, out=90,in=90, distance=0.4cm] (v1) to (v2);
\draw[fermion, out=-90,in=-90, distance=0.4cm] (v2) to (v1);
\draw[fermion] (e2) to (e3);
\draw[fermion] (e4) to (e1);
\draw[fermion] (e5) to (e2);
\draw[fermion] (e1) to (e5);
		
\filldraw (v1) circle (1pt);
\filldraw (v2) circle (1pt);
\filldraw (e1) circle (1pt);
\filldraw (e2) circle (1pt);
\filldraw (e5) circle (1pt);

\draw (1,-0.5) node{$\chi$};
\draw (1.5,-0.8) node{$\Gamma_2$};
\draw (0.6,-0.8) node{$\Gamma_1$};
\draw (1,-1.5) node{(b)};
\end{scope}

\begin{scope}[shift={(0,-3.4)}]
\coordinate[label=left :$\gamma$]  (e1) at (1,1);
\coordinate[label=right :]  (e2) at (1,-1);
\coordinate[label=below :] (v1) at (1,0.4);
\coordinate[label=below :] (v2) at (1,-0.4);
\coordinate[label=below :] (v3) at (1.3,-0.2);
\coordinate[label=above left :] (v4) at (0.7,-0.2);
\coordinate[label=below :] (v5) at (1.7,-0.8);
\coordinate[label=below :] (v6) at (0.3,-0.8);

\coordinate[label=below :$\gamma^{*}$] (e3) at (0.,-0.6);
\coordinate[label=below :$\gamma^{(*)}$] (e3) at (2.1,-0.55);

\draw[photon] (e1) to (v1);
\draw[fermion, out=180,in=120, distance=0.3cm] (v1) to (v4);
\draw[fermion, out=-60,in=-120, distance=0.25cm] (v4) to (v3);
\draw[fermion, out=60,in=0, distance=0.3cm] (v3) to (v1);
\draw[photon, bend angle=10, bend right] (v5) to (v3);
\draw[photon, bend angle=10, bend left] (v6) to (v4);

\filldraw (v1) circle (1pt);
\filldraw (v3) circle (1pt);
\filldraw (v4) circle (1pt);

\draw (1,0) node{$\chi$};
\draw (0.2,-0.1) node{\footnotesize AM};
\draw (1.5,0.6) node{\footnotesize EDM};
\draw (2.3,-0.1) node{\footnotesize \{$\epsilon$Q,CR,MDM\}};
\draw (1,-1.5) node{(c)};
\end{scope}

\begin{scope}[shift={(3.8,-3.4)}]
\coordinate[label=left :$\gamma$]  (e1) at (1,1);
\coordinate[label=right :]  (e2) at (1,-1);
\coordinate[label=below :] (v1) at (1,0.4);
\coordinate[label=below :] (v2) at (1,-0.4);
\coordinate[label=below :] (v3) at (1.4,0);
\coordinate[label=below :] (v4) at (0.6,0);
\coordinate[label=below :] (v5) at (1.7,-1);
\coordinate[label=below :] (v6) at (0.3,-1);

\coordinate[label=below :$e$] (e3) at (-0.4,-1);
\coordinate[label=below :$e$] (e4) at (2.4,-1);
		
\draw[photon] (e1) to (v1);
\draw[photon] (v2) to (e2);
\draw[fermion] (e3) to (v6);
\draw[fermion] (e2) to (v5);
\draw[fermion] (v5) to (e4);
\draw[fermion] (v6) to (e2);
\draw[fermion, out=180,in=90, distance=0.22cm] (v1) to (v4);
\draw[fermion, out=-90,in=180, distance=0.22cm] (v4) to (v2);
\draw[fermion, out=0,in=-90, distance=0.22cm] (v2) to (v3);
\draw[fermion, out=90,in=0, distance=0.22cm] (v3) to (v1);
\draw[photon, bend angle=10, bend right] (v5) to (v3);
\draw[photon, bend angle=10, bend left] (v6) to (v4);

\filldraw (v1) circle (1pt);
\filldraw (v2) circle (1pt);
\filldraw (v3) circle (1pt);
\filldraw (v4) circle (1pt);
\filldraw (e2) circle (1pt);
\filldraw (v5) circle (1pt);
\filldraw (v6) circle (1pt);

\draw (1.5,0.6) node{\footnotesize EDM};
\draw (1,0) node{$\chi$};
\draw (1,-1.5) node{(d)};
\end{scope}

\end{tikzpicture}
\caption{Loop diagrams relevant in the discussion of SM corrections
  discussed in Sec.~\ref{sec:precision-tests}. The vacuum polarization
  diagram (a) contributes to the running of $\alpha$. The leading
  contribution to $g-2$ is from diagram (b), which we evaluate for
  $\Gamma_1=\Gamma_2$. It can be shown that a similar diagram with
  $\Gamma_1={\rm EDM}$ and $\Gamma_1\neq\Gamma_2$ does not contribute
  to an electric dipole moment of the electron.  Diagram (c) shows an
  \textit{a priori} non-vanishing $\chi$-loop with three photons
  attached. Diagram (d) shows a 3-loop contribution to the
  electron electric dipole moment.\label{fig:loop_diag}}
\end{figure}

  Electromagnetic interactions of $\chi$ will contribute to $\Delta r$
  through the running of $\alpha(Q^2)$ they induce. Explicitly,
  $\Delta r_{\rm new} \simeq \Pi(-M_Z^2) - \Pi(0)$ where $\Pi(q^2)$
  is the polarization function (Fig.~\ref{fig:loop_diag}a) at photon momentum $q^2 =
  -Q^2$. Details on the calculation of the photon vacuum polarization
  are found in App.~\ref{sec:phot-vacu-polar}.
  Equation~(\ref{eq:DeltarRange}) then implies for MDM and EDM 
  \begin{align}
       \label{PT:Wboson}
    |\mu_{\chi}|~\text{or}~ |d_{\chi}|~ < 3.2 \times 10^{-6} \mu_{B} , 
  \end{align}
  independent of $m_{\chi}$, saturating the upper limit
  in~(\ref{eq:DeltarRange}) and improving the previously obtained
  limits in~\cite{Sigurdson:2004zp} by half an order of magnitude. In turn, for
  AM and CR we obtain
  \begin{align}
      	|a_{\chi}|~\text{or}~|b_{\chi}|~ < 3.2 \times 10^{-5}\, \GeV^{-2} ,
  \end{align}
  saturating the lower limit in (\ref{eq:DeltarRange}).  The latter
  interactions hence contribute with opposite sign.

  The above limits are at or even stronger than the projected
  sensitivity from the various experiments considered above. As a
  quantum effect, they are, however, susceptible to a model-dependence
  that a direct observation of a particle is not. If the theory
  contains additional states at a mass scale below~$M_Z$ or if $\chi$
  carries more than one EM form factor, cancellations in $\Delta r_{\rm new}$
  can occur. It is for this reason that we do not explicitly show
  those limits in Figs.~\ref{fig:limits_dim5} and
  \ref{fig:limits_dim6}. Finally, for completeness we also record a
  limit obtained from (\ref{eq:DeltarRange}) on the positive
  contribution from $\epsilon$Q, $\epsilon \lesssim 0.1$, weakening
  for increasing $\chi$-mass to $\epsilon \lesssim 0.3$ at
  $m_{\chi} = 10\,\GeV$.

\subsection{Muon \texorpdfstring{\boldmath$g-2$}{g-2}}
\label{sec:muon-g-2}

Here we consider the effect of the various interactions on the
anomalous magnetic moment of the muon
$a_{\mu} = \frac{1}{2}(g-2)_{\mu} $. There is a tantalizing
long-standing discrepancy between the measured
value~\cite{Bennett:2006fi} and prediction in the SM
(see~\cite{Jegerlehner:2009ry} for a summary),
\begin{align}
  \Delta a_{\mu} = a_{\mu}^{\rm exp} - a_{\mu}^{\rm SM}
  = (290 \pm 90 )\times 10^{-11} ,
\end{align}
indicating a $3-4\sigma$ tension. Hence, any positive contribution at
the level of $3\times 10^{-9}$ may solve the $(g-2)$ puzzle, and much
anticipated upcoming experiments~\cite{Grange:2015fou,Saito:2012zz}
will further weigh in on the anomaly in the near future.

In the current context, the lowest-order contribution to $g-2$ enters
through the vacuum polarization diagram, Fig.~\ref{fig:loop_diag}b
where $\chi$ is in the loop of the internal photon line. In QED, the
2-loop leptonic contribution is of course well
known~\cite{Petermann:1957hs,Sommerfield:1957zz}, readily evaluated by
noting the correction from the vacuum polarization to the photon
propagator, and one may in principle proceed in analogy using
(\ref{eq:vacuumpolsdiv}), appropriately renormalized.
A quicker way is to estimate the correction $\Delta a_{\mu}$ from
$\chi$ by using the dispersion relation and optical theorem from the
total cross section of $\chi$-pair creation, $\sigma_{e^+e^-\to \chi\bar\chi}$,
similar to what is done with hadronic contributions in SM,
\begin{align}
  \label{eq:amu-dispersive}
  \Delta a_{\mu} = \frac{1}{4\pi^3} \int_{4m_{\chi}^2}
  ds\, \sigma_{e^+e^-\to \chi\bar\chi}(s) K(s) \,,
\end{align}
where $K(s)/(\pi\alpha)$ is the contribution to the muon anomalous
magnetic moment from a photon of mass $\sqrt{s}$,%
\footnote{$K(s)$ is thereby equivalent to the expression obtained in
  the context of dark photon contributions to
  $g-2$~\cite{Gninenko:2001hx,Boehm:2003hm,Pospelov:2008zw}.}
\begin{align}
  K(s) = \int_0^1 dx\, \frac{x^2(1-x)}{x^2 + \frac{s}{m_{\mu}^2} (1-x)} \,.
\end{align}
The cross section $\sigma_{e^+e^-\to \chi\bar\chi}$ is given in
(\ref{eqn:collider_prod_cs}) with $s=\sdark$. The integral
(\ref{eq:amu-dispersive}) has to be cut off such that the validity of
the effective theory is respected. This yields a reasonable estimate
for dim-5 operators, and we show the resulting 90\% bands in
Figs.~\ref{fig:limits_dim5}. Since $K(s)\sim 1/s$, for dim-6
operators the energy dependence makes the same integral linearly
sensitive on the chosen cut-off so that we do not show any region in
Fig.~\ref{fig:limits_dim6}. However, (\ref{eq:amu-dispersive}) is
sufficient to convince oneself that both AM an CR only yield
interesting contributions in an otherwise deeply excluded region; we
leave a systematic study of this to a dedicated work.

\subsection{Induced electric dipole moments}
\label{sec:electr-dipole-moment}

A sensitive and ``background-free'' probe of new physics are electric
dipole moments of SM particles. Most recently, the ACME collaboration
improved the limit on the electron EDM $d_e$ by an order of
magnitude~\cite{Andreev:2018ayy},
\begin{align}
  \label{eq:de-limit}
  |d_e| <  1.1\times 10^{-29}\,e\, \cm . 
\end{align}
Given the strong limit in (\ref{eq:de-limit}), for simplicity, we
restrict our attention to the electron EDM; for a general discussion
of neutron and atomic EDMs cf.~\cite{Pospelov:2005pr} and references
therein.

Electron electric dipole moments may be generated by the loops of
$\chi$-particles. Since the interaction is T- and P-odd it can only be
induced in a loop diagram with an odd number of insertions of EDM
vertices of $\chi$, as it is the only T-odd interaction among the ones
we consider.  Using dim-5 MDM and EDM operators,
Ref.~\cite{Sigurdson:2004zp} then argued that such dipole moments
appear at the 3-loop level with four photons attached to the
$\chi$-loop, as lower order diagrams vanish by virtue of Furry's
theorem. The interaction $d_e$ hence appears through a combination
$d_{\chi}\mu_{\chi}^3$ and $d_{\chi}^3\mu_{\chi}$.

Furry's theorem relies on C-invariance of the considered interactions
and moving to dim-6 operators, we observe that the AM interaction is
C-odd. Thereby, a $\chi$-loop with 3~photons that are attached through
EDM, AM and one out of the set of interactions
$\{\epsilon\text{Q}, \text{MDM}, \text{CR}\}$ as shown in
Fig.~\ref{fig:loop_diag}c does not vanish \textit{a priori}.%
\footnote{A way to see it is by following a proof for a generalized
  version of Furry's theorem~\cite{1951PThPh...6..614N}. Also note
  that the photons attached via AM and CR need to be virtual.}
However, when attached to the electron line, this does not imply a
2-loop contribution to $d_e$ as the resulting operator would be P-odd;
we also verify the vanishing 2-loop contribution through explicit
calculation using a projector onto $d_e$~\cite{Czarnecki:1996rx}.

Hence, we conclude that $d_e$ is only induced
at the 3-loop level, an example of which is shown in
Fig.~\ref{fig:loop_diag}d. A parametric estimate of its size would
then be,
\begin{align}
  \label{eq:de}
 d_e \sim  e^3 m_e d_{\chi}  \times (\rm mass)^{-1} \,,
\end{align}
where the factor of $m_e$ originates from the necessary helicity-flip
on the electron line to which 3 photons are attached
($e^3$-factor). The factor $(\rm mass)^{-1}$ is built from the other
interactions present in the diagram, saturated by powers of $m_{\chi}$
or, possibly, $m_e$ to obtain the correct dimension, such as
$\epsilon^3e^3/m_{\chi}$, $\mu_{\chi}^3m_{\chi}^2$,
$d_{\chi}^2\mu_{\chi}m_{\chi}^2$, $b_{\chi}d_{\chi}^2m_{\chi}^3$ and so on. We have
omitted any loop factors $(16\pi)^{-2n}$ with $n=3$ as we do not know
if other large combinatorial factors are present.%
\footnote{$d_e$ induced by a possible EDM of the SM $\tau$ lepton,
  $d_{\tau}$, at (finite) 3-loop level has explicitly been calculated
  in \cite{Grozin:2009jq} with the result
  $d_e = a (m_e/m_{\tau}) d_{\tau} (\alpha/\pi)^3 $, where $a$ turned
  out to be a number close to unity. For MDM/EDM interactions of
  $\chi$, Ref.~\cite{Sigurdson:2004zp} saturates the dimensions by
  powers of $m_e$ and includes a double-logarithmic enhancement
  $\log^2(m_{\chi}/m_e)$. }
Saturating the present limit (\ref{eq:de-limit}) we infer that the
suppression mass scale in (\ref{eq:de}) is at the level of
$ \sim 10^{13}\,\GeV \, (d_{\chi}/\mu_B)$ or larger. For example, if
the electron EDM then solely appears through the combination of dim-5
operators, such estimate points to a related inverse effective scale
$(10^{-5}-10^{-6}) \, \mu_B \sqrt{\GeV/m_{\chi}}$, implying an important constraint
on the presence on joint EDM and MDM interactions. A
systematic study of this goes beyond the scope of this work, but
similar limits on other combination of operators are readily evaluated
from the above arguments.

\subsection{Rare meson decays}
\label{sec:meson-decays}

Measurements and searches of the invisible decay width of various
mesons place a constraint on light dark particles~\cite{Bird:2004ts}.
Strong constraints are in particular obtained from data on $B^+$ and
$K^+$ decays.
The decays $K^+ \to \pi^+ \chi\bar\chi$ and $B^+ \to K^+ \chi\bar\chi$
are closely related to the semi-leptonic decay with a charged lepton
pair in the final state, $K^+\to \pi^+ l^+ l^- $ and
$B^+\to K^+ l^+ l^- $, as both are accompanied by the emission of
$\gamma^{*}$ in the flavor changing $s\to d$ and $b\to s$
transitions. Hence, by evaluating explicitly the decay---details are
found in App.~\ref{sec:meson-decays-app}---we find
\begin{align}
  \frac{ \Gamma(K^+\to\pi^+\bar\chi\chi) }{\Gamma(K\to\pi e^+e^-) }
  \simeq 1.9\times 10^4
  \left( \frac{\mu_{\chi}~\text{or}~d_{\chi}}{\mu_B} \right)^2 ,
\end{align}
for the MDM and EDM interactions and
\begin{align}
  \frac{ \Gamma(K^+\to\pi^+\bar\chi\chi) }{\Gamma(K\to\pi e^+e^-) }
  \simeq 2.6\times 10^{4}
  \left( \frac{a_{\chi}~\text{or}~b_{\chi}}{\TeV^2} \right)^2 . 
\end{align}
With the experimental value
${\rm Br}(K\to\pi e^+e^-)= (3.00\pm 0.09)\times
10^{-7}$ \cite{Tanabashi:2018oca} together with the strong constraint
on the branching ratio
$K^+\to \pi^+ + \text{inv} \lesssim 4\times 10^{-10}
$~\cite{Anisimovsky:2004hr, Artamonov:2009sz} which is applicable in
the pion momentum ranges $211\,\MeV < p_{\pi} < 229\,\MeV $ and
$140\,\MeV < p_{\pi} < 199\,\MeV $ allows to set a constraint in the
$\chi$-mass range $m_{\chi} < 58\,\MeV$ and
$76\,\MeV < m_{\chi} < 130\,\MeV$. We obtain, approximately,
  \begin{align}
       \label{PT:Ktopimdm} 
    |\mu_{\chi}|~\text{or}~ |d_{\chi}|~ & \lesssim  3\times 10^{-4} \mu_{B} , \\
    |a_{\chi}| ~ \text{or}~ |b_{\chi}|~ &  \lesssim  0.2 \,\GeV^{-2} ,
  \end{align}
  and show the numerically obtained curves in
  Figs.~\ref{fig:limits_dim5} and \ref{fig:limits_dim6}. 
  Equation (\ref{PT:Ktopimdm}) also shows that only MDM and EDM
  interactions are reasonably well constrained by rare $K$-decays.

  To estimate a constraint on the electromagnetic interactions from a
  limit on
  ${\rm Br }(B^+ \to K^+ + \text{inv}) < 7 \times 10^{-5}
  $~\cite{Aubert:2003yh} one can proceed in a similar fashion and
  relate the branching ratio ${\rm Br}(B^+ \to K^+ \chi \bar\chi)$ to
  ${\rm Br}( B^+ \to K^+ \mu^+ \mu^- ) = (4.41\pm 0.23)\times 10^{-7}
  $~\cite{Tanabashi:2018oca}.  We obtain,
  \begin{align}
  \frac{ \Gamma(B^+\to K^+\bar\chi\chi) }{\Gamma(B^+\to K^+ \mu^+\mu^-) }
  \simeq 4.8\times 10^6
  \left( \frac{\mu_{\chi}~\text{or}~d_{\chi}}{\mu_B} \right)^2 ,
\end{align}
for the MDM and EDM interactions and
\begin{align}
  \frac{ \Gamma(B^+\to K^+\bar\chi\chi) }{\Gamma(B^+\to K^+ \mu^+\mu^-) }
  \simeq 1.6\times 10^3
  \left( \frac{a_{\chi}~\text{or}~b_{\chi}}{\GeV^2} \right)^2 . 
\end{align}
From those numbers we limit the interactions away from the kinematic
threshold as,
  \begin{align}
       \label{PT:BtoKmdm}
    |\mu_{\chi}|~\text{or}~ |d_{\chi}|~ & \lesssim  6\times 10^{-3} \mu_{B} , \\
    |a_{\chi}| ~ \text{or}~ |b_{\chi}|~ &  \lesssim  0.3 \,\GeV^{-2} ,
  \end{align}
  where the latter limit on CR/AM again implies a UV-scale that would
  be well below the electroweak scale. The numerically obtained curves
  are again included in in Figs.~\ref{fig:limits_dim5} and
  \ref{fig:limits_dim6}.

\subsection{Invisible \texorpdfstring{\boldmath$Z$}{Z}-width}
\label{sec:invisible-z}

  A model-dependent constraint arises if $\chi$ has
  fundamental interactions with the hypercharge through similar
  operators as in (\ref{eq:Lagrangians}), but with $F_{\mu\nu}$
  replaced by the hypercharge field strength $F_{\mu\nu}^Y$ and
  $A_{\mu}$ replaced by the associated gauge boson $B_{\mu}$. The
  $Z$-decay into $\chi\bar\chi$ is possible for $m_{\chi}< M_Z/2$ and
  the partial width is given by
\begin{align}
  \Gamma_{Z\to \chi\bar\chi} =  \frac{s_w^2 \, f(M_Z^2)}{16\pi M_Z} \sqrt{1- \frac{4m_{\chi}^2}{M_Z^2}}\,, 
\end{align}
where the functions $f(M_Z^2)$ for the various operators are found in
(\ref{eqn:LDMX_dark_matter_part}). The experimental measurement of the
invisible with
$\Gamma(Z\to {\rm inv})_{\rm exp} = 499.0\pm 1.5 \,
\MeV$ together with the SM prediction
$\Gamma(Z\to {\rm inv})_{\rm SM} = 501.44\pm 0.04 \, \MeV $~\cite{Tanabashi:2018oca}
 limits any additional contribution to
$\Gamma(Z\to {\rm inv})_{\rm new}< 0.56\,\MeV$
 at $95\%$~C.L. In the
kinematically unsuppressed region this implies,
  \begin{subequations}
  \begin{align}
      |\mu^{(Y)}_{\chi}| ~ \text{or}~ |d^{(Y)}_{\chi}|~ & < ~1.7\times 10^{-6}\mu_B , \\
     |a^{(Y)}_{\chi}| ~ \text{or}~ |b^{(Y)}_{\chi}|~ & < ~ 3.8\times 10^{-6}\,\GeV^{-2}.
  \end{align}
\end{subequations}

\subsection{Missing Energy at high-energy colliders}
\label{sec:colliders}
At high-energy colliders, sub-GeV particles $\chi$ can be produced
kinematically unsuppressed, yielding limits on the couplings that are
independent of $m_\chi$. Among them, the strongest bounds on
interactions studied here come from the LEP collider. 
Ref.~\cite{Fortin:2011hv} has studied the L3 data with an
integrated luminosity of 619/pb at CM energies
$\sqrt{s}=188.6-209.2$\,GeV~\cite{Achard:2003tx}. Using the
mono-photon channel, it was found that MDM and EDM models are
constrained to
  \begin{align}
      |\mu_{\chi}| ~ \text{or}~ |d_{\chi}|~ & < ~ 1.3\times 10^{-5}\mu_B .
  \end{align}
  The results can also be generalized to AM and CR interactions. Here
  we only adopt the high-energy single photon selection, which
  requires one photon with transverse momentum larger than
  $0.02\sqrt{s}$. Within the polar angular range
  $14^\circ \le \theta_\gamma \le 166^\circ$, in total 1898 events
  were reported while the SM expectations are 1905.1 events, mostly
  from $e^+e^-\to \nu\bar\nu\gamma(\gamma)$. 
  Following \cite{Fortin:2011hv}, we use the the eight data subsets with different kinematic regions of $\sqrt{s}$ studied in the single photon search in \cite{Achard:2003tx} and place 90\% C.L.~bounds on the couplings from the data subset that leads to the best constraints using the CLs method. %
  We take a flat selection
  efficiency of 71\% and a systematic background uncertainty of $1.1\%$, and obtain
  \begin{align}
      |a_{\chi}| ~ \text{or}~ |b_{\chi}|~ & < ~ 1.5\times 10^{-5}\, \GeV^{-2}
  \end{align}
  for the AM and CR models. Taking into account the low-energy
  selection would not significantly change these results. The limits
  are also shown in Figs.~\ref{fig:cosmo_limits} and
  \ref{fig:cosmo_limits_dim6}. 
While the bounds above are stronger than those derived from LHC data~\cite{Barger:2012pf, Gao:2013vfa}, future experiments, such as ILC and
  HL-LHC, may improve the sensitivity on these couplings by one to two
  orders of magnitude~\cite{Kadota:2014mea, Primulando:2015lfa, Alves:2017uls}.

\section{\texorpdfstring{\boldmath$\chi$}{chi} Dark Matter}
\label{sec:dark-matt-electr}

If $\chi$ is long-lived on cosmological time-scales, it may be
DM. Additional laboratory, astrophysical, and cosmological
limits then apply, which we discuss in this section.

\subsection{Direct detection}
\label{sec:direct-detection}

Our primary interest is in the GeV and sub-GeV mass scale in
$m_{\chi}$. The elastic scattering cross sections on nuclei are given
by the non-relativistic expansion of (\ref{eq:el-cs}) in the incident
energy of $\chi$, $E_\chi \simeq m_\chi+m_\chi v^2/2$.  In the case of the
cross sections for EDM or MDM, the interaction of the dipole with the
charge of the nucleus can be significantly enhanced by a small
relative velocity. Direct detection experiments have put very
stringent constraints on such models for $m_\chi$ above several
GeV~\cite{Banks:2010eh, DelNobile:2012tx, DelNobile:2014eta}.

The sub-GeV region is better probed through DM-electron scattering as
it allows to extend the conventional direct detection down to masses
of $\sim 10\,$MeV~\cite{Essig:2011nj,Essig:2012yx}, below which the
halo DM kinetic energy falls below atomic ionization thresholds. First
limits on even lower mass DM have been achieved by the use of
semiconductor targets~\cite{Essig:2015cda,Crisler:2018gci},  by
utilizing either a solar reflected velocity component~\cite{An:2017ojc} or a cosmic-ray accelerated component~\cite{Bringmann:2018cvk}.

\begin{figure*}[tb]
\centering
\includegraphics[width=0.48\textwidth]{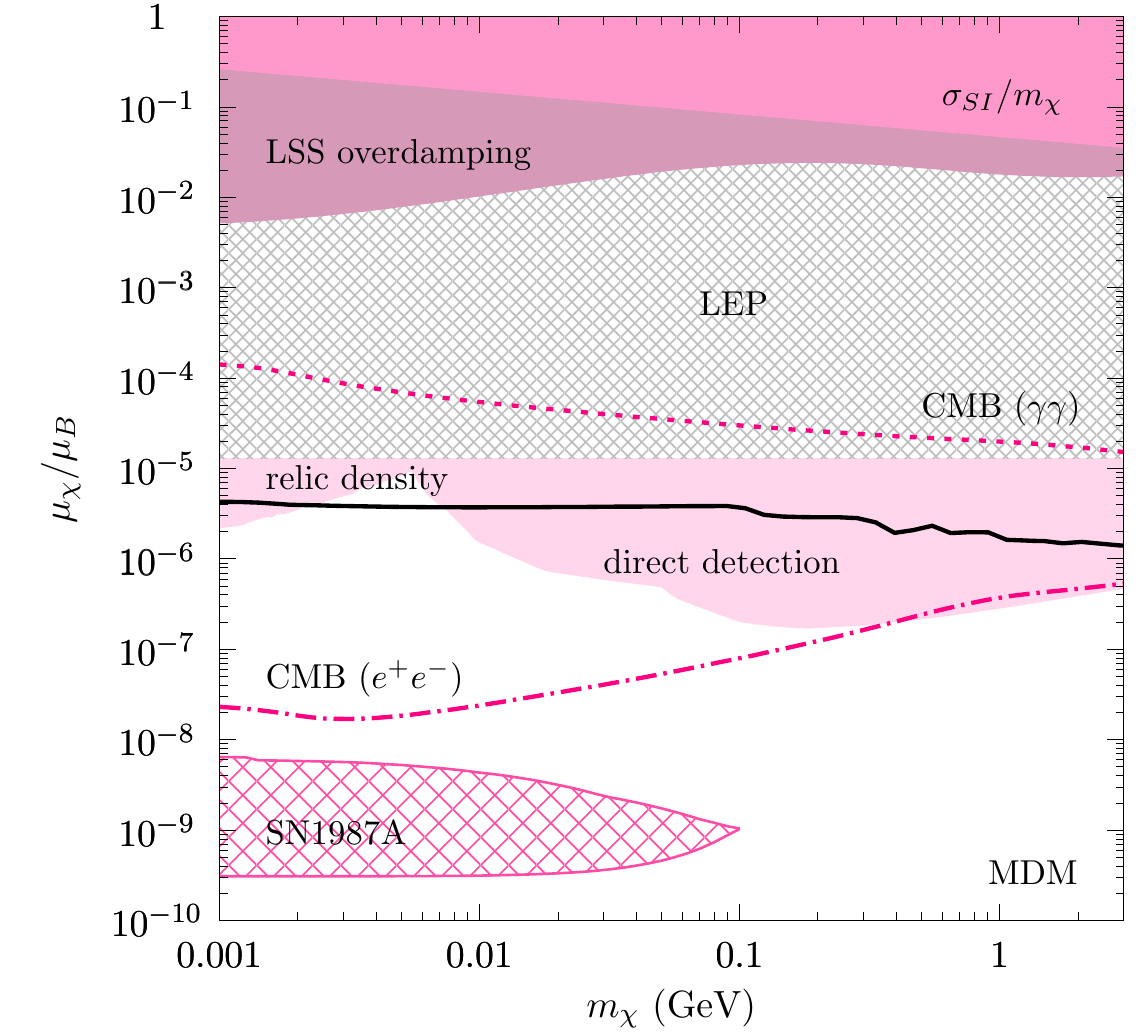} \quad
\includegraphics[width=0.48\textwidth]{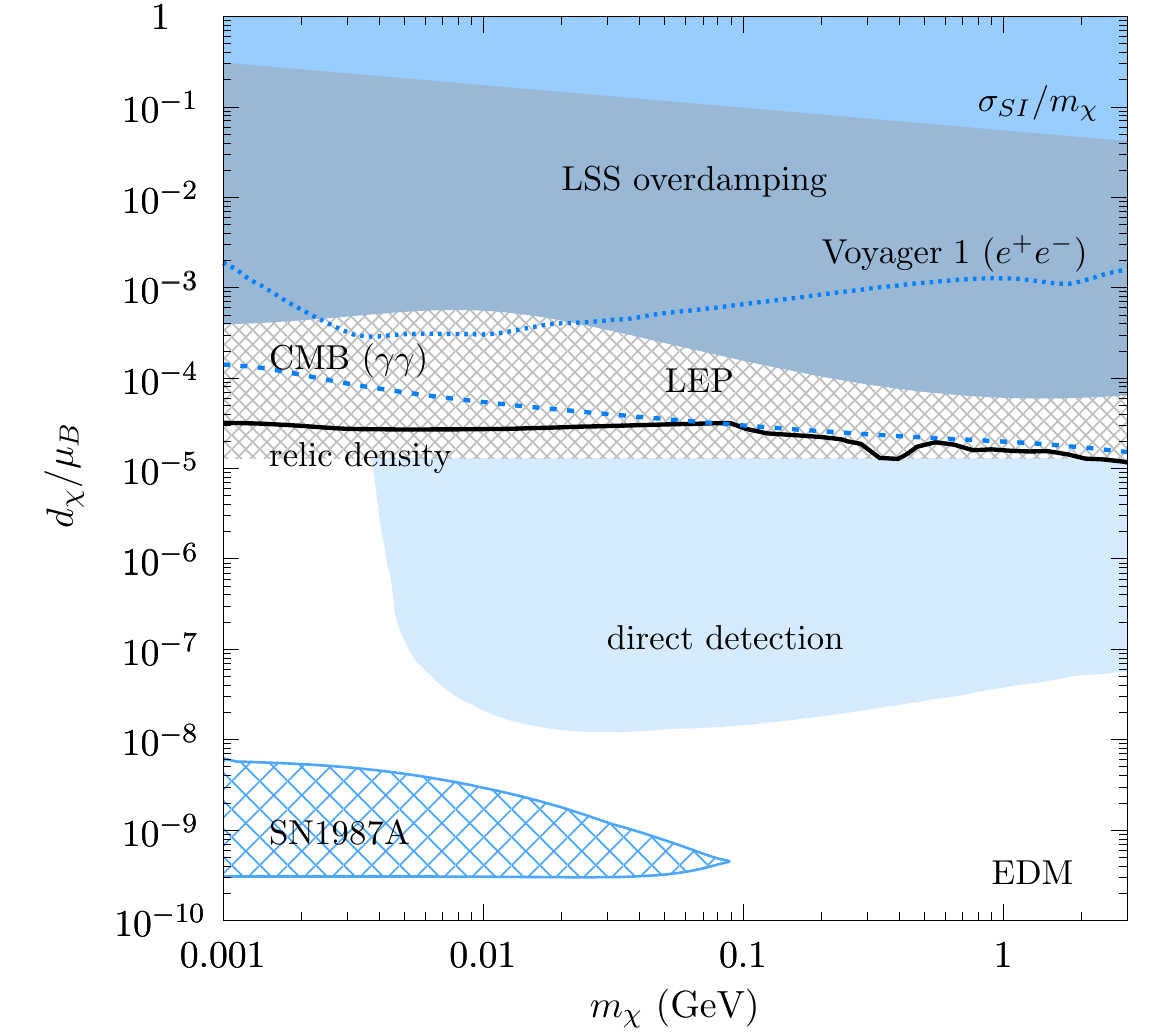}
\caption{Constraints on electromagnetic form
  factors for Dirac fermion DM with dim-5 operators from conventional direct/indirect/collider searches and
  cosmology. %
  Bounds from indirect searches only
  apply to symmetric DM, while others are general. Parameters that generate the
  observed relic abundance via thermal freeze-out are illustrated by  black
  lines.}\label{fig:cosmo_limits}
\end{figure*}

\begin{figure}[tb]
\centering
\includegraphics[width=0.48\textwidth]{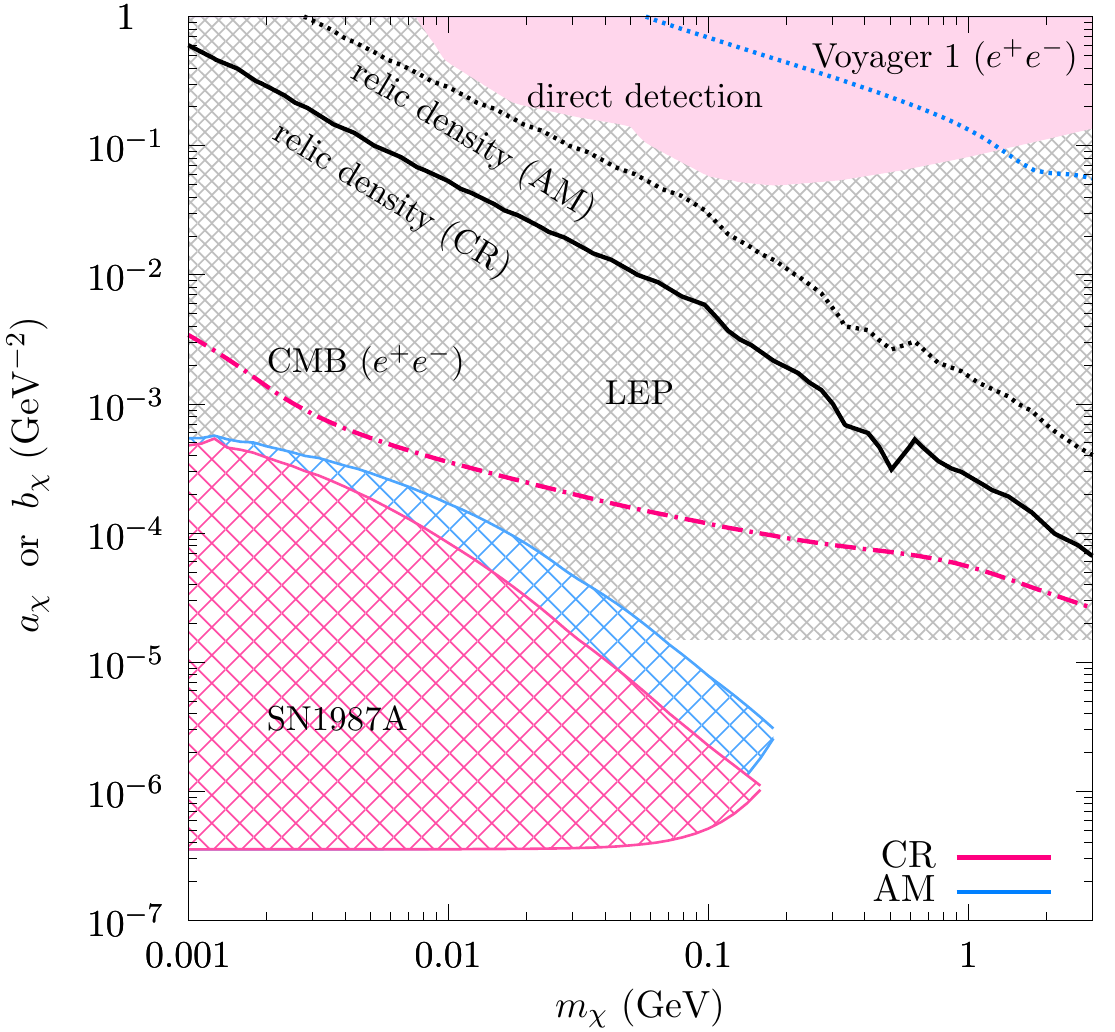} 
\caption{Same as Fig.~\ref{fig:cosmo_limits} for fermions with dim-6
  operators. Constraints for AM are shown in blue and constraints for CR in pink. The LEP constraints are the same for both models. %
  }\label{fig:cosmo_limits_dim6}
\end{figure}

Limits on DM electron scattering are conventionally quoted in terms of
a reference cross section~\cite{Essig:2011nj}
\begin{align}\label{eq:elecRecoil}
  \bar \sigma_e \equiv \frac{1 }{ 16\pi (m_e + m_\chi)^2 }
  \overline{|M_{\chi e}(q)|}_{q^2= \alpha^2 m_e^2}^2\, ,
\end{align}
where $\overline{|M_{\chi e}(q)|}^2$ is the scattering amplitude on a free
electron, evaluated at a typical atomic squared momentum transfer
$q^2= \alpha^2 m_e^2 $; we list  various $\overline{|M_{\chi e}(q)|}^2$ in
App.~\ref{sec:elast-scatt-cross}. The $q$-dependence of the actual
ionization cross sections is moved into a DM-form factor and exclusion
limits on $\bar \sigma_e $ have been derived for a constant
form factor~\cite{Essig:2017kqs, An:2017ojc} and for one proportional to
$1/q$~\cite{Essig:2012yx}. For MDM and CR, which correspond to constant
form factors, we combine the bound from \cite{Essig:2017kqs} based on the XENON10 data~\cite{Angle:2011th} and the 
bound in the low mass region by considering solar reflection~\cite{An:2017ojc} based on the XENON1T data~\cite{Aprile:2017iyp}. For EDM, corresponding to a $1/q$ form factor, the XENON10 bound derived
in \cite{Essig:2012yx} has been adopted here. These limits are shown in Figs.~\ref{fig:cosmo_limits} and \ref{fig:cosmo_limits_dim6}.  For AM, the
$\chi$-electron scattering is velocity suppressed, and we estimate a
relatively weak limit in the range
$a_\chi \lesssim 0.1-10$\,GeV$^{-2}$; we hence omit it in the figures
below.

It should be noted that for the operators in (\ref{eq:Lagrangians}) it
is also possible that they induce transitions between mass-split
(Majorana) states $\chi$ and $\chi^{*}$. Once
$\Delta m \equiv |m_{\chi^*} - m_\chi| \ge 100$\,keV, such DM
candidates can easily avoid any constraints from current direct
detection experiments, see \textit{e.g.}~\cite{Masso:2009mu,
  Chang:2010en}.

\subsection{Dark radiation (\texorpdfstring{\boldmath$N_{\rm eff}$}{Neff})}
\label{sec:dark-radi-boldm}

It is well known that (sub-)MeV mass particles,  when they are thermally
populated in the early Universe, will face severe limits from
cosmology. $\chi$ particles with mass
$m_\chi \lesssim {\rm few}\times \MeV$ and couplings
$\mu_\chi, d_\chi\ge 10^{-9}\mu_B$ or
$a_\chi, b_\chi \ge 5\times 10^{-5}$\,GeV$^{-2}$ reach chemical
equilibrium with the SM thermal bath once the early Universe's
temperature exceeded $T_\gamma \sim m_\chi$.
For example, the annihilation of $\chi$ into electron and photon
pairs---if it continues after neutrino decoupling---heats the photon
bath relative to the one of neutrinos, and hence modifies the
effective number of neutrino species, $N_{\rm eff}$, which in SM is
given by $N_{\rm eff} = 3.045$~\cite{Mangano:2001iu,deSalas:2016ztq}.
A general lower limit $m_\chi \gtrsim 7-10 \,$MeV has been derived for
a Dirac fermion $\chi$~\cite{Heo:2015kra} based on the measurements
by the Planck satellite, yielding
$N_{\rm eff}= 3.15 \pm 0.23$~\cite{Ade:2015xua}; see
also~\cite{Boehm:2013jpa, Nollett:2013pwa, Nollett:2014lwa} for
earlier work.  Since the bound is subject to a model-dependence when
annihilation into other dark states (including neutrinos) is open, we
do not show this bound explicitly in the figures below.

\subsection{Kinetic decoupling}
\label{sec:kinetic-decoupling}

DM is required to kinetically decouple from the thermal bath
after the photon temperature falls below 0.5\,keV in order to avoid
the over-damping of large-scale structures (LSS)~\cite{Boehm:2004th,
  Bringmann:2006mu}.  The kinetic decoupling temperature can be
obtained by considering the transport scattering cross section between
$\chi$ and target $i= e,p$,
\begin{align}
  \label{eq:transport1}
  \sigma^{\chi i}_T = \int (1-\cos\theta )\frac{d\sigma^{\chi i}}{ d\Omega }d\Omega \simeq \sigma^{\chi i}_0 v^n,
\end{align}
where $\theta$ is the CM scattering angle, and the cross section have been factored as a product of a constant $\sigma^{\chi i}_0 $  and  the $n$-th power of the relative velocity $v$. The expressions for the 
various operators in leading order of $v$ 
are listed in App.~\ref{sec:elast-scatt-cross}. 
We find that $n=-2$ for EDM, $n=0$ for MDM and
CR, 
 and $n=2$ for AM. 
Then, assuming $T_\chi = T$ and  demanding that the specific heating rate of DM particles derived in \cite{Dvorkin:2013cea}, 
\begin{align}  \label{eq:transport2}
\sum_i \frac{2^\frac{n+5}{ 2 }\Gamma(3+\frac{n}{ 2}) }{ \sqrt{\pi} }\frac{n_i \, m_i m_\chi }{ (m_i + m_\chi)^2}  \, \sigma^{\chi i}_0 \left(\frac{T}{  m_i} +\frac{T}{  m_\chi} \right)^\frac{n+1}{ 2} \,,
\end{align}
be smaller than the Hubble rate, $H(T)$, from $T\sim 0.5$\,keV to
recombination, we obtain  upper limits on the couplings. For dim-5 operators, 
they are shown as dark shaded regions labeled ``LSS overdamping'' in
Fig.~\ref{fig:cosmo_limits}. As can be seen, the scaling of the cross
section with $v^{-2}$ makes the EDM case more stringently
constrained.  For fermions with dim-6 operators the bounds are too
weak to be visible in Fig.~\ref{fig:cosmo_limits_dim6}. Our results on
MDM and EDM agree with earlier studies~\cite{Dvorkin:2013cea, Gondolo:2016mrz}. %

The scattering processes tend to cool the gas in reverse. One can
estimate the corresponding cooling rate using \eqref{eq:transport2} by replacing $n_i $ with $f_i n_\chi$, where
$f_i$ is the fraction of electron/proton particles involved in the
interaction and $n_\chi$ gives the DM number
density~\cite{Tashiro:2014tsa, Munoz:2015bca}. Therefore, cosmological
21\,cm observations, such as recently reported by the EDGES
experiment~\cite{Bowman:2018yin}, may also constrain the couplings
concerned here.  Nevertheless, due to the minute free-electron fraction
at redshift $z\sim 20$, $f_{e,p}\sim 10^{-4}$, the bounds will be
much weaker than the ones from considering over-damping. Even for the
EDM case with $n=-2$, in order to have the cooling rate comparable to
the corresponding Hubble rate, one needs $d_\chi \sim \mu_B$ for
$m_\chi = 10$\,MeV.\footnote{Note that the scattering cross section of
  $\chi$ with the dipole moment of hydrogen atoms always has $n\ge 0$,
  and thus is not velocity-enhanced at low redshifts. This is different from
  the $\epsilon Q$ case~\cite{Barkana:2018qrx}. }

\subsection{Self-scattering of \texorpdfstring{\boldmath$\chi$}{chi}}
\label{sec:self-scatt-boldm}

At low-redshift, a strong self-interaction among light DM particles
modifies the shape and density profile of DM halos, as well as the
kinematics of colliding clusters.  A fair amount of attention has been
devoted to the study of millicharged DM, which exhibits a strong
velocity dependence of the self-scattering cross section. For the
models studied here, however, the corresponding interactions are not
enhanced at low relative velocities, which can be seen explicitly the
expression of the cross sections listed in
App.~\ref{sec:annih-cross-sect}. Hence, we may directly use the
constraint on the self-scattering cross section obtained from from
Bullet cluster observations,
$\sigma_\text{SI}/m_\chi \lesssim
1.25\,$cm$^2/$g~\cite{Randall:2007ph, Robertson:2016xjh}, to derive
the bounds on MDM, EDM, CR and AM. Again, while the limits on CR and
AM interactions are too weak to be shown, in
Fig.~\ref{fig:cosmo_limits} the ensuing constraints on MDM and EDM are
shown by the shaded region labeled ``$\sigma_\text{SI}/m_\chi$''.  To
be general, we do not take into account the scattering between $\chi$
and $\bar\chi$ particles, which, in any case, does not change our
conclusion.

\subsection{Supernova cooling}
\label{sec:supernova-cooling}

Light $\chi$ particles with $m_{\chi}\lesssim 400\,\MeV$ can be
produced in pairs inside supernovae (SN), and, if such particles
escape from the core, it increases the SN cooling rate. In turn, if
the coupling is only strong enough, the particles may reach thermal
equilibrium, be trapped, and an energy loss argument does not
immediately apply.

We estimate the region in parameter space where $\chi$-particles
affect the SN cooling as follows: if $\chi$ is thermalized within the
SN core, a radius $r_d$ can be estimated as the radius where the
blackbody luminosity of $\chi$ equals the luminosity in neutrinos
$L_{\nu}= 3\times 10^{52}$~erg/s~\cite{Raffelt:1996wa}. This is a
maximum permissible luminosity if $\chi$ were to carry away this
energy freely (``Raffelt criterion''). However, $\chi$ will typically
be further deflected in elastic scatterings for $r > r_d$, engaging in
a random walk before it reaches a free streaming radius that we take
as $r_\text{inf} = 50\,$km. In order to estimate the efficiency of the
trapping, as a rough criterion, we impose
\begin{equation}
  \label{eq:sn1}
  \int^{r_\text{inf}}_{r_d} dr \frac{\rho_p(r) }{ m_p} {\sigma^{\chi p}_T} \le 2 \,, 
\end{equation}
as an upper limit above which energy-loss fails to be efficient; a
similar argument has been used in~\cite{Chang:2018rso} which we
largely follow. We evaluate~(\ref{eq:sn1}) by taking the SN numerical
model of~\cite{Fischer:2016cyd} with $T_\chi = T(r)$, and protons,
with mass density $\rho_p(r)$, are assumed to be at rest as we focus
on $m_\chi \ll m_p$.
The value of $r_d$ is then estimated from the SN model for each value
of $m_\chi$, and it ranges from 22\,km for $m_\chi\sim $MeV to
11.5\,km for $m_\chi \sim 0.4$\,GeV.

On the flip side, the lower boundary in couplings for energy loss to
be efficient can be estimated by calculating the energy production
rate of $\chi$ particles by electron-positron annihilation within the
SN core,
\begin{equation}
  \dot Q = \int \frac{ d^3 p_{e^-} d^3 p_{e^+} }{ (2\pi)^6} f_{e^-}f_{e^+} \, (\sigma_{e^-  e^+ \to \chi\bar\chi} v ) \sqrt{s}\,  , 
\end{equation} 
following \cite{Dreiner:2003wh, Kadota:2014mea,
  Magill:2018jla}.\footnote{For interactions studied here, the
  contribution from plasmon decay is either comparable or sub-leading,
  and neglected for simplicity.} The corresponding momentum distributions of electron and positron, $f_{e^-}$ and $f_{e^+}$, as functions of the temperature and (opposite) chemical potentials for each $r$, are also given by the same SN model~\cite{Fischer:2016cyd}.   Setting the core size
$r_\text{core}=10$\,km, we derive the lower boundary by requiring
\begin{equation}
	\int_0^{r_\text{core}} d^3 r \, \dot Q\le  L_{\nu}. 
\end{equation} 
Both the lower and upper bounds discussed above are presented in
Figs.~\ref{fig:cosmo_limits} and \ref{fig:cosmo_limits_dim6}. While
these results are indicative of the relevant region in parameter
space, a more detailed analysis may moderately but not qualitatively
alter the results
(\textit{e.g.}~\cite{Dreiner:2003wh,Magill:2018jla}); we defer a more
precise study of SN cooling to a dedicated future work.

In passing, we note that, besides SN, DM particles traversing the
solar system may also be captured by the sun, affecting the stellar
evolution. This has been studied in~\cite{Lopes:2013xua, Geytenbeek:2016nfg} for
$m_\chi$ above 4\,GeV. For lighter $\chi$ particles as we are
concerned with in this work, any effect will be strongly suppressed
due to immediate DM evaporation; see also~\cite{An:2017ojc}.

\subsection{Dark matter annihilation and indirect search}
\label{sec:indirect-detection}

Here we consider the case that $\chi$ is a symmetric DM 
candidate, so that $\chi\bar \chi$ annihilation into light charged SM
fermions and photons is operative both at CMB decoupling and at
present day. The associated release of electromagnetic energy modifies
the recombination history at high redshift and generates an excess in
cosmic rays at low redshift.

For sub-GeV DM, the most stringent bounds typically come from CMB
observations~\cite{Liu:2016cnk}.  For MDM and CR DM particles, the
annihilation to $e^+e^-$ is $s$-wave, resulting in stringent limits on
the relevant couplings. In contrast, for EDM and AM, where the
annihilation cross sections to $e^+e^-$ is $p$-wave and hence
velocity suppressed, the leading bounds come from Voyager~1
data~\cite{Boudaud:2016mos, Boudaud:2018oya}, and are relatively
weaker.
Moreover, CMB observations also strongly constrain the other
annihilation channel, $\chi\bar\chi \to \gamma\gamma$. In contrast, gamma-ray
telescope searches lead to bounds that are one to two orders of
magnitude weaker in the sub-GeV region (see \textit{e.g.}~\cite{Boddy:2015efa}), and are thus not
shown in the figures.  This channel is
operative for MDM and EDM; for AM and CR, tree-level annihilation into
physical photons vanishes identically.
Here we will not further consider annihilation at loop-level (see
\textit{e.g.}~\cite{Latimer:2017lwm}).

The bounds are summarized in Figs.~\ref{fig:cosmo_limits} and
\ref{fig:cosmo_limits_dim6}. They show that MDM and CR, as symmetric
DM candidates, are strongly constrained by the current observational
data. Moreover, on-going and future experiments~\cite{Bartels:2017dpb,
  Bowman:2018yin} have the potential to further improve the
constraints on light DM annihilations soon.

\section{Conclusions and Outlook}
\label{sec:conclusions}

In this paper, we consider the phenomenology of a new long-lived Dirac
fermion $\chi$ in the MeV-GeV mass range and which interacts with the
photon through mass dimension 5 and 6 operators MDM, EDM and CR, AM,
respectively, but is otherwise neutral. In a bottom-up approach, the
existence of such states $\chi$ is constrained from intensity frontier
and high energy collider experiments, from SM precision observables,
from flavor physics, and, once the lifetime exceeds roughly one
second, from cosmology and astrophysics. We have considered all these
possibilities in turn and derived constraints on the dimensionful
couplings. When not present in the current literature, we also derive
analogous constraints on a potential millicharge of $\chi$.

At intensity frontier experiments where these fermions can be produced
in pairs, we focus on electron beams and show that the most stringent
constraints on these models are currently set by mono-photon searches
at BaBar setting upper bounds on electric and magnetic dipole moments
of $\sim 4 \times 10^{-5} \, \mu_B$ and on anapole moments and charge
radius interactions of $\sim 2 \times 10^{-3} \,
\GeV^{-2}$. Furthermore, we find that ongoing and future experiments,
such as Belle II, LDMX and BDX, may extend the sensitivity by more than
one order of magnitude.  The projections for $e^+e^-$-colliders
discussed in this work are basically independent of the fermion mass
$m_\chi$ across the whole MeV--GeV mass range, while the constraints from
fixed target experiments exhibit a mass dependence that makes the
interactions better testable in the low MeV-mass range. These results
need to be compared to mono-photon searches at high energy colliders.
We find that LEP is superior in terms of present constraining power,
but Belle-II and LDMX can improve the reach to smaller couplings in
the dim-5 case. In the dim-6 case, however, due to the higher
dimensionality of the coupling, we find that LEP remains the best
probe for AM and CR interactions.

Next, we use a number of SM precision observables as an indirect test
to constrain the parameter space further, albeit in a possibly more
model dependent way than a direct observation can offer. For example,
corrections to the vacuum polarization of the photon from the new
interactions induce a running of the fine structure constant $\alpha$,
and we place a constraint on the respective new physics models that is
even stronger than the bounds from accelerator experiments. However,
we find that some of the operators contribute with a relative sign and
cancellations are in principle possible. We also estimate the
contribution to $(g-2)_{\mu}$ of the muon, concluding that the new
interactions cannot contribute at the level of $3\times 10^{-9}$ ---
reflecting the current tension in this observable --- without being in
conflict with other direct searches, most notably with mono-photon
limits derived from BaBar-data.  In turn, relating
$K^{\pm}\to \pi^{\pm} \chi \bar\chi$ and
$B^{\pm}\to K^{\pm} \chi \bar\chi$ meson decays to analogous channels
with missing energy we find that flavor observables yield only
relatively weaker limits and are superseded by other intensity
frontier searches.
On the other hand, if $\chi$ also couples to the hypercharge field
tensor with similar strength as to $F_{\mu\nu}$, the LEP observation
on the invisible width of the $Z$-boson strongly constrains the whole
parameter space which will be reachable by any of the future
experiments discussed here. Clearly, the existence of such constraint
hinges on the concrete UV-realization that induces the studied dim-5
and dim-6 operators. 

Finally, we have addressed the question on the possibility of $\chi$
being DM. If $\chi$ is to provide 100\% of the DM abundance and has a
mass below the GeV-scale, direct detection constraints derived from
DM-electron scattering are stronger than the limits derived from
accelerators by 2 to 3 orders of magnitude for MDM and EDM. In
contrast, we find that direct detection experiments have presently
little sensitivity to AM and CR interactions when compared to other
direct tests.
In all cases, a combination of cosmological, astrophysical, direct
detection and accelerator constraints rule out fermions with
electromagnetic form factors making up 100\% of the DM assuming a
standard freeze-out scenario. If $\chi$, however, only makes up a
small fraction, $\lesssim 0.1 \%$, of the DM abundance, one can evade
the astrophysical constraints and will only be left with the
accelerator and the supernova cooling constraints, which are
independent of the DM relic density. For couplings smaller than
$\mu_\chi, d_\chi \sim 10^{-9}\mu_B$ or
$a_\chi, b_\chi \sim 5\times 10^{-5}$\,GeV$^{-2}$,  DM particles of MeV mass do not
reach thermal equilibrium with the SM anymore, allowing for
alternative production mechanisms in the early Universe and therefore
in this region of parameter space fermions with EM form factors could
still be DM without violating any constraints. Note that
$\Delta N_\text{eff}$ bounds also do not apply in this region of
parameter space.

This work constitutes a systematic study of sub-GeV mass dark states
carrying electromagnetic form factors. Yet, many further avenues exist, which we are going to address in upcoming works:

First, we have focused on electron beams employed in intensity
frontier searches.  Once we consider protons as projectiles, a whole
number of relevant experiments for similar analyses become
available. Most notable are the proton beam dump facilities, such as
E613 at Fermilab~\cite{Ball:1981nu}, SHIP at CERN's SPS
\cite{Anelli:2015pba} and the proposed MilliQan experiment at the LHC
\cite{Haas:2014dda}, as well as the neutrino detectors like MiniBooNE
\cite{Aguilar-Arevalo:2018gpe}, DUNE \cite{Acciarri:2015uup}, LSND
\cite{Athanassopoulos:1996ds} and CE$\nu$NS
\cite{Akimov:2017ade}. %
Studies of dark sectors along similar lines have already been performed
  in~\cite{Soper:2014ska, Mohanty:2015koa, Magill:2018tbb, Sher:2017wya}.
In order to systematically study $\chi$-pair production from hadron
beams, simulations are needed to generate event spectra because hadron
physics cannot be neglected anymore. For example, in this set of
experiments, contributions of the on-shell production and subsequent
decay of mesons $\pi^0,\eta \to \gamma \chi\bar\chi$ and
$J/\psi, \Upsilon \to \chi\bar\chi$ will be non-negligible for the
production of a $\chi\bar\chi$ pair when kinematically allowed. Working
along the lines of our presented appendices for the 2-to-4 processes,
a systematic write-up of the production process and derivation of the
relevant kinematic quantities that feed into the computation of
observables is in preparation, filling a gap in the existing
literature.

Second, in this work we have considered a Dirac fermion $\chi$ for
concreteness. A complex, but electrically neutral scalar particle
$\phi$ can have dim-6 CR interactions. In addition, at the same mass
dimension, a vector particle may carry electric and magnetic
quadrupole moments. Finally, at mass-dimension 7, the Rayleigh or
susceptibility operators alter the phenomenology: the elementary
interaction involves two photons, so that either an additional
loop-suppressing factor is present in signals with one external
photon, or new signal topologies with two external photons need to be
considered. We leave a systematic study of those possibilities for a
future paper.

Third, in this paper we have chosen a bottom-up approach, leaving the
UV physics that generates the dim-5 and dim-6 operators
unspecified. Such UV completion will result in additional constraints,
most prominently from high energy colliders, in particular when the
value of the effective dimensionful coupling points towards generating
physics that must be at or below the TeV-scale. Such program for
electroweak-scale particles has recently been started
in~\cite{Kavanagh:2018xeh}. Given the relatively weak intensity
frontier limits on dim-6 couplings, our considered parameter space is
likely further constrained by the presence of additional particles
carrying electromagnetic charge and/or hypercharge.

Fourth, a more systematic study of cosmological and astrophysical
constraints can be performed. For example, the limits on supernova
cooling are derived on simplifying assumptions, and a more detailed
study incorporating finite temperature effects and a better simulation
of the efficiency of trapping may alter the inferred regions in
parameter space, albeit only moderately. Finally, we have shown that a
$\chi$ particle that freezes-out at the appropriate relic DM density is
already ruled out. On the other hand, once the couplings diminish,
$\chi$ may freeze-in, and a detailed computation of its abundance as a
function of mass and interaction strength as well as a discussion on
its detectability is still outstanding.

Dark sector particles with mass below the GeV-scale that are perfectly
electrically neutral may still interact with the photon through a
number of higher-dimensional operators. The question how strong such
coupling can be and ``how dark is dark'' is an experimental one. In
this paper we have answered it utilizing currently available
experimental observables and provided forecasts to clarify the
detectability with the future experimental program that is coming
online or is considered.

\begin{acknowledgments}
  We thank M.~Battaglieri, A.~Hoang, Z.~Ligeti, M.~Passera, A.~Ritz for helpful
  discussions.  The authors are supported by the New Frontiers program
  of the Austrian Academy of Sciences. LS is supported by the Austrian
  Science Fund FWF under the Doctoral Program W1252-N27 Particles and
  Interactions. We acknowledge the use of computer packages for
  algebraic~\cite{Mertig:1990an,Shtabovenko:2016sxi} and
  numeric~\cite{Hahn:2004fe} evaluations.

\end{acknowledgments}

\appendix

\section{Matrix elements for emission}
\label{sec:matrix-el}

In this appendix we provide the detailed calculation of the pair
production of two long-lived particles $\chi$ through their
electromagnetic form factors.  From the interaction operators
(\ref{eq:Lagrangians}) we derive the Feynman rules $i\Gamma^{\mu}(q)$
with incoming photon four momentum $q$ and
\begin{subequations}
  \label{eq:FR}
\begin{align}
\label{eq:FRmQ} \text{$\epsilon$Q:\quad }    \Gamma^{\mu}(q) & = +   \epsilon e \gamma^{\mu} , \\
\label{eq:FRmdm} \text{MDM:\quad }  \Gamma^{\mu}(q) & = + i \mu_\chi  \sigma^{\mu\nu} q_{\nu} ,\\   %
\label{eq:FRedm} \text{EDM:\quad }  \Gamma^{\mu}(q) & =  - d_\chi \sigma^{\mu\nu} q_{\nu}  \gamma^5 , \\ %
\label{eq:FRam} \text{AM:\quad }    \Gamma^{\mu}(q) & =  - a_\chi \left[  q^2  \gamma^{\mu} -q^{\mu} \slashed q  \right]\gamma^5 ,\\
\label{eq:FRcr} \text{CR:\quad }    \Gamma^{\mu}(q) & = +  b_\chi \left[  q^2  \gamma^{\mu} -q^{\mu} \slashed q  \right] .
\end{align}
\end{subequations}
Hermiticity of the electromagnetic current implies
$ \bar \Gamma^{\mu}(q) = \Gamma^{\mu}(-q)$ where
$ \bar\Gamma = \gamma^0 \Gamma^{\dag} \gamma^0$ is the Dirac adjoint.

We wish to compute the scattering cross section of $e^-$ on a nucleus
$N$ with emission of a $\chi \bar \chi$ pair and into a potentially
inclusive hadronic $n$-particle final state $X_n$,
\begin{align}
\label{eq:dis-scattering}
  N(p_1) +  e^-(p_2) \to  X_n(p_3) + e^-(p_4) + \chi(p_{\chi}) +   \bar\chi(p_{\bar\chi}) ,
\end{align}
where $p_3 = \sum_{i=1}^n p_{3,i} $ is the total inclusive four
momentum of the recoiling target, potentially fragmented into $n$
states. We introduce the momentum transfer variables
\begin{align}
  q & \equiv p_{\chi}+p_{\bar\chi} ,  \quad  q_1  \equiv p_1 - p_3 ,\quad q_2  \equiv p_2 - p_4   ,
\end{align}
such that $ q = q_1 + q_2$ which follows from overall energy-momentum
conservation. Since $q,q_1,q_2$ are linearly dependent, we express the
scattering by the six independent momenta
$p_1, p_2, q, q_1 , q_2, p_\chi$ and their scalar products.

The matrix elements for the processes depicted in
Fig.~\ref{fig:feyn_diag}b can split into the emission piece of the
$\chi\bar\chi$-pair,
$   \mathcal M_{\chi}^{\nu} =  \bar u(p_\chi) \Gamma^{\nu}(q) v(p_{\bar\chi}) ,$
and the scattering piece on the
nucleus, $\mathcal{M}_N$, 
\begin{subequations}
  \label{eq:mat1}
\begin{align}
  \mathcal M_{N,a}^{\mu} &= \frac{e^2 g_{\rho\sigma} }{q_1^2 \left[ (p_4+q)^2 - m_e^2  \right]}
                           \times \langle p_3 | J^{\rho}(0) | p_1 \rangle \nonumber \\
                         & \times \left[  \bar u(p_4) \gamma^{\mu} (\slashed p_4 + \slashed q + m_e) \gamma^{\sigma} u(p_2)   \right] ,  \\
 \mathcal M_{N,b}^{\mu} &= \frac{e^2 g_{\rho\sigma}}{q_1^2 \left[ (p_2-q)^2 - m_e^2  \right]}  \times \langle p_3 | J^{\rho}(0) | p_1 \rangle  \nonumber \\
                         & \times \left[  \bar u(p_4) \gamma^{\sigma} (\slashed p_2 - \slashed q + m_e) \gamma^{\mu}  u(p_2)   \right] , 
\end{align}
\end{subequations}
where
$ \langle p_3 | J_{\rho}(0) | p_1 \rangle = \langle J_{\rho} \rangle $
is the hadronic matrix element of the electromagnetic current.
The overall matrix element for the emission is hence,
\begin{align}
  \label{eq:M}
  \mathcal{M} = \frac{1}{q^2} ( \mathcal M_{N,a}^{\mu}  +   \mathcal M_{N,b}^{\mu} )   (\mathcal M_{\chi})_{\mu} , 
\end{align}
where we have dropped all gauge-dependent pieces as is manifest for when
$q$ gets dotted into the vertices~(\ref{eq:FR}). In addition, the
$q_1$ gauge-dependent part of the photon propagator in~(\ref{eq:M})
drops out when being dotted into the hadronic matrix element,
$ q_{1,\nu} \langle p_3 | J^{\nu}(0) | p_1\rangle = 0 $, so that we
omitted those terms above as well.

The differential cross section for the process
(\ref{eq:dis-scattering}) then reads,
\begin{align}
   d\sigma & = \frac{1}{4E_1 E_2 |\vec v_1 - \vec v_2|}   | \mathcal{\overline M}|^2 \,  d\Phi\,,
\end{align}
where $v_{1,2}$ are the velocities of the incoming particles and the total phase space 
\begin{align} \label{eqn:phase_space}
	d\Phi &=
	 \prod_{i=1}^n \frac{d^3 p_{3,i}}{(2\pi)^3 2E_{3,i}} 
 \prod_{j=4}^6  \frac{d^3 p_j}{(2\pi)^3 2E_j} \notag\\*
	&\times (2\pi)^4\delta^{(4)}\left(p_1+p_2-\sum_{i={1}}^n p_{3,i} - \sum_{j=4}^6 p_j  \right) \, .
\end{align}
Here $p_{5,6} = p_{\bar \chi, \chi}$. One can further introduce to
\eqref{eqn:phase_space} an integral with respect to
$\shadr = m_X^2 = p_3^2 $ to obtain
\begin{align}
\begin{split}
	d\Phi &= 2ds_X d\Phi_4 \\ &\times \frac{ 1}{ 4\pi }   \prod_{i=1}^n \frac{d^3 p_{3,i}}{(2\pi)^3 2E_{3,i}} (2\pi)^4\delta^{(4)}\left(p_3-\sum_{i={1}}^n p_{3,i} \right) ,
\end{split}
\end{align}
where $ d\Phi_4$ is the 4-body phase space of $p_{j}$ ($j=3,4,5,6$) --- which will explicitly be solved analytically in Appendices~\ref{sec:4-body-phase-ldmx} and \ref{sec:4-body-phase-bdx} ---
and where the second line will be absorbed by the hadronic tensor
$W_{\rho\sigma}$ defined below.

  In the lab-frame where
$|\vec v_1|= 0$,   the cross section can be written as 
\begin{align}
    \label{eq:diffcs}
\begin{split}
 d\sigma & =  
           \frac{(4\pi \alpha)^3}{2 |\vec p_2| m_N  q^4 q_1^4 }     %
    \;  L^{\rho\sigma, \mu\nu} \; \chi_{\mu\nu}(q) \; W_{\rho\sigma}(-q_1) \, ds_X \, d\Phi_4 \,,
\end{split}
\end{align}
where $\chi_{\mu\nu}$ is the $\chi$ emission piece.
The electron scattering, averaged (summed) over the initial
(final) spins is described by the terms
\begin{align}
 \begin{split}
 	 L^{\rho\sigma, \mu\nu} &=
 	\frac{L_a^{\rho\sigma, \mu\nu}  }{\left[ (p_4+q)^2 - m_e^2  \right]^2 } 
    +   \frac{L_b^{\rho\sigma, \mu\nu}  }{ \left[ (p_2-q)^2 - m_e^2 \right]^2 }  \\
      &+   \frac{2L_{ab}^{\rho\sigma, \mu\nu}  }{\left[ (p_4+q)^2 - m_e^2  \right] \left[ (p_2-q)^2 - m_e^2 \right] }
 \end{split}
\end{align}
with
\begin{subequations}\label{eqn:electron_part}
\begin{align}
\begin{split}
	 L_a^{\rho\sigma, \mu\nu} &=  \frac{1}{2} \Tr \Big[  (\slashed p_4 + m_e )  \gamma^{\mu}   (\slashed p_4 + \slashed q + m_e )  \\
  &\times \gamma^{\rho} ( \slashed p_2 + m_e )  \gamma^{\sigma}   (\slashed p_4 + \slashed q + m_e) \gamma^{\nu}      \Big] ,
\end{split} \\ 
\begin{split} 
  L_b^{\rho\sigma, \mu\nu} &=  \frac{1}{2}\Tr \Big[     ( \slashed p_2 + m_e )    \gamma^{\nu}    (\slashed p_2 - \slashed q + m_e)     \\
  &\times \gamma^{\sigma} (\slashed p_4 + m_e )   \gamma^{\rho}    (\slashed p_2 - \slashed q + m_e ) \gamma^{\mu}    \Big] ,
 \end{split}\\
 \begin{split}
  L_{ab}^{\rho\sigma, \mu\nu}& =  \frac{1}{2}\Tr \Big[  (\slashed p_4 + m_e )  \gamma^{\mu}    (\slashed p_4 + \slashed q + m_e ) \\
  &\times \gamma^{\rho} ( \slashed p_2 + m_e )     \gamma^{\nu}      (\slashed p_2 - \slashed q + m_e)    \gamma^{\sigma}    \Big] .
 \end{split}
\end{align}
\end{subequations}
The emission of the $\chi\bar\chi$ pair is captured in the final state
spin-summed matrix element $\chi_{\mu\nu}$ given by,
\begin{align}
  \chi_{\mu\nu}(q)  =  \Tr \left[ (\slashed p_\chi + m_{\chi})
   \Gamma_{\mu}(q) (\slashed p_{\bar\chi} - m_{\chi})  \bar\Gamma_{\nu}(q)   \right] .
\end{align}

\section{Hadronic tensor and form factors}
\label{sec:had-ten}
The response of the nuclear target is described by a hadronic tensor
in (\ref{eq:diffcs}). In its general form, $W_{\sigma\rho}(-q_1)$ is
given by,
\begin{align}
\begin{split} \notag
W_{\sigma\rho}(-q_1) &=
  \frac{1}{4\pi} \left(\frac{1}{2s+1} \sum_{s}\right)
   \sum_{n} \int \prod_{i=1}^n  \frac{d^3 \vec p'_{3,i}}{(2\pi)^3 2 E_{\vec p'_{3,i}}}    \\
 &\quad \times {(2\pi)^4} \delta^{(4)}\left(p_1-q_1-\sum_{i={1}}^n p_{3,i} \right) \\  
 &\quad \times \sum_{s'_{3,i}} \bra{ p_1,s } J^{\dag}_{\sigma}(0) \ket{ X_n } \bra{ X_n } J_{\rho}(0) \ket{  p_1,s } 	
\end{split} \\
&= \frac{ \sum_s \int d^{4}x \, e^{-i
  q_1\cdot x} \langle p_1,s | [ J^{\dag}_{\sigma}(x) , J_{\rho}(0) ] |
p_1,s \rangle }{4\pi (2s+1)},
\label{eqn:hadronic_tensor_int}
\end{align}
where the $\delta$-function is from overall energy-momentum
conservation and in this context guarantees that the fragments have the overall momentum $p_3$.
The most general (parity conserving) form for (\ref{eqn:hadronic_tensor_int}) is
\begin{align}
\begin{split}
	W^{\rho \sigma} &=
	\left[-g^{\rho \sigma}+\frac{q_1^\rho q_1^\sigma}{q_1^2}\right] W_1(q_1^2,\shadr) \\
	&+
	\left[p_1^\rho-\frac{p_1\cdot q_1}{q_1^2}q_1^\rho\right] \left[p_1^\sigma-\frac{p_1\cdot q_1}{q_1^2}q_1^\sigma\right] \frac{W_2(q_1^2,\shadr)}{m_N^2}\,,
\end{split}
\end{align}
where the form factors $W_1$ and $W_2$ are functions of $q_1^2$ and $s_X$.

In the elastic limit $\shadr=m_N^2$ and $q_1^2 = 2 (p_1 \cdot q_1)$ and  $W_{1,2}$ are only functions of $q_1^2$. 
For spin-1/2 fermions, they are related to the usual electric and
magnetic form factors $F_E$ and $F_M$ via
\begin{subequations}
\begin{align} 
	W_1(q_1^2) &= -  q_1^2 F_M^2(q_1^2)\, \frac{ \delta(\shadr-m_N^2)}{2} \,,\\
	W_2(q_1^2) &= \frac{4 m_N^2 F_E^2(q_1^2) - q_1^2 F_M^2(q_1^2)}{1-q_1^2/(4m_N^2)} \frac{ \delta(\shadr-m_N^2)}{2 },
\end{align}	
\end{subequations}
whereas for a scalar target they read,
\begin{subequations}
\begin{align} 
	W_1(q_1^2) &= 0,\\
	W_2(q_1^2) &= 4 m_N^2 F_E^2(q_1^2) \; \frac{\delta(s_X-m_N^2)}{2} .
\end{align}
\end{subequations}

When considering the scattering on protons $p$ or neutrons $n$,
the form factors can be taken in the dipole form~\cite{Perdrisat:2006hj},
\begin{subequations}
\begin{align}
	F^p_{E}(t) &= \frac{ 1 }{ (1+ t /0.71\,\text{GeV}^2)^2} , \\
           	F^n_{E}(t)  &=  - \frac{ \mu_n  t }{ (4m_n^2 +5.6 t) } F^p_{E}(t),
\end{align}	
\end{subequations}
and $ F_M^{p,n} = \mu_{p,n} \, F^{p,n}_{E}$ where $t=-q_1^2$ and
$\mu_{p,n}$ are the magnetic moments of proton and electron in units
of the nuclear magneton $\mu_N = e/2m_p$, respectively; with $\mu_p = 2.79$ and
$\mu_n = -1.91$.

For nuclear targets of charge $Z$, the electric form factor is given by%
\begin{equation}
F_E(t) = Z\cdot  \frac{a^2(Z) t }{ 1+a(Z)^2  t } \frac{1 }{ 1+t/d(A) }\, ,
\end{equation}
where $A$ is the mass number and $a (Z)=111Z^{1/3}/m_e$ and
$d(A)= 0.164\, \GeV^2 A^{-2/3}$~\cite{Kim:1973he, Tsai:1973py}.  The above expression includes a
factor that accounts for the screening of $Z$ by electrons such that
$F_E(t) = 0$ for $t\to 0$. Finally, in our numeric evaluations we
neglect the magnetic form factor for $Z\gg 1 $, as it is generally subleading.  We
note in passing, that for heavy states $\chi$, inelastic scattering
may start to play a significant role~\cite{Bjorken:2009mm}.

\section{4-body phase space for LDMX and NA64}
\label{sec:4-body-phase-ldmx}
NA64 and LDMX measure the energy and momentum of the final state
electron. Therefore we want to show the most important steps in our
calculation of the covariant phase space for the elastic scattering
process in Fig.~\ref{fig:feyn_diag}b with $X_n = N$.

While a 4-body phase space has 12 degrees of freedom, 4 of them can be
reduced by the momentum-energy conservation, and another one, the
rotation along the axis of $\vec p_1 + \vec p_2$, is redundant. That
is, only 7 of them are necessary to express the squared amplitude, and
we obtain here:
 \begin{align} \label{eqn:LDMX_phas_space}
	\frac{d\Phi_4}{d\sXdark \, d q_2^2 } &=
	\frac{|J|}{16(2\pi)^6} \, \frac{d \sdark}{\sdark} \, d q_1^2
	\frac{\lambda^{1/2}(\sdark,m_\chi^2,m_\chi^2)}{\lambda^{1/2}(\sXdark,m_N^2,q_2^2) }\notag \\
	&\times d u_{2q} \left|\frac{\partial \phi_3^{\RXdark}}{\partial u_{2q}}\right|
	 \, \frac{d\Omega_\chi^{\Rdark}}{4\pi} .
\end{align}
The phase space can be described with the 5 independent Lorentz invariant integration variables $\sXdark=(p_3+p_\chi+p_{\bar\chi})^2$, $\sdark= q^2=(p_\chi+p_{\bar\chi})^2$ and $u_{2q}=p_2\cdot q$  as well as $q_1^2=(p_1-p_3)^2$ and $q_2^2=(p_2-p_4)^2$ with all momenta defined as in App.~\ref{sec:matrix-el}. The other 2 degrees of freedom come from $\Omega_\chi^{\Rdark}$, the solid angle of $\vec p_\chi$ in the CM frame of the $\chi \bar \chi$ pair (which will be integrated out). $\phi_3^{\RXdark}$ is the azimuthal angle of $p_3$ in the frame where $\vec p_3 + \vec p_\chi + \vec p_{\bar\chi} = 0$, $\lambda(a,b,c)=a^2+b^2+c^2-2(ab+ac+bc)$ is the triangle function. The Jacobian of the transformation from $E_4$ and $\cos\theta_4$ to the invariant variables $\sXdark$ and $q_2^2$ is given by
\begin{align}
\frac{\partial (E_4, \cos\theta_4) }{\partial (\sXdark, q_2^2)} =	\frac{|J|}{ |\vec p_4|} \equiv  \frac{\lambda^{-1/2} (s, m_N^2,m_e^2)}{2|\vec p_4|}  .
\end{align} 
The integration boundaries for the invariant masses $\sdark$ and $\sXdark$ are
\begin{align}
	\begin{matrix}
		\left(m_N+2m_\chi\right)^2 &\le & \sXdark &\le & \left(\sqrt{s}-m_e\right)^2 ,\\
		\left(2m_\chi\right)^2 &\le & \sdark &\le & \left(\sqrt{\sXdark}-m_N\right)^2 .
	\end{matrix} \label{eqn:inv_mass_boundaries}
\end{align}
The integration boundaries of the $t$-channel variables $q_1^2$ and $q_2^2$  are
\begin{subequations}
\begin{align}
	\left[q_1^2\right]^\pm &= 2m_N^2 - \frac{(\sXdark+m_N^2-q_2^2)(\sXdark+m_N^2-\sdark)}{2\sXdark} \notag \\ &\mp \frac{\lambda^{1/2}(\sXdark,m_N^2,q_2^2) \; \lambda^{1/2}(\sXdark,m_N^2,\sdark)}{2\sXdark},\\
	\left[q_2^2\right]^\pm &= 2m_e^2 - \frac{(s+m_e^2-m_N^2)(s+m_e^2-\sXdark)}{2s}\notag \\ 
	&\mp \frac{\lambda^{1/2}(s,m_e^2,m_N^2) \; \lambda^{1/2}(s,m_e^2,\sXdark)}{2s}. \label{eqn:q2_bound}\end{align}
\end{subequations}		
At last, the angular variable $u_{2q}$ is given by 
\begin{align}
u_{2q}  &=  \frac{ (p_1 \cdot p_2)G_2(p_1,q_2;\,q_2,q)  }{ -\Delta_2(p_1,q_2)} \notag \\
	&-\frac{  (q_2 \cdot p_2) G_2(p_1,q_2;\,p_1,q) }{ -\Delta_2(p_1,q_2)} \notag \\
	& +  \frac{ \sqrt{ \Delta_3 (p_1, q_2, p_2) \Delta_3 (p_1, q_2, q)  } }{ -\Delta_2(p_1,q_2)} \cos \phi_3^{\RXdark} \,, \label{eqn:boundaries_u23}
\end{align}
from which one can easily obtain the boundaries of $u_{2q}$ with $\phi_3^{\RXdark} \in \{0, \pi\}$, as well as the expression of 
$	\left|\frac{\partial \phi_3^{\RXdark}}{\partial u_{2q}}\right| $.
In (\ref{eqn:boundaries_u23}), $G_n$ is the Gram determinant of dimension $n$ and $\Delta_n$ the respective Cayley determinant (or symmetric Gram determinant)~\cite{Byckling:1969sx}.
The double differential cross section in the lab frame obviously follows from (\ref{eqn:LDMX_phas_space}) as
\begin{align} \label{eqn:diff_ldmx_cs}
	\frac{d\sigma}{dE_4 d\cos\theta_4} = \frac{|\vec p_4|}{4 E_2 m_N |\vec v_2|}  \int \frac{d\Phi_4}{d\sXdark dq_2^2} \frac{1}{|J|} |\mathcal M|^2.
\end{align}
For LDMX and NA64 where we are not interested in the direction of the $\chi$ pair, we can integrate out the  part $\Omega_{\chi}^{\Rdark}$ immediately, using the quantity defined below
\begin{subequations}
\begin{align}
	X_{\mu\nu}(q)
	&\equiv   \int \frac{d\Omega_{\chi}^{\Rdark}}{4\pi} \chi_{\mu\nu}(q) \\
	&= f(\sdark) \; \left(-g_{\mu\nu}+\frac{q_\mu q_\nu}{\sdark}\right) 
	\label{eqn:int_DM_piece}
\end{align}	
\end{subequations}
with $f(\sdark)$ found in the main text~(\ref{eqn:LDMX_dark_matter_part}).
This is because the rest of the squared amplitude is independent of $\Omega_{\chi}^{\Rdark}$.

\section{4-body phase space for mQ and BDX}
\label{sec:4-body-phase-bdx}

Since  mQ and BDX ought to detect the produced dark fermions directly via elastic scattering on nuclei and electrons, we are interested in the direction of one of the fermions --- denoted as $\theta_\chi$ below --- instead of the direction of the final state electron. 

Note that there are seven independent Lorentz invariants  for the full four-body phase space.  In this case, the phase space  can be given by 
\begin{align} \label{eqn:bdx_phase_space}
\begin{split}
\frac{ d\Phi_4 }{d \MCDE\, d\tBF } 	  & =   d \MDE \, dq^2_1 \, d\tAD \,\frac{ d\phi_3 }{ 2\pi}  \,   \frac{  d\phi_4}{ 2\pi }   \frac{|J| }{ (4\pi)^5 }   \\
  \times  \lambda^{-1/2} &(\MCDE^2, m_N^2, \tBF)\,  \lambda^{-1/2} (\MDE^2, m_N^2, \qBCF^2)   \,, 
\end{split} 
  \end{align}
where the Lorentz invariant integration variables are $\MCDE= (p_3+p_4+p_{\bar\chi})^2$, $\tBF=(p_2-p_\chi)^2$, $\MDE= (p_4+p_{\bar\chi})^2$,  $q_1^2 = (p_1-p_3)^2$ and $\tAD = (p_1-p_4)^2$,  as well as the four-vector $\qBCF= p_2 -p_3- p_\chi$, with $\qBCF^2= m_N^2 + m_X^2  + \MDE + \tAD - \MCDE -q_1^2$. Here we fix the value of the effective mass of hadronic final states $X_n$, denoted as $m_X$. The other two independent Lorentz invariants are chosen to be $p_2\cdot p_3$ and $p_3\cdot p_4$, which can be calculated as
\begin{align}
	 p_2\cdot p_3 &=  \frac{ (p_1\cdot p_2)\, G_2 (p_1, p_2-p_\chi; p_2-p_\chi, p_3) }{ -\Delta_2(p_1, p_2-p_\chi)} \notag\\ 
	&-\frac{  (m^2_e- p_2 \cdot p_\chi)  \,  G_2(p_1, p_2-p_\chi; p_1, p_3) }{ -\Delta_2(p_1, p_2-p_\chi)} \\ 
	&-  \frac{ \sqrt{ \Delta_3 (p_1, p_2-p_\chi , p_2) \Delta_3 (p_1, p_2-p_\chi , p_3)  } }{ -\Delta_2(p_1,p_2-p_\chi)} \cos\phi_3 \notag
\end{align}		
and 
\begin{align}
	p_3 \cdot p_4 &=  \frac{ (p_1\cdot p_3) \, G_2 (p_1, \qBCF ; \qBCF, p_4) }{ -\Delta_2(p_1, \qBCF )} \notag\\
	&- \frac{(\qBCF \cdot p_3) \, G_2(p_1, \qBCF; p_1, p_4) }{ -\Delta_2(p_1, \qBCF)} \\
	&- \frac{ \sqrt{ \Delta_3 (p_1, \qBCF , p_3) \Delta_3 (p_1,\qBCF , p_4)  } }{ -\Delta_2(p_1,\qBCF )} \cos\phi_4 \,, \notag
\end{align}
where $\phi_3$ and $\phi_4$ are the angle between the two planes
defined by ($\vec p_2- \vec p_\chi$, $ \vec p_2$) and
($\vec p_2- \vec p_\chi$, $ \vec p_3$) in the frame where
$\vec p_1 + \vec p_2 -\vec p_\chi =0$, and the angle between the two
planes defined by ($\vec \qBCF$, $ \vec p_3$) and ($\vec \qBCF $,
$ \vec p_4$) in the frame where $\vec p_1 + \vec \qBCF =0$,
respectively. Both scalar products are also functions of $\MCDE^2$,
$\tBF$, $\MDE^2$, $q^2_1$, and $\tAD$ defined above. Note that the
values of $\phi_3$ and $\phi_4$ remain the same after being
transformed into the lab frame.

The integration ranges of $\MCDE$ are given by 
\begin{equation}
	(m_X + m_e + m_\chi)^2 \le \MCDE \le (\sqrt{s}-m_\chi)^2,
\end{equation}
and 
that of $\tBF$ are 
\begin{align}
\left[\tBF\right]^\pm &=	m_e^2+m_\chi^2 -\frac{(s+m_e^2 - m_N^2 )(s+m_\chi^2 - \MCDE ) }{ 2s} \notag\\
&\mp  \frac{\lambda^{1/2}(s,  m_e^2,  m_N^2 ) \lambda^{1/2}(s, m_\chi^2, \MCDE) }{ 2s}\,.
\end{align}
For $ \MDE$, we have 
\begin{equation}
(m_4 + m_\chi)^2	\le  \MDE  \le  {(\sqrt{\MCDE^2} - m_X)^2}\, . 
\end{equation}
And the upper and lower bounds of $t_{13}$ are given by
\begin{align}
\left[q^2_1 \right]^\pm &= m_N^2   - \frac{(\MCDE + m_N^2 - \tBF )(\MCDE + m_X^2 - \MDE) }{ 2 \MCDE} \notag  \\ 
	&+ m_X^2 \mp  \frac{ \lambda^{1/2}(\MCDE, m_N^2, \tBF  ) \lambda^{1/2} (\MCDE, m_X^2, \MDE) }{ 2 \MCDE } \,,
\end{align}
and  the upper and lower bounds of $\tAD$ are 
\begin{align}
\left[\tAD\right]^\pm  &= m_N^2  - \frac{[\MDE  + m_N^2 -\qBCF^2] [\qBCF^2 +m_e^2 - m_\chi^2]  }{ 2 \MDE } \notag\\   
	&+m_e^2 \mp  \frac{  \lambda^{1/2}(\MDE, m_N^2,\qBCF^2) \,\lambda^{1/2}(\MDE , m_e^2, m_\chi^2) }{ 2 \MDE }  \,. 
\end{align}
At last, the integration ranges of $\phi_3$ and  $\phi_4$ are both from $0$ to $2\pi$.

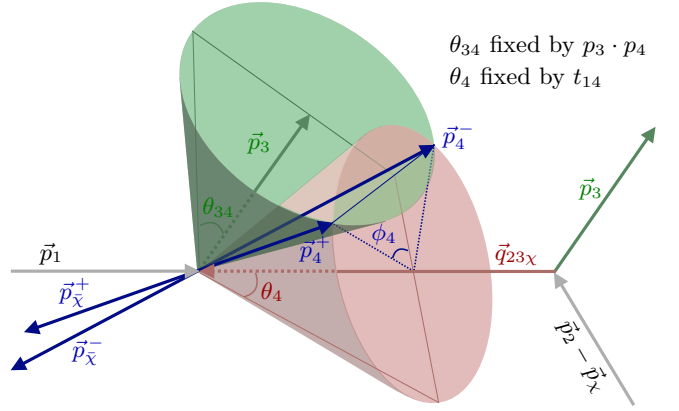
\begin{figure}[tb]
\centering
\setlength{\unitlength}{0.01\columnwidth}
\begin{picture}(100,65)
\put(0,0){\includegraphics[width=\columnwidth]{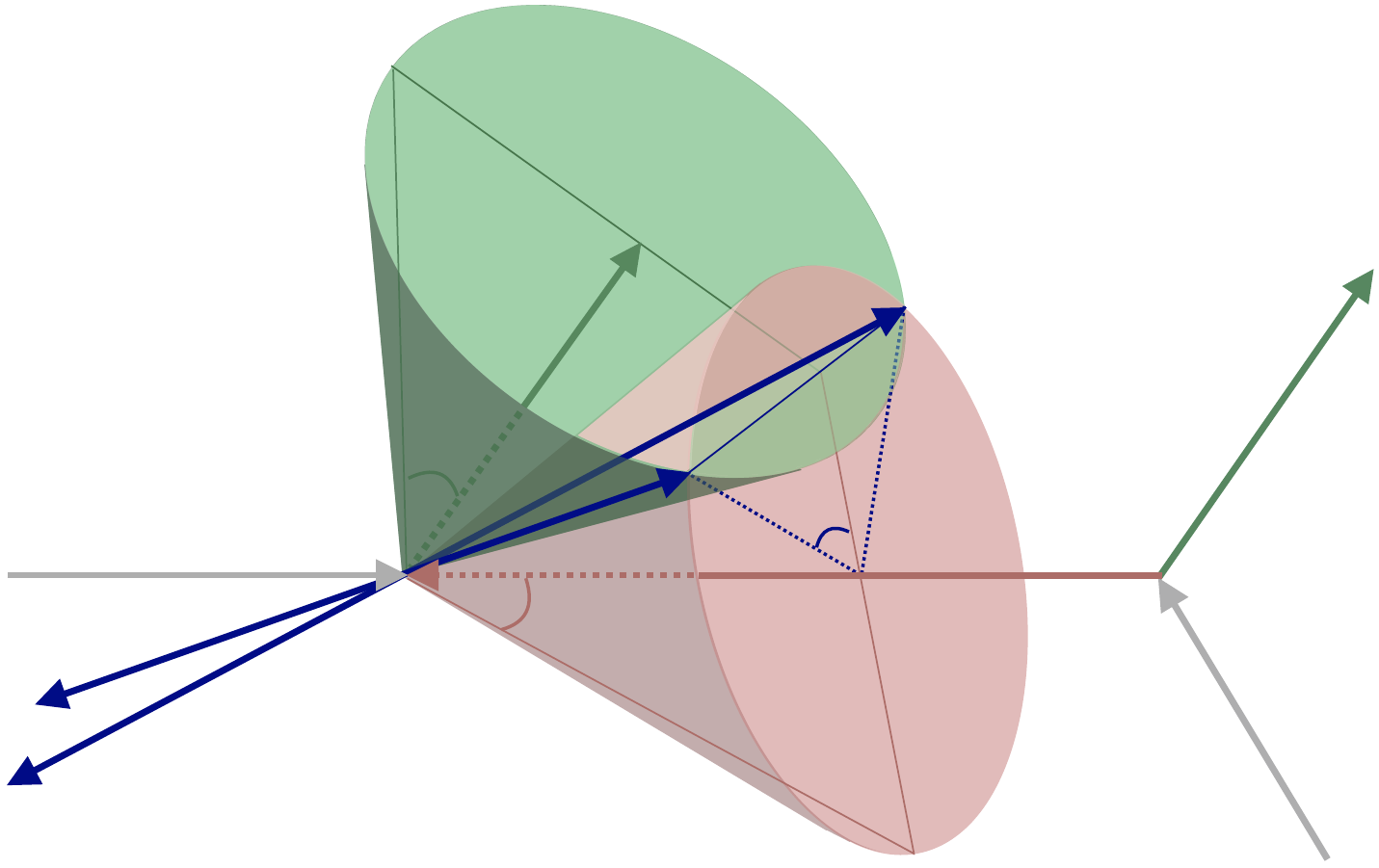}}
\put(68,55) {$\theta_{34}$ fixed by $p_3\cdot p_4$}
\put(68,50) {$\theta_4$ fixed by $t_{14}$}
\put(5,23) {$\vvec p_1$}
\put(84,13) {\rotatebox{-60}{$\vvec p_2 - \vvec p_\chi$}}
\put(8,17) {\textcolor{blue!70!black}{$\vvec p_{\bar\chi}^{\,+}$}}
\put(10,8) {\textcolor{blue!70!black}{$\vvec p_{\bar\chi}^{\,-}$}}
\put(45,23.5) {\textcolor{blue!70!black}{$\vvec p_{4}^{\,+}$}}
\put(67,41) {\textcolor{blue!70!black}{$\vvec p_{4}^{\,-}$}}
\put(56,26) {\textcolor{blue!70!black}{$\phi_4$}}
\put(88,33) {\textcolor{green!50!black}{$\vvec p_{3}$}}
\put(37,40) {\textcolor{green!50!black}{$\vvec p_{3}$}}
\put(30,30) {\textcolor{green!50!black}{$\theta_{34}$}}
\put(39,17) {\textcolor{red!60!black}{$\theta_{4}$}}
\put(75,23) {\textcolor{red!60!black}{$\vvec q_{23\chi}$}}
\end{picture}
\caption{Illustration of the solutions of $p_\chi\cdot p_{\bar\chi}$ to $\det[{\mathcal M}]=0$ in the frame where   $\vec p_1 + \vec \qBCF =0$. The direction of $\phi_4$, the angle between the two planes defined by ($\vec \qBCF$, $ \vec p_3$) and ($\vec \qBCF $, $ \vec p_4$),  breaks the degeneracy. }
\label{4ps:p5p6}
\end{figure}

Note that one Lorentz invariant needs to be given by the equation
$\det[{\mathcal M}]=0$, where the $(i,\,j)$ entry of the $5\times 5$
matrix ${\mathcal M}$ is the scalar product of $p_i$ and $p_j$. This
equality is due to the requirement that in four-dimensional
space-time, any five 4-vectors cannot be linearly independent. We use
this requirement to obtain the value of $(p_\chi\cdot p_{\bar\chi})$
from the seven Lorentz invariants above. Moreover, there are two
solutions of $p_\chi\cdot p_{\bar\chi}$ from $\det[{\mathcal
  M}]=0$. This can be understood in the frame where
$\vec p_1 + \vec \qBCF =0$, as shown in Fig.~\ref{4ps:p5p6}. The two
constraints derived from $t_{14}$ and $p_3\cdot p_4$, illustrated by
the circles at the bottom of the two cones, cannot uniquely fix
$\vec p_4$ (and thus $\vec p_{\bar\chi}$). Nevertheless, the
degeneracy can be broken by fixing the rotation direction of $\phi_4$,
as can be seen from Fig.~\ref{4ps:p5p6}. Here we take one solution of
$p_{\chi}\cdot p_{\bar\chi}$ for $\phi_4 \in [0,\pi)$ and the other
for $\phi_4 \in [\pi,2\pi)$. Other Lorentz invariants are not affected
by $\phi_4 \leftrightarrow 2\pi - \phi_4$.

In the lab frame, we have 
\begin{align} \label{eqn:BDX_cross_section}
\frac{ d\sigma  }{  d E_\chi \, d\cos\theta_{\chi}}   = \frac{|\vec p_\chi|}{4 E_2 m_N |\vec v_2|}  \int \frac{ d \Phi_4  }{   d \MCDE \, d\tBF  } \frac{1}{|J|}  |\mathcal M|^2,
\end{align}
where in the limit of $m_e\to 0$ the available range of $E_\chi$ is
\begin{subequations}\label{eq:inputmax}
\begin{align}
	E_\chi &\ge \frac{m_\chi (E_2 + m_N) }{ \sqrt{m_N(2E_2 + m_N)} },	\\
\begin{split}
E_\chi &\le \frac{(E_2 + m_N)} {2}\left(1 - \frac{m_X(2  m_\chi + m_X )} { m_N(2E_2 + m_N)} \right)\\
&+\frac{E_2 \sqrt{  m_N (2E_2 +m_N) - (2 m_\chi + m_X )^2}} {2 m_N(2E_2 + m_N)} \\
&~~~~~\times \sqrt{ m_N (2E_2 +m_N)   -m_X^2 },
\end{split}
\end{align}
and the corresponding range of $\cos\theta_\chi$ is 
\begin{align}
\begin{split}
1\ge &	\cos\theta_\chi \ge \max \left\{-1, \,\,\,\frac{E_\chi   -m_N   }{\sqrt{E_\chi^2 -m_\chi^2} }\right. \\
	&\left. +\frac{ m_N( 2E_\chi - m_N) + m_X(2m_\chi + m_X) }{2 E_2 \sqrt{E_\chi^2 -m_\chi^2} } \right\} \,.	
      \end{split}
\end{align}
\end{subequations}
It is straightforward to also take into account the contribution of
inelastic scattering between electron beam and the target by setting
$m_X$, or equivalently $s_X$, variable, and by introducing the inelastic
form factor of the target. Nevertheless, such a contribution is in
general sub-leading, so following \cite{Prinz:2001qz,
  Battaglieri:2016ggd} we neglect it here and take $m_X= m_N$ to
obtain our numerical results .

\section{Elastic scattering cross sections}
\label{sec:elast-scatt-cross}
Here, we present the full relativistic form of the differential recoil
cross section of a nucleus with charge $Z$, electric and magnetic form
factors $F_E$ and $F_M$, mass $m_N$ and spin $I_N$ when scattering on
a fermion $\chi$ with mass $m_\chi$ and energy $E_\chi$,
\begin{align}
\begin{split}
	\frac{d\sigma}{dE_R} 
		&=  \frac{1}{A}\bigg[g_E(E_R) \; \alpha Z^2 F_E^2(t) \\
		&+g_M(E_R) \; \frac{\mu_N^2m_N^2}{\pi} \frac{I_N+1}{3I_N} F_M^2(t)\bigg],	
\end{split}
\end{align}
where $A=\left(E_{\chi }^2-m_{\chi}^2\right) \left(2 m_N+E_R\right)$ and the functions $g_E(E_R)$ and $g_M(E_R)$ for all models are given by
\begin{subequations}
    \label{eq:el-cs}
\begin{align}
\begin{split} \label{eq:el_cs_mq}
 \text{$\epsilon$Q:\quad }    
	\frac{g_E}{\epsilon^2 e^2} &=  
		\frac{1}{2 E_R^2 }\left(2 E_{\chi }^2-m_N E_R-2 E_{\chi } E_R \right)		
		 \\
	\frac{g_M}{\epsilon^2 e^2} &=
		\frac{1}{4 m_N E_R} \left(2E_{\chi }^2-2m_{\chi }^2 + E_R^2\right)\\
		&\quad+ \frac{1}{4 m_N^2} \left(m_N^2-m_{\chi }^2- 2 m_N E_\chi \right),
\end{split} \\
\begin{split} \label{eq:el_cs_mdm}
 \text{MDM:\quad }    
	\frac{g_E}{\mu_\chi^2} 	&=
		\frac{ m_N }{2 E_R}	\left(4E_\chi^2-4m_\chi^2 + E_R^2\right) \\
		&\quad-\left(  m_\chi^2 + 2 m_N E_\chi \right) \\
	\frac{g_M}{\mu_\chi^2} 	&=
		 \frac{E_R}{2 m_N} \left(m_\chi^2-m_N^2 -2m_N E_\chi \right) \\
		 &\quad+\left(E_\chi^2+m_\chi^2\right),
\end{split} \\
\begin{split} \label{eq:el_csdm}
 \text{EDM:\quad }    
	\frac{g_E}{d_\chi^2}&=
		\frac{ m_N}{ 2 E_R} \left(E_R-2 E_\chi\right)^2 \\
	\frac{g_M}{d_\chi^2}&=
		- \frac{E_R}{2
   		 m_N }\left(m_N^2+m_\chi^2+2 m_N E_\chi\right) \\
		 &\quad +\left(E_{\chi }^2-m_{\chi }^2\right) ,
\end{split} \\
\begin{split} \label{eq:el_cs_am}
 \text{AM:\quad }    
	\frac{g_E}{a_\chi^2}&=
		-2 m_N E_R\left(m_N^2+m_\chi^2+2 m_N E_\chi\right) \\
		&\quad+ 4 m_N^2 \left(E_{\chi }^2-m_{\chi }^2\right) \\
	\frac{g_M}{a_\chi^2}&=
		m_N E_R  \left(2E_\chi^2+2m_{\chi }^2+E_R^2\right) \\
		&\quad+E_R^2
   		\left(m_N^2 +m_{\chi }^2-2 m_N E_\chi\right) ,
\end{split} \\
\begin{split} \label{eq:el_cs_cr}
 \text{CR:\quad }    
	\frac{g_E}{b_\chi^2}&= 
		2 m_N^2 \left(2 E_{\chi }^2-m_N E_R-2 E_{\chi } E_R\right)	 \\
	\frac{g_M}{b_\chi^2}&=
		 m_N E_R \left(2E_{\chi }^2-2m_{\chi }^2 + E_R^2\right) \\
		 &\quad+ E_R^2 \left(m_N^2-m_{\chi }^2-2 m_N E_{\chi }\right) .
\end{split}
\end{align}
\end{subequations}
where $t = 2m_NE_{R}$; the form factors $F_{E,M}(t)$ are given in
App.~\ref{sec:matrix-el}.  The expressions above also apply to
$\chi e^-$ scattering with $F_{E,M}=1$ and $Z=-1$ and for which the
pre-factors of the electric and magnetic pieces agree by setting
$I_N = 1/2$ and $\mu_N = \mu_B$.  For millicharged particle scattering
in the ultra relativistic limit, \textit{i.e.}~for $m_\chi \to 0$, we
recover the Rosenbluth formula. Finally, the maximal recoil energy
induced by $\chi$ carrying energy $E_\chi$ is given by
\begin{align}\label{eq:recoilmax}
	E_R^\text{max}= \frac{2m_i \left(E_\chi^2 -m_\chi^2 \right) }{ m_i (2 E_\chi + m_i ) +m_\chi^2 },
\end{align}
where $m_i=m_{N}$ or $m_e$. Equivalently, the minimal value of
$E_\chi$ to deposit energy $E_R$ in the target at rest is
\begin{align}
E_\chi^\text{min}=\frac{E_R }{ 2} + \frac{1}{ 2}\sqrt{\left(2 + \frac{E_R }{ m_i}\right)\left(2m_\chi^2 + E_R m_i\right)}.	
\end{align}

For completeness, we also list the squared scattering amplitudes
$\overline{|M_{\chi e}(q)|^2}$   on free electrons, averaged over initial states and summed over final states, as they
go into the definition of a reference cross section
$\bar \sigma_e$~(\ref{eq:elecRecoil}) for estimating the direct
detection limits. The leading terms are%
\footnote{For detectable electron recoils the typical momentum
  transfer is $q^2 \sim \alpha^2 m_e^2 \gg m_e^2 v^2$, so that we
  have neglected terms that scale as $m_e^2 v^2/q^2$.}
 \begin{subequations}
  \label{eq:elecDD}
\begin{align}
\label{eq:DDmdm} &\text{ MDM: }  & & \overline{|M_{\chi e}(q)|^2} =  {16\pi \mu_\chi^2 \alpha m_\chi^2 },\\
\label{eq:DDedm} &\text{ EDM: }  & & \overline{|M_{\chi e}(q)|^2} =\frac{64\pi d_\chi^2 \alpha m_e^2 m_\chi^2}{q^2},\\
\label{eq:DDam}  &\text{ AM:   }  & & \overline{|M_{\chi e}(q)|^2} =16\pi a_\chi^2 \alpha m_\chi^2 \, q^2 ,\\
\label{eq:DDcr}  &\text{ CR:    }  & & \overline{|M_{\chi e}(q)|^2} ={64\pi b_\chi^2 \alpha m_e^2 m_\chi^2}.
\end{align}
\end{subequations}

Finally, in the limit of small velocities the transport cross sections $\sigma^{\chi i}_T$ as they are
used in \eqref{eq:transport1} read,
 \begin{subequations}
  \label{eq:transportapp}
\begin{align}
\label{eq:TsPmdm} &\text{ MDM: }  & &\!\!\!\!\!\!\! \sigma^{\chi i}_T =  \frac{ {\mu_\chi}^2 \alpha \left[(2 m_i m_{\chi }+2 m_i^2+m_{\chi }^2 )F^{2}_E + 2 m_\chi^2 F^{2}_M \right]}{\left(m_i+m_{\chi }\right)^2},\\
\label{eq:TsPedm} &\text{ EDM: }  & & \!\!\!\!\!\!\! \sigma^{\chi i}_T = \frac{ 2 d^2_\chi \alpha  F_E^2 }{  v^{2} } ,\\
\label{eq:TsPam}  &\text{ AM:   }  & &\!\!\!\!\!\!\!  \sigma^{\chi i}_T = \frac{ 4 a_\chi^2 \alpha  m_i^2 m_\chi^2 \left[(m_i + m_\chi)^2 F_E^2 + 4m_\chi^2 F_M^2 \right]v^2 }{ 3 (m_i + m_\chi)^4 },\\
\label{eq:TsPcr}  &\text{ CR:    }  & &\!\!\!\!\!\!\!  \sigma^{\chi i}_T = \frac{4 b_\chi^2 \alpha  m_i^2 m_\chi^2 F_E^2 }{ (m_i + m_\chi)^2 }.
\end{align}
\end{subequations}
Here the momentum transfer of interest is, on average, at the same
order as the temperature $T$. Hence we use $F_E =1 $ and $F_M = -2.79$
for the proton, neglecting the dependence on $t$, as we are interested in $T=O(\keV)$ in this context.

\section{Photon Vacuum Polarization}
\label{sec:phot-vacu-polar}

Virtual $\chi$-loops contribute to the photon self-energy at
four-momentum $q^2$, 
\begin{align}
  i \Pi_{\mu\nu}(q)  = i (q^2 g_{\mu\nu} - q_{\mu} q_{\nu}) \Pi(q^2) .
\end{align}
The polarization function $\Pi(q^2) $ for the operators considered in
this paper is given by
\begin{align}
   \Pi(q^2) = \int_0^1 dx\, A(x,q^2) L_0 . 
\end{align}
In dimensional regularization the space-time dimension is written as
$d=4-\epsilon$ and 
$L_0 = 2/\epsilon + \log ( \tilde \mu / \Delta) $ with
$\Delta = m_{\chi}^2 - (1-x) x q^2$ and $\tilde\mu = 4\pi e^{-\gamma_E}\mu$
where $\mu$ is the renormalization scale; $\gamma_E = 0.577\dots$
is the Euler-Mascheroni constant. For the function $A(x,q^2)$ we find
 \begin{subequations}
  \label{eq:vacuumpolsdiv}
\begin{align} 
\label{eq:VPmQ}  &\text{ $\epsilon$Q:      }  & & - \frac{2\epsilon^2 \alpha}{\pi} (1-x) x ,\\
\label{eq:VPmdm} &\text{ MDM: }  & & - \frac{\mu^2_{\chi}}{4\pi^2}\left[  q^2 (1-x) x + m_{\chi}^2  \right] ,\\
\label{eq:VPedm} &\text{ EDM: }  & & - \frac{d^2_{\chi}}{4\pi^2} \left[   q^2 (1-x) x - m_{\chi}^2  \right],\\
\label{eq:VPam}  &\text{ AM:   }  & &  - \frac{a^2_{\chi}}{2\pi^2} q^2 \left[   q^2 (1-x) x - m_{\chi}^2  \right],\\
\label{eq:VPcr}  &\text{ CR:    }  & &- \frac{b^2_{\chi}}{2\pi^2} q^4   (1-x) x .
\end{align}
\end{subequations}
Keeping the finite correction, in the limit of $|q^2|/m_{\chi}^2\ll 1$
we find for the difference $\Pi(q^2)-\Pi(0)$,%
\footnote{We are at variance with the MDM and EDM expressions obtained
  in~\cite{Sigurdson:2004zp} which are, however, numerically
  of little relevance. }
 \begin{subequations}
\begin{align} 
\label{eq:VPmQ2}  &\text{ $\epsilon$Q:      }  & &  - \frac{\epsilon^2 \alpha}{15\pi } \frac{q^2}{m_{\chi}^2}  ,\\
\label{eq:VPmdm2} &\text{ MDM: }  & &  - \frac{\mu_\chi^2}{24 \pi ^2} q^2  \left[ 1 +  \log \left(\frac{\tilde\mu^2}{m_{\chi}^2}\right)\right], \\
\label{eq:VPedm2} &\text{ EDM: }  & &  + \frac{d_\chi^2}{24 \pi ^2} q^2  \left[ 1 -  \log \left(\frac{\tilde\mu^2}{m_{\chi}^2}\right)\right] , \\
\label{eq:VPam2}  &\text{ AM:   }  & & + \frac{a_\chi^2}{2 \pi ^2} q^2 m_{\chi}^2  \log \left(\frac{\tilde\mu^2}{m_{\chi}^2}\right), \\
\label{eq:VPcr2}  &\text{ CR:    }  & & -\frac{b_\chi^2}{12 \pi ^2} q^4 \log \left(\frac{\tilde\mu^2}{m_{\chi}^2}\right) .
\end{align}
\end{subequations}
In deriving the limit on the running of $\alpha$ in the main text, we
set $\tilde\mu=1\,\TeV$ and use the full (finite)
forms~(\ref{eq:vacuumpolsdiv}) that are also valid for
$|q^2|\gg m_{\chi}$. Also note that the finite part of $\Pi(0) = 0$
for AM and CR and hence those operators do not contribute to a
constant shift in the fine structure constant.

\section{Meson decays}
\label{sec:meson-decays-app}

The decay $K^+\to \pi^+ \chi\bar\chi$ is closely related to the
process of $K$-decay with emission of a charged lepton pair
$K^+\to \pi^+ l^+ l^- $ as both are accompanied by the emission of
$\gamma^{*}$ in the $s\to d$ transition. The decay width reads,
\begin{align}
    \label{eq:GammaKtopichibarchi}
  \Gamma_{K^+\to\pi^+\bar\chi\chi} &= \frac{G_F^2 \alpha  m_K^3 }{4 (4\pi)^6}
  \int_{4m\chi^2}^{(m_K-m_{\pi})^2} d\sdark\, \frac{f(\sdark)}{\sdark} \notag\\
  &\times \sqrt{ 1 - \frac{4 m_{\chi}^2}{\sdark}} \lambda^{3/2}\!\left(1, \frac{m_{\pi}^2}{m_K^2}, \frac{\sdark}{m_K^2}  \right)
  |f_V(\sdark)|^2
\end{align}
and the model dependence for the various interactions we consider is
entirely captured in the factor $f(\sdark)$, found
in~(\ref{eqn:LDMX_dark_matter_part}). 
For the kaon form factor $f_V$ we use~\cite{Batley:2009aa},
\begin{align}
  f_V(q^2) =  - 0.578  - 0.779 (q^2/m_K^2) . 
\end{align}
The decay $B^+ \to K^+ \chi \bar\chi$ is treated analogously, with the
obvious replacement of masses in (\ref{eq:GammaKtopichibarchi}) and
using instead the
form factor~\cite{Ball:2004ye},
  \begin{align}
    f_V(q^2) = \frac{0.161}{ 1- {q^2}/{(5.41\, \GeV)^2}} 
    + \frac{0.198}{ \left[ 1-{q^2}/{(5.41\, \GeV)^2} \right]^2 } .
  \end{align}

\section{Annihilation and self-interaction cross sections}
\label{sec:annih-cross-sect}

Here we collect, for completeness, all $2\to 2$ annihilation and self-scattering cross
sections in the non-relativistic velocity expansion.%
\footnote{In the calculation of the relic density of $\chi$-states, we
  use the full expression as a function of CM-energy $\sqrt{s}$ and
  follow~\cite{Gondolo:1990dk} for the thermal average and in the
  solution of the Boltzmann equation.} Annihilation into charged lepton
pairs $l^+ l^-$ is given by
\begin{align}
  \sigma_{\chi \bar \chi \to l^+ l^-} v  = B \frac{ \alpha }{ m_{\chi}^2} \left(1 + \frac{m_l^2}{2m_{\chi}^2}  \right) \sqrt{1- \frac{m_l^2}{m_{\chi}^2}} \,,
\end{align}
where the factor $B$ reads
 \begin{subequations}
  \label{eq:annlep}
\begin{align} 
\label{eq:AllmQ}  &\text{ $\epsilon$Q:      }  & &   \pi \epsilon^2 \alpha   ,\\
\label{eq:Allmdm} &\text{ MDM: }  & &  \mu_{\chi}^2 m_{\chi}^2  ,\\
\label{eq:Alledm} &\text{ EDM: }  & &  \frac{1}{12}   d_{\chi}^2 m_{\chi}^2 v^2 ,\\
\label{eq:Allam}  &\text{ AM:   }  & &  \frac{2}{3}   a_{\chi}^2 m_{\chi}^4  v^2 ,\\
\label{eq:Allcr}  &\text{ CR:    }  & & 4 b_{\chi}^2 m_{\chi}^4 .
\end{align}
\end{subequations}
Since we are interested in DM masses below few GeV and hence
freeze-out below (or at) the QCD phase-transition, we relate the
annihilation of hadronic final states to the experimentally measured
$R$-ratio,
\begin{align}
  \sigma_{\chi \bar \chi \to {\rm had }}(s)  =
  \sigma_{\chi \bar \chi \to \mu^+ \mu^-}(s)\times R(\sqrt{s}) ,
\end{align}
and we use the tabulated data from~\cite{Tanabashi:2018oca}.%
\footnote{The data starts at $\sqrt{s} = 0.3\,\GeV$, somewhat above
  the di-pion threshold. The intermediate regime can be accounted for
  by using the cross section data
  $e^+ e^- \to \pi^+\pi^-$~\cite{Ezhela:2003pp,Davier:2002dy}. We neglect
  this complication as it is of minor importance for our purposes.} 
The annihilation to photon-pairs is given by
\begin{align}
  \sigma_{\chi \bar \chi \to \gamma\gamma} v  = C /  m_{\chi}^2
\end{align}
with $C = \pi \epsilon^4 \alpha^2$ ($\epsilon Q$),
$\mu_{\chi}^4m_{\chi}^4/4\pi$ (MDM), $d_{\chi}^4m_{\chi}^4/4\pi$ (EDM). The
cross section is identically zero for CR and AM.

Finally,  for the self-scattering process $\chi\chi \to \chi \chi$, we adopt the viscosity cross section~\cite{Tulin:2013teo},  defined as 
\begin{align}
	\sigma_\text{SI}^{\chi\chi} &= \frac{1}{2}\int_{-1}^1 d\cos\theta \left(1-\cos^2\theta\right)\frac{d\sigma}{d\cos\theta}\,,
\end{align}
which, to leading order in the relative velocity, is
 \begin{subequations}
  \label{eq:self_scattering}
\begin{align} 
\label{eq:SImdm} &\text{ MDM: }  & & \sigma_\text{SI}^{\chi\chi}=  \frac{\mu_\chi ^4 m_\chi^2}{2 \pi } ,\\
\label{eq:SIedm} &\text{ EDM: }  & &  \sigma_\text{SI}^{\chi\chi}= \frac{d_\chi ^4 m_\chi^2}{4 \pi } ,\\
\label{eq:SIam}  &\text{ AM:   }  & &  \sigma_\text{SI}^{\chi\chi}= \frac{2a_\chi ^4 m_\chi^6 v^4}{15 \pi },\\
\label{eq:SIcr}  &\text{ CR:    }  & & \sigma_\text{SI}^{\chi\chi}= \frac{b_\chi ^4 m_\chi^6 v^4}{30 \pi } ,
\end{align}
\end{subequations}
where we do not show the self-scattering cross section for millicharged particles here, since it is formally divergent and depends on a cut-off of the interaction range, see \cite{Ackerman:mha, Feng:2009mn, McDermott:2010pa,Agrawal:2016quu}. %
For particle-antiparticle scattering  $\chi \bar \chi \to \chi \bar \chi$, we adopt the transport cross section,  defined as 
\begin{align}
	\sigma_\text{SI}^{\chi\bar\chi} &= \int_{-1}^1 d\cos\theta \left(1-\cos\theta\right)\frac{d\sigma}{d\cos\theta}\,,
\end{align}
which, to leading order in the relative velocity, is
 \begin{subequations}
  \label{eq:self_scattering_anti}
\begin{align} 
\label{eq:SIAmdm} &\text{ MDM: }  & & \sigma_\text{SI}^{\chi\bar\chi} =  \frac{7\mu_\chi ^4 m_\chi^2}{4 \pi } ,\\
\label{eq:SIAedm} &\text{ EDM: }  & &  \sigma_\text{SI}^{\chi\bar\chi} = \frac{d_\chi ^4 m_\chi^2}{4 \pi } ,\\
\label{eq:SIAam}  &\text{ AM:   }  & & \sigma_\text{SI}^{\chi\bar\chi} = \frac{a_\chi ^4 m_\chi^6 v^4}{2 \pi }  ,\\
\label{eq:SIAcr}  &\text{ CR:    }  & & \sigma_\text{SI}^{\chi\bar\chi} = \frac{12 b_\chi ^4 m_\chi^6}{\pi } .
\end{align}
\end{subequations}

\bibliography{refs}

\end{document}